\documentclass[
reprint,
superscriptaddress,
amsmath,
amssymb,
aps,
pra]{revtex4-2}

\usepackage{hyperref}
\usepackage{float}
\usepackage{cleveref}
\usepackage{graphicx}
\usepackage{xcolor}

\begin{document}

\title{Tensor networks for quantum computing}

\author{Aleksandr Berezutskii}
\thanks{Equal contribution}
\email{albe@terraquantum.swiss}
\affiliation{Terra Quantum AG, Kornhausstrasse 25, St. Gallen, 9000, Switzerland}
\affiliation{Institut Quantique \& D\'epartement de Physique, Universit\'e de Sherbrooke, Sherbrooke, QC J1K 2R1, Canada}

\author{Minzhao Liu}
\thanks{Equal contribution}
\email{minzhao.liu@jpmchase.com}
\affiliation{Global Technology Applied Research, JPMorganChase, New York, NY 10001, USA}

\author{Atithi Acharya}
\affiliation{Global Technology Applied Research, JPMorganChase, New York, NY 10001, USA}

\author{Roman Ellerbrock}
\affiliation{Terra Quantum AG, Kornhausstrasse 25, St. Gallen, 9000, Switzerland}

\author{Johnnie Gray}
\affiliation{Division of Chemistry and Chemical Engineering, California Institute of Technology, Pasadena, CA 91125, USA}

\author{Reza Haghshenas}
\affiliation{Quantinuum, Broomfield, CO 80021, USA}

\author{Zichang He}
\affiliation{Global Technology Applied Research, JPMorganChase, New York, NY 10001, USA}

\author{Abid Khan}
\affiliation{Global Technology Applied Research, JPMorganChase, New York, NY 10001, USA}

\author{Viacheslav Kuzmin}
\affiliation{Terra Quantum AG, Kornhausstrasse 25, St. Gallen, 9000, Switzerland}

\author{Dmitry Lyakh}
\affiliation{NVIDIA Corporation, 2788 San Tomas Expressway, Santa Clara, CA 95051, USA}

\author{Danylo Lykov}
\affiliation{NVIDIA Corporation, 2788 San Tomas Expressway, Santa Clara, CA 95051, USA}

\author{Salvatore Mandr\`a}
\affiliation{Google Quantum AI, Venice, CA 90291, USA}
\affiliation{Quantum Artificial Intelligence Laboratory, NASA Ames Research Center, Moffett Field, CA 94035, USA}
\affiliation{KBR Inc., Houston, TX 77002, USA}

\author{Christopher Mansell}
\affiliation{Terra Quantum AG, Kornhausstrasse 25, St. Gallen, 9000, Switzerland}

\author{Alexey Melnikov}
\affiliation{Terra Quantum AG, Kornhausstrasse 25, St. Gallen, 9000, Switzerland}

\author{Artem Melnikov}
\affiliation{Terra Quantum AG, Kornhausstrasse 25, St. Gallen, 9000, Switzerland}

\author{Vladimir Mironov}
\affiliation{Terra Quantum AG, Kornhausstrasse 25, St. Gallen, 9000, Switzerland}

\author{Dmitry Morozov}
\affiliation{Terra Quantum AG, Kornhausstrasse 25, St. Gallen, 9000, Switzerland}
\affiliation{Nanoscience Center and Department of Chemistry, University of Jyväskylä, Jyväskylä, 40014, Finland}

\author{Florian Neukart}
\affiliation{Terra Quantum AG, Kornhausstrasse 25, St. Gallen, 9000, Switzerland}

\author{Alberto Nocera}
\affiliation{Department of Physics and Astronomy and Quantum Matter Institute, The University of British Columbia, Vancouver, BC V6T 1Z4, Canada}

\author{Michael A. Perlin}
\affiliation{Global Technology Applied Research, JPMorganChase, New York, NY 10001, USA}

\author{Michael Perelshtein}
\affiliation{Terra Quantum AG, Kornhausstrasse 25, St. Gallen, 9000, Switzerland}

\author{Matthew Steinberg}
\affiliation{Global Technology Applied Research, JPMorganChase, New York, NY 10001, USA}

\author{Ruslan Shaydulin}
\affiliation{Global Technology Applied Research, JPMorganChase, New York, NY 10001, USA}

\author{Benjamin Villalonga}
\affiliation{Google Quantum AI, Venice, CA 90291, USA}

\author{Markus Pflitsch}
\affiliation{Terra Quantum AG, Kornhausstrasse 25, St. Gallen, 9000, Switzerland}

\author{Marco Pistoia}
\affiliation{Global Technology Applied Research, JPMorganChase, New York, NY 10001, USA}

\author{Valerii Vinokur}
\affiliation{Terra Quantum AG, Kornhausstrasse 25, St. Gallen, 9000, Switzerland}

\author{Yuri Alexeev}
\email{yalexeev@nvidia.com}
\affiliation{NVIDIA Corporation, 2788 San Tomas Expressway, Santa Clara, CA 95051, USA}

\begin{abstract}
In the rapidly evolving field of quantum computing, tensor networks serve as an important tool due to their multifaceted utility. In this paper, we review the diverse applications of tensor networks and show that they are an important instrument for quantum computing. Specifically, we summarize the application of tensor networks in various domains of quantum computing, including simulation of quantum computation, quantum circuit synthesis, quantum error correction and mitigation, and quantum machine learning. Finally, we provide an outlook on the opportunities and the challenges of the tensor-network techniques.
\end{abstract}

\maketitle

\tableofcontents

\section{Introduction}

\begin{table*}[ht]
    \centering
    \footnotesize
    \renewcommand{\arraystretch}{1.5}
    \begin{tabular}{|p{2.5cm}|p{3.5cm}|p{3.5cm}|p{4cm}|p{3.5cm}|}
    \hline
    \textbf{Field} & \textbf{Methods} & \textbf{Applications} & \textbf{Advantages} & \textbf{Challenges} \\
    \hline
    Simulation of quantum computation  (Section~\ref{sec:simulation}) & 
    Unstructured TNs, matrix product states (MPS), projected entangled pair states (PEPS), tree tensor network states (TTNS), multi-scale entanglement renormalization ansatz (MERA), density matrix renormalization group (DMRG), various renormalization methods & 
    Simulation of quantum circuits, analog quantum processors, boson sampling, quantum algorithm benchmarking, quantum many-body physics simulation &
    Simulation complexity reduction, efficiency in representation of some quantum states &
    Representing highly-entangled quantum states, simulating long-time dynamics, simulating deep quantum circuits \\
    \hline
    Quantum circuit synthesis (QCS)  (Section~\ref{sec:qcs}) & 
    Encoding of TN quantum states into quantum circuits, realizing TN-inspired quantum circuits & 
    Preparation of quantum states relevant for quantum computation and quantum simulation &
    Efficiency and interpretability in quantum state preparation & 
    Requirement of many-qubit gates whose number normally scales with the bond dimension \\
    \hline
    Quantum error correction (QEC) and mitigation (QEM) (Section~\ref{sec:qec}) & 
    Tensor-network decoders, TN inspired quantum error-correcting codes & 
    Tensor-network decoders of error-correcting codes &
    Efficiency\,and interpretability in decoding, interpretability in creating and studying error-correcting codes, reduction in error mitigation overhead & 
    Computational resources overhead, speed, lack of open-source software \\
    \hline
    Quantum machine learning (QML)   (Section~\ref{sec:qml}) & 
    Variational tensor-network circuits as quantum neural networks (QNNs) & 
    Generative and discriminative models, quantum data classification, quantum data encoding &
    Analytical interpretability, qubit efficiency, noise resilience, absence of barren plateaus, potential reduction in parameters number, avoiding costly TN contraction on classical computer &
    Requirement for nonlocal interactions in quantum hardware for many TN circuits, lack of theoretical guarantees for advantage \\
    
    \hline
    \end{tabular}
    \vspace{5pt}
    \caption{Summary of tensor network applications in quantum computing.}
    \label{tab:general}
\end{table*}

Tensor networks (TNs) have become a useful tool in many areas of physical and mathematical sciences, especially in the field of quantum information science. The interest in quantum computing (QC) has driven a lot of the development in TNs because they are used to represent and manipulate quantum states and processes.

TNs were initially applied to quantum many-body simulations~\cite{schollwock2011density}, for which they offer substantial advantages over alternative methods for simulating weakly coupled quantum systems and quantum systems with significant locality~\cite{orus2019tensor}. Over time, TNs have broadened their application scope to quantum information theory and quantum chemistry~\cite{orus2014practical,orus2019tensor,chubb2021general}, and they have become indispensable to the QC field. In particular, TNs are employed to simulate real quantum experiments that were previously believed to be beyond the capability of classical computers~\cite{pan2022simulation,fu2024achieving,oh2024classical, tindall2024efficient}. This demonstrates that TNs successfully address the so-called \textit{curse of dimensionality}---the problem that the size of the state space increases exponentially with the number of degrees of freedom.

Additionally, TNs provide a general framework of analyzing mathematical objects frequently encountered in quantum information science, which makes them attractive tools to address many other challenges faced by QC. Table\,\ref{tab:general} provides an overview of TN applications in QC that we will discuss in this paper and spans four subdomains: simulation of quantum computation, quantum circuit synthesis (QCS), quantum error correction (QEC) and mitigation (QEM), and quantum machine learning (QML). The Table \ref{tab:general} outlines the specific methods used, their primary applications, key advantages, and the challenges they encounter.

We structure the rest of the paper as follows. Section \ref{sec:methods} briefly introduces tensor networks, key building blocks, and commonly used methods. Sections \ref{sec:simulation}, \ref{sec:qcs}, \ref{sec:qec} and \ref{sec:qml} discuss the aforementioned domains of application. Finally, we conclude the paper with discussions on the overall advantages of tensor networks, and provide an outlook of the future on how tensor networks may benefit quantum computing in Section \ref{sec:discussion}.

\section{Tensor-network methods}\label{sec:methods}

\begin{figure*}[t]
    \centering
    \includegraphics[width=0.7\linewidth]{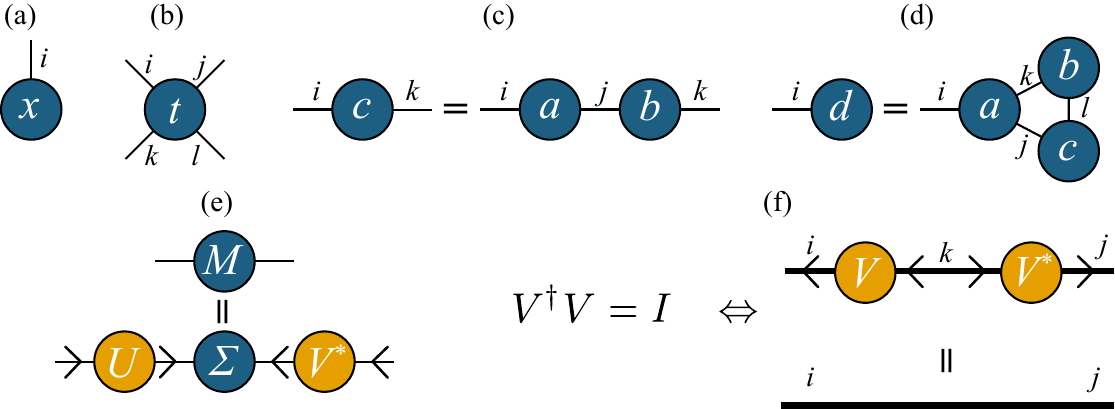}
    \caption{Graphical representations of (a) a vector, (b) a 4th order tensor, (c) matrix multiplication, (d) \Cref{eq:toy_tn}, (e) SVD of a tensor, and (f) the definition of isometry.}
    \label{fig:graphical_representation}
\end{figure*}

Tensors are mathematical objects that describe multilinear relationships between other objects. They can be commonly thought of as multidimensional arrays of complex numbers, where the numerical values of the arrays are coefficients describing the relationships. Each index of a tensor indicates mapping to or from an object, and describing the relationship between $m$ objects requires $m$ indices, resulting in a so-called $m$-th order tensor. For example, a vector $(x_1, \dots, x_n)$ can be compactly represented in index notation as $x_i$ with $i\in(1,\dots, n)$, and is a first-order tensor. A matrix is thus a second-order tensor $x_{i,j}$ with two indices, among other examples.

One can compose the multilinear relationships through tensor contraction. For example, if a tensor $a_{i,j}$ describes the relationship between the objects corresponding to the $i,j$ indices, and a tensor $b_{j,k}$ for those corresponding to the $j,k$ indices, then the resulting tensor describes the relationship between objects corresponding to the $i,k$ indices. This can be described as
\begin{equation}
    c_{i,k} = \sum_{j} a_{i,j} b_{j,k},
\end{equation}
where the $j$ index runs over all possible values. We call this \textit{contraction} of tensors $a$, $b$ at the $j$ index. Matrix multiplication is thus a particular case of contraction between two 2nd-order tensors at one shared index. In general, we could have arbitrary contraction between tensors such as
\begin{equation}
    d_{i} = \sum_{j,k,l} a_{i,j,k} b_{k,l} c_{j,l}.
    \label{eq:toy_tn}
\end{equation}
In this example, a 1st-order tensor $d$ is obtained by contracting a 3rd-order tensor $a$ with two 2nd-order tensors $b$ and $c$.

Tensors and operations upon them are often represented using visual diagrams. An $m$-th order tensor is represented as a node with $m$ edges emerging from the node, each representing an index. Sometimes, the shape and direction of an edge may denote specific properties of the tensor or its indices.

Contraction between two tensors at index $j$ is represented by joining two nodes at the shared edge corresponding to index $j$. \Cref{fig:graphical_representation} provides examples of graphical representations. Since representations of complex contractions between multiple tensors are a network of connected nodes like in \Cref{fig:graphical_representation}(d), such mathematical expressions are referred to as \textit{Tensor Networks}.

Besides composing tensors through contraction, one can also decompose them. This can be done using, for example, the singular value decomposition (SVD).
For any matrix $M$, the SVD yields
\begin{equation}
    M=U\Sigma V^\dag,
\end{equation}
where $\Sigma$ is a diagonal matrix of the singular values and could be absorbed into $U$ or $V$, and $U,V$ are isometries ($U^\dag U=I,V^\dag V=I$). This decomposition presents an opportunity for approximate representation of the original tensor by trimming the singular values, either by keeping only the $k$ largest singular values or discarding singular values smaller than some threshold. Furthermore, the Eckart–Young theorem \cite{eckart1936approximation} states that for approximations with a fixed rank, the solution provided by SVD is optimal. There exist other matrix decompositions which we do not discuss for the sake of brevity~\cite{stewart1998matrix}.

Graphically, the SVD decomposition is shown in \Cref{fig:graphical_representation}(e). There, isometries are represented as orangle nodes with directional bonds shown in \Cref{fig:graphical_representation}(f). For isometries $U$, although $U^\dagger U=I$, but $UU^\dagger\neq I$ unless $U$ is also unitary. Therefore, the arrow notation is used to differentiate the indices of the isometry. Unitaries, being a special case of isometries, are represented as magenta nodes with directional bonds as shown in \Cref{fig:tns}.

\subsection{Common ansatzes}

\begin{figure*}[ht]
    \centering
    \includegraphics[width=1.\linewidth]{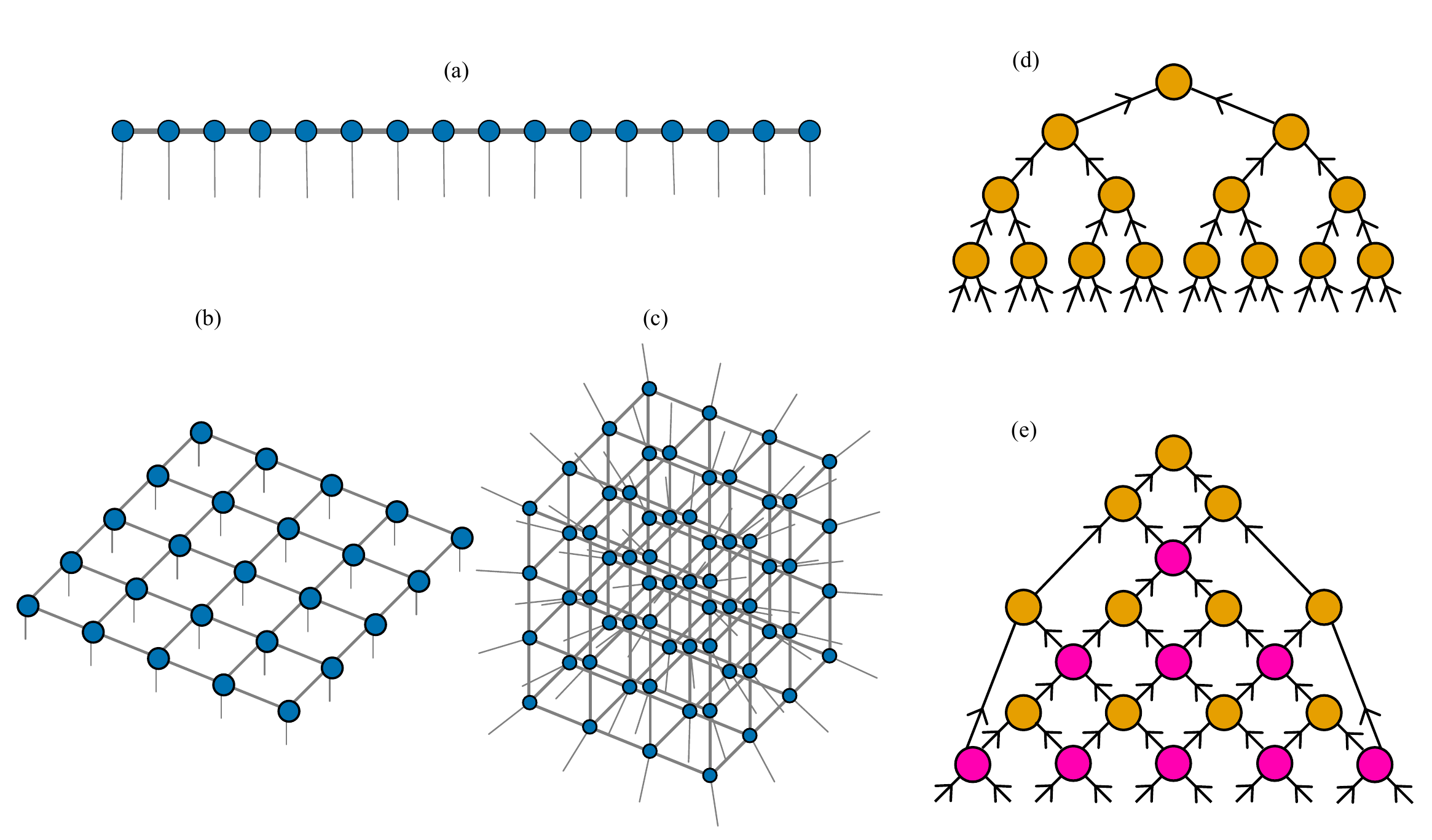}
    \caption{Different types of TNs: (a) the matrix product state, (b, c) the projected entangled pair states in 2D and 3D, (d) the tree tensor network state, and (d) the multiscale entanglement renormalization ansatz. The orange circles represent isometric tensors and the magenta circles represent unitary disentanglers.}
    \label{fig:tns}
\end{figure*}

The matrix product state (MPS) is the simplest yet the most widely used ansatz (in this context, this term refers to the structure as well as constraints on the TN's tensors). In general, the amplitudes of an $N$-body quantum wavefunction form an $N$th-order tensor, the MPS provides an approximate representation of this tensor as
\begin{equation}\label{eq:MPS}
c_{i_1,\dots,i_N}=\sum_{\alpha_1,\dots,\alpha_{N}=1}^{\chi}\Gamma_{\alpha1}^{[1]i_1}\Gamma_{\alpha_1\alpha2}^{[2]i_2}\dots\Gamma_{\alpha_{N-1}}^{[N]i_N},
\end{equation}
which is a contraction of the $\Gamma_i$ tensors, each corresponding to a site in the system as shown in Fig.\,\ref{fig:tns}~(a). Here, $\chi$ is the parameter called the bond dimension which controls the accuracy of compression. Closely related to MPS is the matrix product operator (MPO) concept, which uses a very similar representation for many-body operators instead of vectors (which is the case for the MPS). This is accomplished by adding an additional dual index to each $\Gamma_i$ tensor.

This TN is especially effective for simulating one-dimensional (1D) quantum systems. It follows the area law of entanglement\,\cite{verstraete2006matrix} (the entanglement between two parts of a many-body systems growing at most as the size of the boundary), but has been used for other systems as well \cite{banuls2023tensor, oh2024classical}. Its contraction cost is polynomial in the system size, meaning the physical observables can be evaluated efficiently. This is because local physical observables can be expressed as a tensor network and can be evaluated by contracting it. The concept of MPS has independently emerged in computational mathematics, where it is called the tensor train\,\cite{oseledets2011tensor}.

Projected entangled pair states (PEPS), shown in Fig.\,\ref{fig:tns}~(b) and Fig.\,\ref{fig:tns}~(c), aim at generalizing the 1D locality of the MPS to higher dimensions, making them capable of reproducing large entanglement and suitable for simulating high dimensional quantum systems. However, algorithms for PEPS are typically much more computationally demanding \cite{orus2014practical, schuch2007computational}.

Tree Tensor Networks States (TTNS)\cite{vidal2002ttn, shi2006ttn}, shown in Fig.\,\ref{fig:tns}~(d) generalize the MPS from local to nonlocal correlations. MPS can be understood as the special case of a TTNS with a maximally unbalanced tree topology. Balanced tree topologies provide a worst-case correlation length that is logarithmic in the number of leaves and thereby allow to capture long-range interactions. The scaling of TTNS
is comparable to that of MPS\cite{shi2006ttn}, however, the rank required for accuratly describing the state might scale rapidly. TTNS have been independently developed in the chemical physics community under the name multilayer multiconfigurational Hartree\cite{wang2003ttn}.

The multiscale entanglement renormalizaton ansatz (MERA), shown in Fig.\,\ref{fig:tns}~(e), is designed to capture the entanglement structure of quantum systems across different length scales and can reproduce logarithmic violation of the area law.

The above TNs can also be equipped with the periodic boundary conditions. Furthermore, they can be generalized for providing the description of operators such as Hamiltonians and quantum channels. One can even define TNs with arbitrary geometries for special applications. For example, the quantum state of a quantum circuit with single- and two-qubit gates can be represented by a network reflecting the connectivity~\cite{markov2008simulating}. Consequently, the circuits with gates laid out in a fixed geometric pattern (e.g., a 2D grid) may be easier to simulate~\cite{markov2008simulating, tindall2024efficient, decross2024computational}.

\subsection{Manipulation of TNs}

While TNs can be used to represent a broad range of mathematical objects in quantum information science, their popularity stems primarily from their power as a tool for obtaining meaningful, interpretable quantities such as operator expectation values, reduced density matrices, amplitudes, samples, and more. Due to the high interest in the TN techniques, a plethora of software libraries suitable for different types of TN manipulations are available, such as cuQuantum\,\cite{cuquantum}, quimb\,\cite{gray2018quimb}, ITensor\,\cite{fishman2022itensor}, and many more\,\cite{gray2021hyper, javadi2024quantum, pastaq, hauschild2018efficient, alvarez2009density, rams2024yastn, luo2020yao, brennan2022qxtools, strano2024qrack, zhang2023tensorcircuit, lykov2021qtensor, zhang2019alibaba, villalonga2020establishing, salvatore2021hybridq, lyakh2022exatn, zhai2023block2}.

\subsubsection{Update}

One of the important tasks is optimizing a TN with respect to a cost function, such as the energy of a quantum state. A pioneering approach to this problem was introduced by Steven White in 1992 in which the density matrix renormalization group (DMRG)\,\cite{white1992density} is used. For a cost function defined by a Hamiltonian on a lattice, DMRG is a variational optimization technique that finds the best MPS approximation of the many-body wavefunction of the ground state. This algorithm performs global optimization by sequentially optimizing each local tensor (or pairs of them, depending on the particular version of the algorithm), which is repeated across several sweeps.

In addition to the DMRG algorithm, gradient-based optimization methods can also be employed if the gradient of the objective function (e.g., energy) is accessible, for instance, through automatic differentiation\,\cite{wang2023tensor,liao2019differentiable}. Additionally, Riemannian optimization \cite{hauru2021riemannian, luchnikov2021qgopt} can be particularly helpful when one needs to maintain the isometric properties of the TN \cite{berezutskii2025simulating, luchnikov2021riemannian}.

Another class of methods relies on Monte Carlo sampling. In the variational Monte Carlo (VMC) approach, observables are evaluated by sampling configurations of the many body system instead of exact calculations~\cite{sandvik2007variational, wang2011monte}.

One may also consider the time evolution of a quantum state or an operator in the TN representation. For discrete time steps, this can be done by applying local gates or Kraus operators through contraction, and restore the original form of the TN with SVD. These approaches are called time-evolving block decimation (TEBD). Continuous time evolution under a Hamiltonian can be approximated by discrete steps and the discrete method can be applied if the Hamiltonian is sufficiently short ranged. For long-range Hamiltonians, TEBD becomes inefficient, and methods such as MPO $W^{II}$, TDVP, and Krylov approaches are preferred; see \cite{paeckel2019time} for a recent review. Krylov methods use Lanczos diagonalization (an approximate method) to compute the action of the time evolution operator, while TDVP evolves the state directly on the MPS manifold for fixed bond dimension\,\cite{feiguin2005time, alvarez2011time, haegeman2011time, haegeman2016unifying}.

\subsubsection{Contraction}

A crucial part of many TN algorithms is the \emph{contraction} -- the evaluation of a single scalar or tensor represented by the network. Naively, the cost of the contraction
scales exponentially in the number of indices. In practice, the contraction can be performed by using a sequence of intermediate tensors using pairwise contractions, known as an ordering or a contraction tree. The \emph{optimal} contraction cost of this approach still generally scales exponentially~\cite{markov2008simulating}, but the approach nonetheless offers a dramatic reduction in the cost at the expense of some intermediate memory. Notably, certain tree- and fractal-like geometries can be contracted with only polynomial cost\,\cite{ran2020tensor}. 

Finding the optimal contraction scheme is itself an NP-complete problem\,\cite{arnborg1987complexity, pfeifer2014faster}. However, much progress has recently been made using heuristic approaches targeting the \emph{total} cost of the contraction operation. Specifically, these methods have been based on recursive graph partitioning\,\cite{kourtis2019fast, gray2021hyper, pan2021simulating}, simulated annealing~\cite{kalachev2021classical, morvan2023phase} and now reinforcement learning~\cite{meirom2022optimizing, liu2023classical}. The extraordinary sensitivity of the cost to the contraction tree quality has led to improvements by many orders of magnitude for some problems such as, for example, quantum circuit simulation \cite{kalachev2021multi,gray2021hyper,huang2020classical}.

For exact simulation of quantum circuits, two additional techniques are central to achieving a state-of-the-art performance. First, whilst the time cost of a contraction might be acceptable, the space cost in terms of intermediate memory might still be enormous.\,\emph{Slicing} (also known as cutting or projecting) splits a contraction into many smaller independent contractions~\cite{aaronson2016complexity, chen2018classical, markov2018quantum, villalonga2019flexible, pednault2017pareto}, each of which for example can fit on a GPU and be performed in a massively parallel way~\cite{huang2021efficient, kalachev2021classical}. Depending on the geometry, this can sometimes be performed with very little overhead. Secondly, if one wants to evaluate many related tensor networks differing only in some entries (for example, a set of amplitudes of different output basis states of a single quantum circuit), one can employ \emph{multi-contraction}~\cite{kalachev2021multi, pan2022simulation} to cleverly avoid repeating the same computation. The largest exact simulations of quantum circuits use both of these techniques in tandem~\cite{pan2022simulation, pan2022solving, liu2024verifying, zhao2024leapfrogging}.

To go beyond exactly treatable network sizes, one must use approximate contraction. While even this is not expected to be viable in the general case~\cite{roth1996hardness}, evidence obtained from simulating many-body physics systems shows that many real world tensor networks are tractable approximately~\cite{nishino1996corner, levin2007tensor, xie2012coarse, evenbly2015tensor, chen2024sign, jiang2024positive}. Time evolved MPS~\cite{vidal2008class, haegeman2016unifying} and PEPS~\cite{verstraete2004renormalization, jiang2008accurate, corboz2010simulation, lubasch2014unifying, tindall2024efficient} can be thought of as approximate contraction, in which case one is limited by buildup of entanglement. These methods have historically been handcrafted, but recent work has focused on automatic approximation~\cite{pan2020contracting, chubb2021general, ma2024approximate} and contraction sequence-optimized\,\cite{gray2024hyperoptimized} approximate contraction. An outstanding question is in which classes of quantum circuits do such techniques admit a polynomial or exponential cost reduction~\cite{zhou2020limits}.

\section{Simulation of quantum computation}\label{sec:simulation}

Many quantum objects—statevectors, operators, channels, and others—can be represented as tensors. Given that quantities of interest often derive from these objects, TNs are often useful for simulating quantum systems.

\subsection{Gate-based quantum computation}

TNs can be used for simulating quantum algorithms in the gate-based model. One of the approaches includes evolving an ansatz, such as an MPS or PEPS, using TEBD-like techniques. Alternatively, a quantum state can be represented as a TN of contracted circuit gates with fixed input indices and open output indices. By contracting appropriate TNs constructed from the state network, amplitudes and expectation values can be computed. Another example is the evaluation of the trace of quantum circuits, which enables the estimation of the circuit ensemble randomness\,\cite{liu2022estimating}.

A good example is the simulation of quantum circuits used to simulate many-body physics, such as the experiment on the dynamics of a kicked Ising model~\cite{kim2023evidence}. While the problem was considered initially intractable using state-vector MPS and isometric-TN\,\cite{zaletel2020isometric} approaches, it was later demonstrated that many other TN methods \cite{tindall2023gauging,anand2023classical,liao2019differentiable,patra2024efficient,rudolph2023classical} can solve the same problem more efficiently, and can even produce more accurate results than the quantum processor itself.

Another important class of experiments suitable for TN techniques is random circuit sampling (RCS). RCS is a computational task where a quantum computer executes a random $n$-qubit quantum circuit and output the measurement results (a length-$n$ bitstring) of the resulting quantum state. This task is classically hard because the existence of an efficient classical algorithm for estimating the probability of a given output would lead to the collapse of the polynomial hierarchy\,\cite{bouland2022noise, krovi2022average}, which theoretical computer scientists believe to be highly unlikely. As a result, RCS has been experimentally demonstrated many times as a proof of quantum computational advantage\,\cite{arute2019quantum, wu2021strong, zhu2022quantum, morvan2023phase, decross2024computational}. Additionally, applications based on RCS in which classical simulation using TNs constitute an integral part have been proposed\,\cite{aaronson2023certified,liu2025certified,amer2024certified,amer2025applications}.

The leading approach for simulating RCS with the lowest computational cost and the highest fidelity is contracting the TN representing the quantum circuit as a network of contracted tensors that correspond to the applied gates\,\cite{markov2008simulating}. Contracting a fraction of slices effectively performs a finite fidelity simulation\,\cite{kalachev2021classical}. Approximation and simplification of the original TN could also be implemented alongside slicing, which reduces the fidelity\,\cite{pan2022solving}.

Additionally, conventional RCS experiments usually request millions of samples from a single circuit. One can reduce the cost by leaving some indices open\,\cite{pednault2019quantum, schutski2020adaptive, pan2022simulation}, using a sparse output state\,\cite{pan2022solving}, or reuse intermediate results\,\cite{kalachev2021classical, kalachev2021multi, pan2022simulation}. One can also perform some post processing on the samples to spoof the quality metric\,\cite{zhao2024leapfrogging}.

Another commonly analyzed approach is the MPS based on DMRG\,\cite{ayral2023density}. This approach also breaks the TN corresponding to a sample amplitude into three parts: the beginning of the circuit with fixed input indices, the middle, and the end with fixed output indices. The first and third parts can be represented using an MPS, and the amplitude can be obtained by contracting the two MPSs with the middle part in between. Noteworthy, this method has been recently used to prove quantum utility in simulating the quantum Ising model's dynamics~\cite{haghshenas2025digital}.

Finally, prior to the development of the aforementioned methods, other TN techniques such as MPO\,\cite{noh2020efficient}, MPDO\,\cite{cheng2021simulating}, PEPS\,\cite{guo2019general}, conversion of 3D to 2D networks\,\cite{villalonga2019flexible} and TTNS\,\cite{ellerbrock2020multilayer, dumitrescu2017tree} were utilized to simulate gate-based quantum computations. Overall, while the earlier RCS experiments \cite{arute2019quantum, wu2021strong, zhu2022quantum} can now be simulated, the more recent experiments with better fidelity and larger circuit volume still remain hard \cite{morvan2023phase,decross2024computational,gao2025establishing}.

\subsection{Analog evolution}

Analog quantum computers are devices where a specific class of models is implemented ``natively" on the quantum hardware. These systems include optical lattices of neutral atoms\,\cite{gross2017quantum}, trapped ions\,\cite{blatt2012quantum}, Rydberg tweezers\,\cite{browaeys2020many}, photonic waveguides\,\cite{aspuru2012photonic}, as well as superconducting circuits\,\cite{houck2012chip}. Contrary to digital quantum processors, analog quantum devices allow the simultaneous, time-dependent continuous control of pairwise interactions of all the qubits available on the quantum chip. The major drawbacks of analog quantum simulators are the calibration errors of Hamiltonian parameters and decoherence. One also needs to make sure that the quantum processor is faithfully behaving according to closed-system Schr\"{o}dinger time-evolution for the largest number of qubits possible. \emph{TNs provide an invaluable benchmarking tool in this regard.}

In adiabatic quantum annealing\,\cite{kadowaki1998quantum} (AQA), the lowest-energy state of a complex Hamiltonian is sought by starting from a simple and well-defined initial low-energy state of a well-controlled Hamiltonian. Then the parameters of the established Hamiltonian are changed very slowly to arrive at the more complex Hamiltonian. In a coherent QA, the annealing process is performed faster, i.e., on the time scale well less than the expected qubit decoherence time. Therefore, rather than seeking for optimization of the final complex Hamiltonian, one simulates the dynamics of a closed quantum system swept through a quantum critical point. In particular, Ref.\,\cite{king2024computational} claimed that in the studied the parametric range, the approximate classical TN methods such as MPS and PEPS cannot match the solution quality of the quantum simulator in solving the Schr\"{o}dinger dynamics for a transverse field Ising spin glass system in 3D and biclique (all-to-all) lattice geometries, despite the limited correlation length and finite experimental precision. Time evolution of the MPS was performed using the GPU accelerated TDVP on snake-like unfolded 2D, 3D, and biclique lattices (which was found superior against other methods, such as TEBD as well as local and global Krylov methods~\cite{paeckel2019time}) performed on Summit and Frontier ORNL supercomputers. In this study, the MPS methods played a crucial role via the estimation of an equivalent ``QPU bond dimension"\,\cite{shaw2024benchmarking}, defined by matching the sampled QPU distribution quality against converged MPS simulations at simulatable scales.

A recent simulation technique was able to achieve comparable accuracies to the quantum annealer \cite{tindall2025dynamics} for two- and three-dimensional systems. It used belief propagation for time evolution, and more sophisticated variants of belief propagation for calculating expectation values. However, no such simulation was performed for the infinite dimensional biclique lattice studied by the quantum annealer~\cite{king2025comment}.

\subsection{Boson sampling}
Boson sampling\,\cite{aaronson2011computational} is a computational model based on linear optical elements, non-linear input, and measurements. This model of computation is non-universal, but it is, nonetheless, hard to simulate classically under plausible complexity theoretic assumptions. Similar to RCS, boson sampling is a computational task of producing samples from a probability distribution corresponding to measurements of the outputs of a linear-optical interferometer.

Since the transmission matrices (matrices describing the relationship between input and output optical modes) of boson sampling experiments are approximately Haar-random\,\cite{zhong2020quantum, zhong2021phase, madsen2022quantum, deng2023gaussian}, using an MPS approach is despite the high dimensionality of the optical interferometer\,\cite{huang2019simulating, oh2021classical}. Additionally, for simulations without photon loss, photon number conservation can be exploited to reduce the cost of the MPS\,\cite{huang2019simulating, oh2021classical}.

In reality, various experimental noises, especially photon loss, scale with the system size, and the complexity-theoretic argument is not applicable in this case. Therefore, directly approximating the mixed state in the Fock basis using an MPO can be efficient.
In this approach, if we have $N$ input photons and $N_{\rm out}\propto N^\gamma$ output photons (where $\gamma$ is the scaling exponent), the MPO entanglement entropy (MPO EE, which roughly characterizes the simulation hardness) grows as $S=O(N^{2\gamma-1}\log N)$, which indicates a logarithmic growth of the MPO EE when $N_{\rm out}\propto\sqrt{N}$\,\cite{oh2021classical, liu2023simulating}.

Further, for Guassian boson sampling\,\cite{hamilton2017gaussian}, the lossy output states can be modeled by applying random classical operations on a pure state\,\cite{oh2024classical, quesada2022quadratic}, allowing the state to be simulated with an MPS. Further, one can optimize this pure state to have significantly fewer photons compared to the original state\,\cite{oh2024classical}. The bond dimension grows logarithmically when $N_{\rm out}\propto \sqrt{N}$, and this theoretically guarantees polynomial growth of the bond dimension to fixed fidelity. This enabled the largest boson sampling experiments to be simulated on up to $288$ GPUs in under two hours\,\cite{oh2024classical}, meaning that no boson sampling experiments demonstrate clear evidence for beyond classical hardness.

Additionally, one can simulate boson sampling in the Heisenberg picture in a way similar to the simulation of quantum circuits. Instead of evolving the quantum state using a TN, this approach evolves the Fock state projector of the desired measurement outcome\,\cite{cilluffo2023simulating}. For Gaussian boson sampling, the Schr\"odinger picture approach evolves a Gaussian state (an infinite superposition of Fock states), which usually has a larger bond dimension, whereas the Heisenberg picture evolves a Fock state.
\section{Quantum circuit synthesis}\label{sec:qcs}

Quantum circuit synthesis is the process of decomposing a target quantum operation into a sequence of executable gate operations that are compatible with a specific quantum computing architecture\,\cite{nielsen2010quantum}. When considering modern quantum processors, the problem of quantum circuit synthesis faces two challenges: (1) the decomposition algorithms must adhere to the native connectivity of the quantum device and (2) the circuit depth allowed to faithfully prepare a quantum operation is limited by the characteristic noise in the quantum device. A universal solution of these problems, either in terms of scalability or precision, remains, at the present stage, out of reach. 

TNs offer a promising general pathway for addressing these challenges\,\cite{biamonte2017tensor}. TN architectures, such as MPS and PEPS, feature an inherently geometrical layout provided by the prescribed decomposition/representation of general tensor formats. It turns out that for circuit descriptions, which inherently are TNs, such layouts are particularly attractive due to the ease in attaining connectivity within adjacent sets of qubits. Although this feature is exploited in studying long/short range interactions in condensed matter systems, it naturally provides a partial solution to the problem of circuit synthesis in terms of realizing native hardware connectivity. By representing quantum states and operations as TNs and further casting them as circuits, one can simplify both circuit design as well as address the problem of compiling arbitrary unitaries into natively realizable gates. This can reduce the complexity of quantum circuits and enable more efficient synthesis.

\subsection{Promoting TNs to quantum gates}

Quantum states and channels, represented as TNs can be mapped to quantum circuits. The process of promoting TNs to quantum gates requires four steps: (i)\,transforming the original TN into a TN of isometric tensors; (ii)\,embedding spatial and temporal directions to the network; (iii)\,promoting each isometric tensor into a unitary; and (iv)\,decomposing each unitary as quantum gates.

Quantum operations comprise unitary operations across multiple qubits. In order to map a TN to a quantum circuit, each tensor in the network is mapped to a unitary tensor. An intermediate step in this approach is to convert every tensor in the network into an isometric tensor. Ensuring that each tensor is an isometry is possible due to the gauge freedom in TNs\,\cite{tindall2023gauging, vidal2003efficient}. Additionally, quantum circuits are inherently directional in time, whereas TNs have no directionality. Therefore, we are free to translate and rotate each tensor in the diagram. In order to turn TN diagrams into QC diagrams, we need to place an arrow of time onto the TN diagram, and specify incoming and outgoing wires. Once a TN comprises only isometric tensors, we promote the individual tensors into unitaries. This procedure is depicted in Fig.\,\ref{fig:mps_to_qc}(a).
            
\subsection{Examples of a TN state and operator preparation}

Using the introduced prescription, we provide examples of important TN architectures in the literature, and demonstrate how they can be mapped to quantum circuits.

\begin{figure*}
\centering
\includegraphics[width=\linewidth]{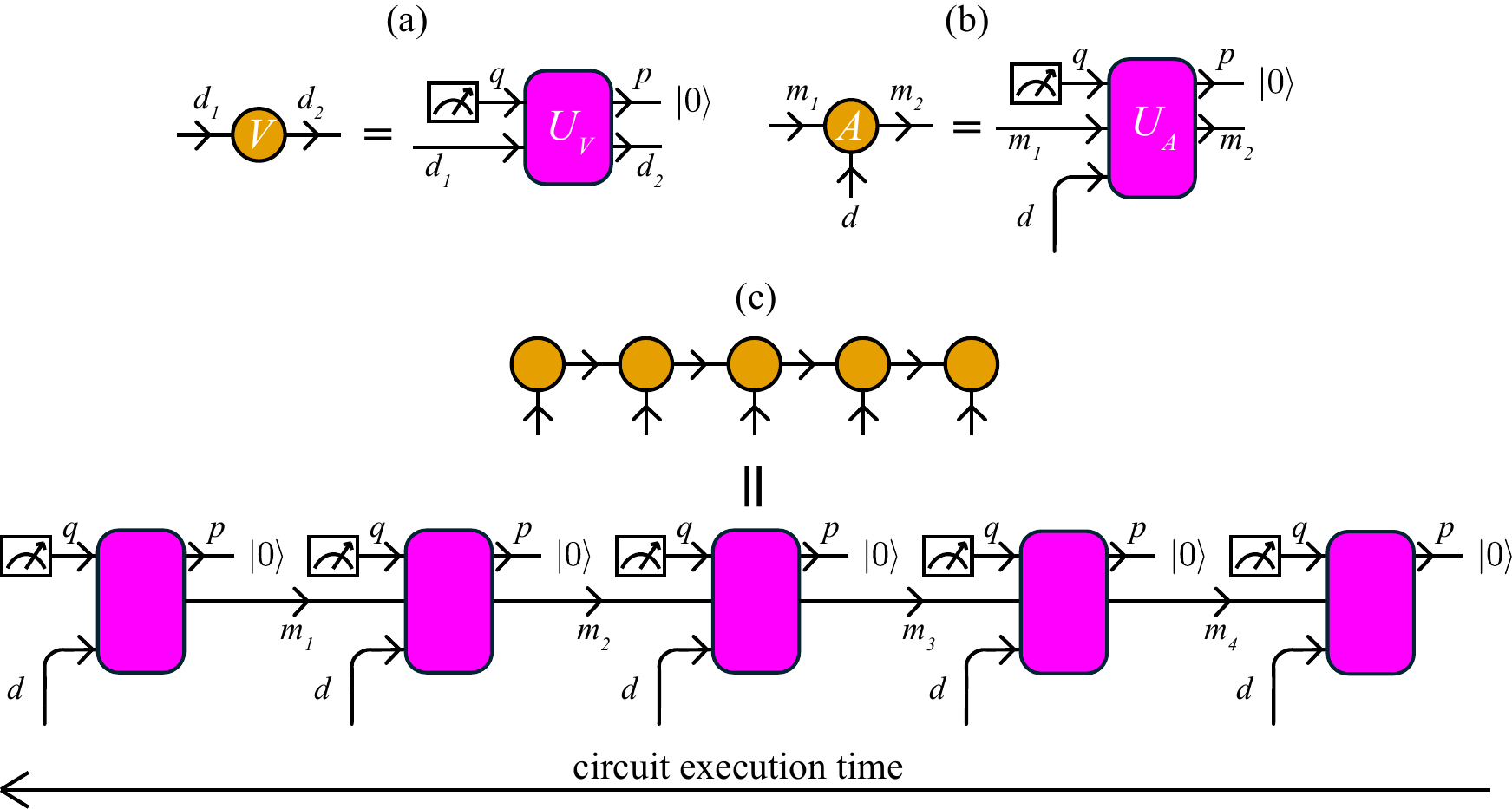}
\caption{(a) Promoting an isometric $d_1\times d_2$ matrix $V$ into a unitary gate $U_V$, with an incoming $p$-dimensional ancilla qudit and an outgoing $q$-dimensional qudit that is measured. The dimensions of $p$ and $q$ are $p=\text{lcm}(d_1,d_2)/d_1$ and $q=\text{lcm}(d_1,d_2)/d_2$, where $\text{lcm}$ denotes the least common multiple operation. (b) Same as (a), but the isometry is taken from an MPS tensor. (c) Converting a (canonized) MPS into a quantum circuit by promoting isometric tensors into unitaries. Each individual unitary is then decomposed into one and two-qubit gates by encoding the qudit wires to qubit wires followed by methods such as Gray codes to decompose arbitrary unitary matrices into two-level unitaries. The arrows in the figure represent the isometric conditions of the tensors, as defined in the convention in \Cref{fig:graphical_representation}, and do not indicate circuit time.}
\label{fig:mps_to_qc}
\end{figure*}

Figure\,\ref{fig:mps_to_qc} depicts how one converts an MPS into a quantum circuit using the steps from the above subsection. The key element in this procedure is that an MPS can be exactly transformed into a canonical form, including only isometric tensors, which is a unique feature of an MPS---this is not the case for the higher-dimensional TNs like PEPS. Preparing an MPS as a quantum circuit was first introduced in\,\cite{schon2005sequential}, and has since been theoretically explored by many others\,\cite{perez2006matrix, ran2020encoding, wei2023efficient, malz2024preparation, haghshenas2022variational}. An experimental realization of an MPS has been extensively explored across various quantum computer architectures\,\cite{foss2021holographic, barratt2021parallel, smith2022crossing, meth2022probing}.

Although such methods are accurate, an exact MPS preparation requires unitary operations acting immediately on the $(\lfloor \log(m) \rfloor + 1)$ qubits, where $m$ is an MPS bond dimension. This may be undesirable for modern quantum computers, given their limited connectivity and the restricted set of native hardware gates. One approach to address this issue is to variationally fit a network of local circuits composed of native gates in order to approximate the original unitary tensor network. Alternatively, iterative methods can be used to remove short-range entanglement through native-gate disentanglers\,\cite{barratt2021parallel, lin2021real, haghshenas2022variational, ran2020encoding}.

The PEPS, however, generally lack a canonical exact isometric form due to the higher connectivity inherent in their underlying structure. In other words, one cannot exactly map a PEPS onto a quantum circuit without an exponential amount of postselection. However, one can create a subclass of PEPS, referred to as isoPEPS, to establish a proper connection to quantum circuits. In the isoPEPS class, each tensor, or set of tensors, respects the isometric condition, making it equivalent to a quantum circuit\,\cite{zaletel2020isometric, haghshenas2019conversion, wei2022sequential}. The question of understanding the variational power of isoPEPS and the way how it is compared to PEPS and to 2dMERA and how they can be realized experimentally on a digital quantum computer is now an active area of research.    

Among all tensor network methods for classical simulation of quantum systems, the MERA stands out as the most natural representation of a quantum circuit: it was originally envisioned in the reverse direction of a quantum circuit in which all isometries are embedded in unitaries. It features an intrinsic robustness to noise and does not suffer from barren plateaus\,\cite{kim2017robust, martin2023barren, barthel2025absence}. The MERA has recently been demonstrated on the ion-trapped digital quantum computer to probe the criticality of many-body systems\,\cite{sewell2021preparing, anand2023holographic, haghshenas2023probing}\cite{miao2024probing}.

\subsection{Implementation techniques} 

As mentioned in Sec.\,4.1, TNs lack the inherent directionality compared to real quantum circuits. Because of this, we must choose how tensors are executed in time, which then is related to the quantum circuit architecture. This freedom allows TN states to be prepared \textit{holographically}. Introduced first in\,\cite{foss2021holographic}, a holographic preparation of a TN state takes a spatial dimension of the physical state and prepares it sequentially in time, with the important feature that once a circuit is done preparing one site, the qubit can be reused to prepare the next site.

In addition to the holographic implementation techniques where each site is sequentially prepared after the next site, adaptive circuits have recently been explored to prepare TN states\,\cite{zhang2024characterizing, larsen2024feedback, stephen2024preparing, smith2023deterministic, sahay2024classifying, sahay2024finite, smith2024constant}. Adaptive circuits are quantum circuits that allow mid-circuit measurements followed by gate operations that are determined by those measurements. While pure unitary circuits require a circuit depth scaling with system size to prepare states with long-range correlations, adaptive circuits can prepare certain states in constant-depth. 

Beyond the exact implementations of mapping a TN to a quantum circuit described above, various variational methods are employed to prepare an approximate TN state\,\cite{ben2024approximate,rudolph2023decomposition,melnikov2023quantum}\cite{termanova2024tensor, jaderberg2025variational}. The basic idea is to numerically optimize a variational ansatz to approximate the target state, typically resulting in a shallower quantum circuit. In addition to preparing a TN state,\,the work\,\cite{kukliansky2023qfactor} used variational methods to instantiate a generic quantum circuit where the TN formulation is leveraged as an efficient back-end in the compilation workflows.

\section{Quantum error correction and mitigation}\label{sec:qec}

Quantum error correction (QEC) plays a critical role in safeguarding quantum information from errors caused by decoherence, dissipation, and control inaccuracies~\cite{nielsen2010quantum,lidar2013quantum}. In this section, we discuss two main topics: (i)\,the parallels between quantum error-correcting codes and TNs, along with (ii) the application of TNs in decoding, including the formal reduction of optimal decoding to a TN contraction\,\cite{ferris2014tensor}.

\subsection{Tensor-network codes}

The first major intersection between TNs and QEC was the establishment of a formal correspondence between certain QEC codes and the associated TNs\,\cite{ferris2014tensor}. Convolutional codes, concatenated block codes, and topological codes, for example, can be respectively represented by the MPS, tree TNs, and PEPS. This insight allows for importing well-established tools of TNs to represent and analyze QEC codes. In particular, the problem of an optimal decoding of an QEC code (i.e. diagnosing errors, see the discussion in Section\,\ref{sec:decoding}) was shown to be formally equivalent to contracting an associated TN\,\cite{ferris2014tensor}. For convolutional and concatenated block codes, which have efficiently contractable TNs, this correspondence thereby provides the efficient optimal decoders.

The QEC-TN correspondence was later generalized by the introduction of TN stabilizer codes \cite{farrelly2021tensor,farrelly2022local} and the `quantum LEGO' formalism\,\cite{cao2022quantum, cao2024quantum, fan2024analyzing} for constructing large codes out of a finite set of smaller `seed' codes. The seed codes are represented by small tensors that are combined into a TN to represent a larger code. This construction was shown to be universal in the sense that \textit{any} qudit QEC code can be represented by a TN built out of three elementary seed tensors \cite{cao2022quantum}. A further advantage to constructing codes in this fashion is due to the fact that TN codes come naturally equipped with optimal decoders that are evaluated by contracting a TN \cite{farrelly2021tensor}.

Quantum codes generated using tensor networks generalize code concatenation, a fact that has been observed in Refs.~\cite{cao2022quantum,cao2024quantum}. Several examples of code constructions hailing from TN structures can be seen in topological \cite{farrelly2022local} and holographic \cite{pastawski2015holographic,jahn2021holographic,steinberg2023holographic} tensor-network codes, as well as examples of non-Abelian stabilizer \cite{shen2023quantum}, non-additive topological \cite{verstraete2022fibonacci}, and approximate quantum codes \cite{bettaque2024nora,cao2021approximate}. Apart from their usefulness in studying topologically-ordered matter or the Anti-de Sitter/ Conformal Field Theory (AdS/CFT) correspondence, many of the codes proposed exhibit favorable properties for practical QEC \cite{steinberg2024far,pastawski2017code,fan2024analyzing}. Indeed, it has been shown that machine-learning methods can even be utilized in order to search for QEC codes possessing desirable code parameters, entirely within the tensor-network formalism \cite{su2023discovery,mauron2024optimization}.

\subsection{Syndrome decoding}
\label{sec:decoding}

Decoding a quantum error correcting code involves processing syndrome measurements using a given  theoretical error model. This process yields appropriate correction operations and involves solving the inference problem of finding the set of errors that occurred in the computation with the highest likelihood based on the syndrome information measured and on the error model at hand. Such a decoding method is known as maximum-likelihood decoding, and has been shown to be formally equivalent to contracting a TN\,\cite{ferris2014tensor,bravyi2014efficient}, even in the case of more invasive noise models, such as those employing correlated noise \cite{chubb2021statistical}.

The error model of a quantum error correcting code provides two sets of information: 1) the probability of occurrence of the different physical errors and 2) the recognition of what syndrome measurements each of these errors affects and how each one affects the logical qubit(s) encoded.
Given a set of syndrome measurements, there are exponentially many configurations of physical errors compatible with it.
This set of error configurations can be further decomposed into several subsets, defined after their effect on the logical information encoded; these subsets are often called cosets.
The sum of the joint probabilities of the error configurations of each coset is proportional to the likelihood of the coset.
Optimally, a decoder infers the maximum likelihood coset as the most probable logical outcome of the code, given a set of syndrome measurements, resulting in an exponentially large sum of probabilities computed in good approximation by the contraction of a TN.

There are two approaches to carry out this summation. The first and most popular one consists of identifying a complete set of symmetries of the coset, i.e., all sets of errors that leave the logical information and the syndrome measurements unchanged. In this setting, an enumeration of all error configurations included in the coset is achieved by finding a particular configuration in the set and applying to it all combinations of the symmetries~\cite{ferris2014tensor,bravyi2014efficient,chubb2021general, darmawan2022low, farrelly2022parallel}.

A second approach consists of building a TN directly from the error model connectivity between errors and syndrome detectors.
In this approach all error configurations outside a coset are not considered by explicitly zeroing out tensor entries corresponding to them.
This approach has the benefit of working with more general error models directly, without the need to find a set of symmetries that is computationally beneficial~\cite{google2023suppressing,piveteau2023tensor,shutty2024efficient}. One can also utilize tensor networks in order to enumerate error paths that appear in the realization of fault-tolerant syndrome extraction, albeit with significantly reduced computational complexity in certain specialized cases \cite{kukliansky2024quantum,cao2023quantum,cao2024quantum}.

Both approaches to TN decoding can be successfully leveraged with diverse noise models, such as erasure \cite{harris2018calderbank,harris2020decoding}; fractal noise \cite{bao2022code}; depolarizing noise \cite{fan2024overcoming,harris2020decoding}; biased noise, and explorations of quantum channel capacity via the hashing bound \cite{fan2024overcoming,tuckett2019tailoring,qian2023tailoredxzzx,tuckett2018ultrahigh,bonilla2021xzzx,dua2024clifforddeformed}; as well as non-Markovian noise sources \cite{kobayashi2024tensor}. In addition to those noise models mentioned, circuit-level noise models \cite{google2023suppressing,piveteau2024tensor} can also be decoded, albeit approximately.

Optimized decoding algorithms benefit from the computational strategies inherent in TN theory, such as the use of approximate contraction methods for large-scale networks, which are directly applicable to decoding large and complex QEC codes~\cite{chubb2021general, darmawan2022low, farrelly2022parallel, piveteau2023tensor, google2023suppressing, shutty2024efficient}.
TN decoders are, up to date, considered as highly accurate but not practical for real time decoding~\cite{battistel2023real}.
The need for fast decoders in real quantum error correction experiments makes the use of heuristic decoders optimized for performance more popular and practical.
This leaves TN decoders as useful tools for the benchmark of experiments and other less accurate decoders.

\subsection{Quantum error mitigation}

Error mitigation methods are expected to play a crucial role in near-term quantum experiments until fault-tolerance is fully achieved~\cite{temme2017error, li2017efficient}. One of the promising methods in this area is probabilistic error cancellation (PEC). However, it requires precise knowledge of the underlying noise model at the gate level. PEC creates an ensemble of circuits, with a denoiser, which on average replicates a noise-inverting channel, inserted according to a quasi-probability distribution, effectively canceling out the noise. A quasi-probability distribution is a probabilistic framework that assigns a sign to each sample, allowing for the representation of non-physical channels. The number of samples (measurement overhead) required to address sign cancellations can be effectively bounded by Hoeffding’s inequality, growing exponentially with the number of qubits and circuit depths. Another approach of error mitigation is zero-noise extrapolation (ZNE). ZNE artificially introduces noise to obtain results at different noise levels before extrapolating to the noiseless case \cite{temme2017error}, and was experimentally demonstrated \cite{kim2023evidence}.

Despite the overhead, it is anticipated that error mitigation will remain practical for circuits with qubit counts in the hundreds and equivalent circuit depth~\cite{kim2023evidence}. The tensor-network error mitigation (TEM) algorithms represent a similar paradigm for reducing overhead. As investigated in a series of recent works\,\cite{guo2022quantum, tepaske2023compressed,filippov2023scalable,filippov2024scalability}, TEM provides a universal lower cost bound for error mitigation, specifically a quadratic cost reduction. TEM is a post-processing procedure that acts on a set of randomized local measurements, as designed to cancel errors by incorporating classically noise-inverting channels. The mitigated estimations are given by contracting the circuit-level TN. Notably, it was experimentally demonstrated on a 91-qubit circuit with 4095 two-qubit gates. \cite{fischer2024dynamical} For systems containing more than a thousand qubits with equivalent circuit depth, we foresee the development of hybrid approaches combining quantum error correction with error mitigation techniques~\cite{piveteau2021error}. It remains an open question whether error mitigation alone, or in combination with quantum error correction, can lead to utility in near-term quantum computation or provide a potential practical advantage in quantum algorithms.
\section{Tensor networks for quantum machine learning}\label{sec:qml}

Over the past decade, TNs have gained significant interest in machine learning (ML). By compressing high-dimensional linear layers in neural networks, TNs reduce memory usage and the number of training parameters~\cite{novikov2015tensorizing, novikov2016exponential, stoudenmire2016supervised}. Moreover, TNs offer strong analytical interpretability and are related to various established ML techniques \cite{chen2018equivalence,li2021boltzmann,han2018unsupervised,liu2023tensor,glasser2019expressive}. The structural flexibility of TNs allows the introduction of inductive biases (constraints on the learning algorithm to restrict the space of possible models to plausible ones) into ML models based on intrinsic data topology and correlations~\cite{stoudenmire2016supervised, cheng2019tree, vieijra2022generative, reyes2021multi}, ultimately enhancing training efficiency and model generalization to data unseen during training. This synergy has paved the way for employing TNs across various ML applications, also sometimes referred to as quantum-inspired ML.

Although the contraction complexity of many TNs grows polynomially with the bond dimension, the degree of the polynomials is higher than three in some cases (such as for MERA), which is often regarded as inefficient in practical ML. Quantum computers can potentially address this problem of TN-based ML by implementing parameterized quantum circuit (PQC) ans\"atze reflecting specific TN architectures such as MPS~\cite{huggins2019towards, wall2021generative}, TTN~\cite{huggins2019towards, grant2018hierarchical, lazzarin2022multi}, and MERA~\cite{grant2018hierarchical, lazzarin2022multi}, resulting in TN-based quantum ML\,\cite{rieser2023tensor} (TN-QML).

In regression and classification tasks~\cite{huggins2019towards}, TN-QML obtains the output of a TN neural layer by measuring observables on the corresponding quantum state rather than performing costly tensor contractions on classical hardware. In turn, in generative tasks~\cite{wall2021generative}, TN-QML generates data by directly sampling from the quantum state encoding the TN, thereby avoiding expensive classical sampling that would otherwise again rely on tensor contractions.

Moreover, in the context of quantum simulation, parameterizing TNs with quantum gates has been shown to offer an exponential reduction in the number of parameters required to achieve results similar to those of classical TNs\,\cite{barratt2021parallel, lin2021real}, which could potentially translate to ML tasks as well. On the other hand, such variational TN circuits offer advantages over commonly used quantum hardware-efficient ans\"atze, such as being free from barren plateaus\,\cite{zhao2021analyzing, pesah2021absence, martin2023barren} (flat regions in the optimization landscape that impede training). Additionally, the TN structure yields qubit-efficient implementations~\cite{huggins2019towards} and robustness against decoherence\,\cite{huggins2019towards, grant2018hierarchical, liao2023decohering}.

The TN-QML approach also inherently enables quantum applications, such as quantum phase classification directly on quantum devices using quantum convolutional neural networks (QCNNs) based on the TTN and MERA architectures \cite{cong2019quantum, lazzarin2022multi}. However, although the QCNN was initially considered a candidate for exponential quantum advantage, recent work \cite{bermejo2024quantum} has argued against this, suggesting it can be dequantized using classical shadow techniques.

Additionally, TNs are promising in QML as a tool for encoding classical data into quantum circuits\,\cite{dilip2022data,iaconis2023tensor} and for pre-training the TN-PQC on classical computers~\cite{huggins2019towards,dborin2022matrix,rudolph2022synergy,khan2023pre} before extending to quantum ans\"atze that cannot be simulated classically. Finally, TN surrogate modeling can be employed to test QML models for dequantization~\cite{shin2024dequantizing}. For an in-depth review of using TNs in the QML, we refer the reader to\,\cite{rieser2023tensor}.

\section{Discussion and Outlook}\label{sec:discussion}

TNs are a valuable tool for the field of quantum computing. Moreover, they can be important for analyzing computational complexity classes. For example, finding a polynomial-time tensor-network algorithm for a problem~\cite{de2020tensor} proves that the problem is in the complexity class $\mathcal{P}$. Similarly, one can empirically argue that a problem is hard by trying existing state-of-the-art TN techniques and observing exponential scaling. Along the same vein, TNs can help validating quantum advantage claims of finite-size quantum experiments, which may not admit asymptotic quantum advantage. Here, rigorous benchmarking of TN techniques as well as other classical techniques allows to decide whether a quantum experiment is out of reach for classical computers.

In particular, recent developments in TNs and other classical simulation techniques have significantly improved our ability to simulate quantum many-body systems~\cite{rakovszky2022dissipation}, as well as near-Clifford circuits. These methods arise from representing TN states directly in the Pauli basis~\cite{tarabunga2024nonstabilizerness}. Recent advancements in representing TN states in both a Pauli basis as well as a computational basis have shown promising results~\cite{masot2024stabilizer}, and it remains an open and interesting question on what classes of circuits can be classically simulated exactly or approximately. Additionally, there is potential in understanding the quantum circuit synthesis of TN states given that they are in the Pauli basis.

On quantum error correction, one future direction may be utilizing the quantum LEGO formalism to write down families of quantum low-density parity-check codes~\cite{breuckmann2021} in a systematic fashion. This is an important direction since quantum low-density parity-check codes represent a promising avenue in quantum error correction~\cite{panteleev2022asymptotically}. Additionally, the first explorations of non-Abelian stabilizer and non-additive codes have already been constructed \cite{verstraete2022fibonacci,shen2023quantum}; it would be interesting to see whether or not the quantum LEGO formalism or further extensions can provide insight into systematic constructions of such codes. Moreover, highly accurate TN decoding is too slow for real-time hardware decoding \cite{battistel2023real}; as such, it could be valuable to design heuristic algorithms that can form approximate TN decoding well. The alternative could be using AI for decoding \cite{bausch2024learning, zhou2025nvidia}, which may benefit from noisy simulations using tensor networks for model training.

Finally, using TNs in QML requires both further investigation into their use in classical ML and the development of new advances in QML itself. Although recent studies question the feasibility of achieving a practical exponential quantum advantage with QML~\cite{cerezo2023does,gil2024relation}, it remains to be seen whether TN-QML can offer any polynomial advantage in this context.

TNs continue to drive progress across multiple areas of quantum computing as demonstrated throughout this review. As we transition towards fault-tolerant quantum devices, TNs will likely maintain their significance, particularly in potential hybrid quantum-classical devices which are anticipated to integrate quantum computers with specialized classical accelerators like the GPUs, TPUs, and FPGAs. Looking forward, we foresee that TNs will evolve to include new algorithmic approaches, integration with AI, and techniques to meet the demands of fault-tolerant quantum computers. As quantum hardware capabilities improve, the synergy between TNs and quantum computing may deepen, keeping TNs an important tool for further development of future quantum computing systems.

\section*{Acknowledgments}
A.\,B., R.\,E., V.\,K., C.\,M., Al.\,M., Ar.\,M., V.\,M., D.\,M., F.\,N., M.\,Pere., M.\,Pf. and V.\,V. were supported by Terra Quantum AG. A.\,N. was supported by Natural Sciences and Engineering Research Council of Canada (NSERC) Alliance Quantum Program (Grant ALLRP-578555), and the Canada First Research Excellence Fund, Quantum Materials and Future Technologies Program. S.\.M. is partially supported by the Prime Contract No. 80ARC020D0010 with the NASA Ames Research Center and acknowledges funding from DARPA under IAA 8839. A.\,B. thanks Ilia Luchnikov, Mikhail Litvinov and Stefanos Kourtis for valuable discussions. A.\,A., Z.\,H., A.\,K., M.\,L., M.\,Perl., M.\,Pi., M.\,S. and R.\,S. thank their colleagues at the Global Technology Applied Research center of JPMorganChase for support and helpful discussions.

\section*{Author contributions}
Yu.\,A., A.\,B., M.\,L. and V.\,V. conceptualized the work. A.\,B., R.\,E., J.\,G., R.\,H., A.\,K., M.\,L., A.\,N., F.\,N. and  V.\,V. contributed to the Tensor-network methods section. M.\,L., A.\,N., F.\,N. contributed to the Simulation of quantum computation section. A.\,A., R.\,H., Z.\,H., A.\,K., M.\,L., M. Pere., V.\,V. contributed to the Quantum circuit synthesis section. A.\,B., R.\,H., M.\,Perl., M.\,S., B.\,V. contributed to the Quantum error correction and mitigation section. A.\,A., V.\,K. contributed to the Tensor networks for quantum machine learning section. All authors contributed to the Introduction and the Discussion and outlook sections. All authors were involved in shaping the direction of the manuscript, as well as in its discussion, review, and editing.

\bibliography{bibliography}

\begin{thebibliography}{255}%
\makeatletter
\providecommand \@ifxundefined [1]{%
 \@ifx{#1\undefined}
}%
\providecommand \@ifnum [1]{%
 \ifnum #1\expandafter \@firstoftwo
 \else \expandafter \@secondoftwo
 \fi
}%
\providecommand \@ifx [1]{%
 \ifx #1\expandafter \@firstoftwo
 \else \expandafter \@secondoftwo
 \fi
}%
\providecommand \natexlab [1]{#1}%
\providecommand \enquote  [1]{``#1''}%
\providecommand \bibnamefont  [1]{#1}%
\providecommand \bibfnamefont [1]{#1}%
\providecommand \citenamefont [1]{#1}%
\providecommand \href@noop [0]{\@secondoftwo}%
\providecommand \href [0]{\begingroup \@sanitize@url \@href}%
\providecommand \@href[1]{\@@startlink{#1}\@@href}%
\providecommand \@@href[1]{\endgroup#1\@@endlink}%
\providecommand \@sanitize@url [0]{\catcode `\\12\catcode `\$12\catcode
  `\&12\catcode `\#12\catcode `\^12\catcode `\_12\catcode `\%12\relax}%
\providecommand \@@startlink[1]{}%
\providecommand \@@endlink[0]{}%
\providecommand \url  [0]{\begingroup\@sanitize@url \@url }%
\providecommand \@url [1]{\endgroup\@href {#1}{\urlprefix }}%
\providecommand \urlprefix  [0]{URL }%
\providecommand \Eprint [0]{\href }%
\providecommand \doibase [0]{https://doi.org/}%
\providecommand \selectlanguage [0]{\@gobble}%
\providecommand \bibinfo  [0]{\@secondoftwo}%
\providecommand \bibfield  [0]{\@secondoftwo}%
\providecommand \translation [1]{[#1]}%
\providecommand \BibitemOpen [0]{}%
\providecommand \bibitemStop [0]{}%
\providecommand \bibitemNoStop [0]{.\EOS\space}%
\providecommand \EOS [0]{\spacefactor3000\relax}%
\providecommand \BibitemShut  [1]{\csname bibitem#1\endcsname}%
\let\auto@bib@innerbib\@empty
\bibitem [{\citenamefont {Schollw{\"o}ck}(2011)}]{schollwock2011density}%
  \BibitemOpen
  \bibfield  {author} {\bibinfo {author} {\bibfnamefont {U.}~\bibnamefont
  {Schollw{\"o}ck}},\ }\bibfield  {title} {\bibinfo {title} {The density-matrix
  renormalization group in the age of matrix product states},\ }\href@noop {}
  {\bibfield  {journal} {\bibinfo  {journal} {Annals of physics}\ }\textbf
  {\bibinfo {volume} {326}},\ \bibinfo {pages} {96} (\bibinfo {year}
  {2011})}\BibitemShut {NoStop}%
\bibitem [{\citenamefont {Or{\'u}s}(2019)}]{orus2019tensor}%
  \BibitemOpen
  \bibfield  {author} {\bibinfo {author} {\bibfnamefont {R.}~\bibnamefont
  {Or{\'u}s}},\ }\bibfield  {title} {\bibinfo {title} {Tensor networks for
  complex quantum systems},\ }\href@noop {} {\bibfield  {journal} {\bibinfo
  {journal} {Nature Reviews Physics}\ }\textbf {\bibinfo {volume} {1}},\
  \bibinfo {pages} {538} (\bibinfo {year} {2019})}\BibitemShut {NoStop}%
\bibitem [{\citenamefont {Or{\'u}s}(2014)}]{orus2014practical}%
  \BibitemOpen
  \bibfield  {author} {\bibinfo {author} {\bibfnamefont {R.}~\bibnamefont
  {Or{\'u}s}},\ }\bibfield  {title} {\bibinfo {title} {{A practical
  introduction to tensor networks: Matrix product states and projected
  entangled pair states}},\ }\href@noop {} {\bibfield  {journal} {\bibinfo
  {journal} {Annals of physics}\ }\textbf {\bibinfo {volume} {349}},\ \bibinfo
  {pages} {117} (\bibinfo {year} {2014})}\BibitemShut {NoStop}%
\bibitem [{\citenamefont {Chubb}(2021)}]{chubb2021general}%
  \BibitemOpen
  \bibfield  {author} {\bibinfo {author} {\bibfnamefont {C.~T.}\ \bibnamefont
  {Chubb}},\ }\bibfield  {title} {\bibinfo {title} {{General tensor network
  decoding of 2D Pauli codes}},\ }\href@noop {} {\bibfield  {journal} {\bibinfo
   {journal} {arXiv preprint arXiv:2101.04125}\ } (\bibinfo {year}
  {2021})}\BibitemShut {NoStop}%
\bibitem [{\citenamefont {Pan}\ and\ \citenamefont
  {Zhang}(2022)}]{pan2022simulation}%
  \BibitemOpen
  \bibfield  {author} {\bibinfo {author} {\bibfnamefont {F.}~\bibnamefont
  {Pan}}\ and\ \bibinfo {author} {\bibfnamefont {P.}~\bibnamefont {Zhang}},\
  }\bibfield  {title} {\bibinfo {title} {Simulation of quantum circuits using
  the big-batch tensor network method},\ }\href@noop {} {\bibfield  {journal}
  {\bibinfo  {journal} {Physical Review Letters}\ }\textbf {\bibinfo {volume}
  {128}},\ \bibinfo {pages} {030501} (\bibinfo {year} {2022})}\BibitemShut
  {NoStop}%
\bibitem [{\citenamefont {Fu}\ \emph {et~al.}(2024)\citenamefont {Fu},
  \citenamefont {Su}, \citenamefont {Zhong}, \citenamefont {Zhao},
  \citenamefont {Zhang}, \citenamefont {Pan}, \citenamefont {Zhang},
  \citenamefont {Zhao}, \citenamefont {Chen}, \citenamefont {Lu} \emph
  {et~al.}}]{fu2024achieving}%
  \BibitemOpen
  \bibfield  {author} {\bibinfo {author} {\bibfnamefont {R.}~\bibnamefont
  {Fu}}, \bibinfo {author} {\bibfnamefont {Z.}~\bibnamefont {Su}}, \bibinfo
  {author} {\bibfnamefont {H.-S.}\ \bibnamefont {Zhong}}, \bibinfo {author}
  {\bibfnamefont {X.}~\bibnamefont {Zhao}}, \bibinfo {author} {\bibfnamefont
  {J.}~\bibnamefont {Zhang}}, \bibinfo {author} {\bibfnamefont
  {F.}~\bibnamefont {Pan}}, \bibinfo {author} {\bibfnamefont {P.}~\bibnamefont
  {Zhang}}, \bibinfo {author} {\bibfnamefont {X.}~\bibnamefont {Zhao}},
  \bibinfo {author} {\bibfnamefont {M.-C.}\ \bibnamefont {Chen}}, \bibinfo
  {author} {\bibfnamefont {C.-Y.}\ \bibnamefont {Lu}}, \emph {et~al.},\
  }\bibfield  {title} {\bibinfo {title} {{Achieving energetic superiority
  through system-level quantum circuit simulation}},\ }\href@noop {} {\bibfield
   {journal} {\bibinfo  {journal} {arXiv preprint arXiv:2407.00769}\ }
  (\bibinfo {year} {2024})}\BibitemShut {NoStop}%
\bibitem [{\citenamefont {Oh}\ \emph {et~al.}(2024)\citenamefont {Oh},
  \citenamefont {Liu}, \citenamefont {Alexeev}, \citenamefont {Fefferman},\
  and\ \citenamefont {Jiang}}]{oh2024classical}%
  \BibitemOpen
  \bibfield  {author} {\bibinfo {author} {\bibfnamefont {C.}~\bibnamefont
  {Oh}}, \bibinfo {author} {\bibfnamefont {M.}~\bibnamefont {Liu}}, \bibinfo
  {author} {\bibfnamefont {Y.}~\bibnamefont {Alexeev}}, \bibinfo {author}
  {\bibfnamefont {B.}~\bibnamefont {Fefferman}},\ and\ \bibinfo {author}
  {\bibfnamefont {L.}~\bibnamefont {Jiang}},\ }\bibfield  {title} {\bibinfo
  {title} {Classical algorithm for simulating experimental gaussian boson
  sampling},\ }\href@noop {} {\bibfield  {journal} {\bibinfo  {journal} {Nature
  Physics}\ ,\ \bibinfo {pages} {1}} (\bibinfo {year} {2024})}\BibitemShut
  {NoStop}%
\bibitem [{\citenamefont {Tindall}\ \emph {et~al.}(2024)\citenamefont
  {Tindall}, \citenamefont {Fishman}, \citenamefont {Stoudenmire},\ and\
  \citenamefont {Sels}}]{tindall2024efficient}%
  \BibitemOpen
  \bibfield  {author} {\bibinfo {author} {\bibfnamefont {J.}~\bibnamefont
  {Tindall}}, \bibinfo {author} {\bibfnamefont {M.}~\bibnamefont {Fishman}},
  \bibinfo {author} {\bibfnamefont {E.~M.}\ \bibnamefont {Stoudenmire}},\ and\
  \bibinfo {author} {\bibfnamefont {D.}~\bibnamefont {Sels}},\ }\bibfield
  {title} {\bibinfo {title} {{Efficient tensor network simulation of IBM’s
  eagle kicked ising experiment}},\ }\href@noop {} {\bibfield  {journal}
  {\bibinfo  {journal} {PRX Quantum}\ }\textbf {\bibinfo {volume} {5}},\
  \bibinfo {pages} {010308} (\bibinfo {year} {2024})}\BibitemShut {NoStop}%
\bibitem [{\citenamefont {Eckart}\ and\ \citenamefont
  {Young}(1936)}]{eckart1936approximation}%
  \BibitemOpen
  \bibfield  {author} {\bibinfo {author} {\bibfnamefont {C.}~\bibnamefont
  {Eckart}}\ and\ \bibinfo {author} {\bibfnamefont {G.}~\bibnamefont {Young}},\
  }\bibfield  {title} {\bibinfo {title} {The approximation of one matrix by
  another of lower rank},\ }\href@noop {} {\bibfield  {journal} {\bibinfo
  {journal} {Psychometrika}\ }\textbf {\bibinfo {volume} {1}},\ \bibinfo
  {pages} {211} (\bibinfo {year} {1936})}\BibitemShut {NoStop}%
\bibitem [{\citenamefont {Stewart}(1998)}]{stewart1998matrix}%
  \BibitemOpen
  \bibfield  {author} {\bibinfo {author} {\bibfnamefont {G.~W.}\ \bibnamefont
  {Stewart}},\ }\href@noop {} {\emph {\bibinfo {title} {Matrix algorithms:
  volume 1: basic decompositions}}}\ (\bibinfo  {publisher} {SIAM},\ \bibinfo
  {year} {1998})\BibitemShut {NoStop}%
\bibitem [{\citenamefont {Verstraete}\ and\ \citenamefont
  {Cirac}(2006)}]{verstraete2006matrix}%
  \BibitemOpen
  \bibfield  {author} {\bibinfo {author} {\bibfnamefont {F.}~\bibnamefont
  {Verstraete}}\ and\ \bibinfo {author} {\bibfnamefont {J.~I.}\ \bibnamefont
  {Cirac}},\ }\bibfield  {title} {\bibinfo {title} {Matrix product states
  represent ground states faithfully},\ }\href@noop {} {\bibfield  {journal}
  {\bibinfo  {journal} {Physical Review B—Condensed Matter and Materials
  Physics}\ }\textbf {\bibinfo {volume} {73}},\ \bibinfo {pages} {094423}
  (\bibinfo {year} {2006})}\BibitemShut {NoStop}%
\bibitem [{\citenamefont {Ba{\~n}uls}(2023)}]{banuls2023tensor}%
  \BibitemOpen
  \bibfield  {author} {\bibinfo {author} {\bibfnamefont {M.~C.}\ \bibnamefont
  {Ba{\~n}uls}},\ }\bibfield  {title} {\bibinfo {title} {{Tensor network
  algorithms: A route map}},\ }\href@noop {} {\bibfield  {journal} {\bibinfo
  {journal} {Annual Review of Condensed Matter Physics}\ }\textbf {\bibinfo
  {volume} {14}},\ \bibinfo {pages} {173} (\bibinfo {year} {2023})}\BibitemShut
  {NoStop}%
\bibitem [{\citenamefont {Oseledets}(2011)}]{oseledets2011tensor}%
  \BibitemOpen
  \bibfield  {author} {\bibinfo {author} {\bibfnamefont {I.~V.}\ \bibnamefont
  {Oseledets}},\ }\bibfield  {title} {\bibinfo {title} {Tensor-train
  decomposition},\ }\href@noop {} {\bibfield  {journal} {\bibinfo  {journal}
  {SIAM Journal on Scientific Computing}\ }\textbf {\bibinfo {volume} {33}},\
  \bibinfo {pages} {2295} (\bibinfo {year} {2011})}\BibitemShut {NoStop}%
\bibitem [{\citenamefont {Schuch}\ \emph {et~al.}(2007)\citenamefont {Schuch},
  \citenamefont {Wolf}, \citenamefont {Verstraete},\ and\ \citenamefont
  {Cirac}}]{schuch2007computational}%
  \BibitemOpen
  \bibfield  {author} {\bibinfo {author} {\bibfnamefont {N.}~\bibnamefont
  {Schuch}}, \bibinfo {author} {\bibfnamefont {M.~M.}\ \bibnamefont {Wolf}},
  \bibinfo {author} {\bibfnamefont {F.}~\bibnamefont {Verstraete}},\ and\
  \bibinfo {author} {\bibfnamefont {J.~I.}\ \bibnamefont {Cirac}},\ }\bibfield
  {title} {\bibinfo {title} {Computational complexity of projected entangled
  pair states},\ }\href@noop {} {\bibfield  {journal} {\bibinfo  {journal}
  {Physical review letters}\ }\textbf {\bibinfo {volume} {98}},\ \bibinfo
  {pages} {140506} (\bibinfo {year} {2007})}\BibitemShut {NoStop}%
\bibitem [{\citenamefont {Vidal}(2003{\natexlab{a}})}]{vidal2002ttn}%
  \BibitemOpen
  \bibfield  {author} {\bibinfo {author} {\bibfnamefont {G.}~\bibnamefont
  {Vidal}},\ }\bibfield  {title} {\bibinfo {title} {Efficient classical
  simulation of slightly entangled quantum computations},\ }\href
  {https://doi.org/10.1103/PhysRevLett.91.147902} {\bibfield  {journal}
  {\bibinfo  {journal} {Phys. Rev. Lett.}\ }\textbf {\bibinfo {volume} {91}},\
  \bibinfo {pages} {147902} (\bibinfo {year} {2003}{\natexlab{a}})}\BibitemShut
  {NoStop}%
\bibitem [{\citenamefont {Shi}\ \emph {et~al.}(2006)\citenamefont {Shi},
  \citenamefont {Duan},\ and\ \citenamefont {Vidal}}]{shi2006ttn}%
  \BibitemOpen
  \bibfield  {author} {\bibinfo {author} {\bibfnamefont {Y.-Y.}\ \bibnamefont
  {Shi}}, \bibinfo {author} {\bibfnamefont {L.-M.}\ \bibnamefont {Duan}},\ and\
  \bibinfo {author} {\bibfnamefont {G.}~\bibnamefont {Vidal}},\ }\bibfield
  {title} {\bibinfo {title} {Classical simulation of quantum many-body systems
  with a tree tensor network},\ }\href
  {https://doi.org/10.1103/PhysRevA.74.022320} {\bibfield  {journal} {\bibinfo
  {journal} {Phys. Rev. A}\ }\textbf {\bibinfo {volume} {74}},\ \bibinfo
  {pages} {022320} (\bibinfo {year} {2006})}\BibitemShut {NoStop}%
\bibitem [{\citenamefont {Wang}\ and\ \citenamefont
  {Thoss}(2003)}]{wang2003ttn}%
  \BibitemOpen
  \bibfield  {author} {\bibinfo {author} {\bibfnamefont {H.}~\bibnamefont
  {Wang}}\ and\ \bibinfo {author} {\bibfnamefont {M.}~\bibnamefont {Thoss}},\
  }\bibfield  {title} {\bibinfo {title} {Multilayer formulation of the
  multiconfiguration time-dependent hartree theory},\ }\href
  {https://doi.org/10.1063/1.1580111} {\bibfield  {journal} {\bibinfo
  {journal} {The Journal of Chemical Physics}\ }\textbf {\bibinfo {volume}
  {119}},\ \bibinfo {pages} {1289} (\bibinfo {year} {2003})},\ \Eprint
  {https://arxiv.org/abs/https://pubs.aip.org/aip/jcp/article-pdf/119/3/1289/19007271/1289\_1\_online.pdf}
  {https://pubs.aip.org/aip/jcp/article-pdf/119/3/1289/19007271/1289\_1\_online.pdf}
  \BibitemShut {NoStop}%
\bibitem [{\citenamefont {Markov}\ and\ \citenamefont
  {Shi}(2008)}]{markov2008simulating}%
  \BibitemOpen
  \bibfield  {author} {\bibinfo {author} {\bibfnamefont {I.~L.}\ \bibnamefont
  {Markov}}\ and\ \bibinfo {author} {\bibfnamefont {Y.}~\bibnamefont {Shi}},\
  }\bibfield  {title} {\bibinfo {title} {Simulating quantum computation by
  contracting tensor networks},\ }\href@noop {} {\bibfield  {journal} {\bibinfo
   {journal} {SIAM Journal on Computing}\ }\textbf {\bibinfo {volume} {38}},\
  \bibinfo {pages} {963} (\bibinfo {year} {2008})}\BibitemShut {NoStop}%
\bibitem [{\citenamefont {DeCross}\ \emph {et~al.}(2024)\citenamefont
  {DeCross}, \citenamefont {Haghshenas}, \citenamefont {Liu}, \citenamefont
  {Alexeev}, \citenamefont {Baldwin}, \citenamefont {Bartolotta}, \citenamefont
  {Bohn}, \citenamefont {Chertkov}, \citenamefont {Colina}, \citenamefont
  {DelVento} \emph {et~al.}}]{decross2024computational}%
  \BibitemOpen
  \bibfield  {author} {\bibinfo {author} {\bibfnamefont {M.}~\bibnamefont
  {DeCross}}, \bibinfo {author} {\bibfnamefont {R.}~\bibnamefont {Haghshenas}},
  \bibinfo {author} {\bibfnamefont {M.}~\bibnamefont {Liu}}, \bibinfo {author}
  {\bibfnamefont {Y.}~\bibnamefont {Alexeev}}, \bibinfo {author} {\bibfnamefont
  {C.~H.}\ \bibnamefont {Baldwin}}, \bibinfo {author} {\bibfnamefont {J.~P.}\
  \bibnamefont {Bartolotta}}, \bibinfo {author} {\bibfnamefont
  {M.}~\bibnamefont {Bohn}}, \bibinfo {author} {\bibfnamefont {E.}~\bibnamefont
  {Chertkov}}, \bibinfo {author} {\bibfnamefont {J.}~\bibnamefont {Colina}},
  \bibinfo {author} {\bibfnamefont {D.}~\bibnamefont {DelVento}}, \emph
  {et~al.},\ }\bibfield  {title} {\bibinfo {title} {The computational power of
  random quantum circuits in arbitrary geometries},\ }\href@noop {} {\bibfield
  {journal} {\bibinfo  {journal} {arXiv preprint arXiv:2406.02501}\ } (\bibinfo
  {year} {2024})}\BibitemShut {NoStop}%
\bibitem [{\citenamefont {Bayraktar}\ \emph {et~al.}(2023)\citenamefont
  {Bayraktar}, \citenamefont {Charara}, \citenamefont {Clark}, \citenamefont
  {Cohen}, \citenamefont {Costa}, \citenamefont {Fang}, \citenamefont {Gao},
  \citenamefont {Guan}, \citenamefont {Gunnels}, \citenamefont {Haidar},
  \citenamefont {Hehn}, \citenamefont {Hohnerbach}, \citenamefont {Jones},
  \citenamefont {Lubowe}, \citenamefont {Lyakh}, \citenamefont {Morino},
  \citenamefont {Springer}, \citenamefont {Stanwyck}, \citenamefont
  {Terentyev}, \citenamefont {Varadhan}, \citenamefont {Wong},\ and\
  \citenamefont {Yamaguchi}}]{cuquantum}%
  \BibitemOpen
  \bibfield  {author} {\bibinfo {author} {\bibfnamefont {H.}~\bibnamefont
  {Bayraktar}}, \bibinfo {author} {\bibfnamefont {A.}~\bibnamefont {Charara}},
  \bibinfo {author} {\bibfnamefont {D.}~\bibnamefont {Clark}}, \bibinfo
  {author} {\bibfnamefont {S.}~\bibnamefont {Cohen}}, \bibinfo {author}
  {\bibfnamefont {T.}~\bibnamefont {Costa}}, \bibinfo {author} {\bibfnamefont
  {Y.-L.~L.}\ \bibnamefont {Fang}}, \bibinfo {author} {\bibfnamefont
  {Y.}~\bibnamefont {Gao}}, \bibinfo {author} {\bibfnamefont {J.}~\bibnamefont
  {Guan}}, \bibinfo {author} {\bibfnamefont {J.}~\bibnamefont {Gunnels}},
  \bibinfo {author} {\bibfnamefont {A.}~\bibnamefont {Haidar}}, \bibinfo
  {author} {\bibfnamefont {A.}~\bibnamefont {Hehn}}, \bibinfo {author}
  {\bibfnamefont {M.}~\bibnamefont {Hohnerbach}}, \bibinfo {author}
  {\bibfnamefont {M.}~\bibnamefont {Jones}}, \bibinfo {author} {\bibfnamefont
  {T.}~\bibnamefont {Lubowe}}, \bibinfo {author} {\bibfnamefont
  {D.}~\bibnamefont {Lyakh}}, \bibinfo {author} {\bibfnamefont
  {S.}~\bibnamefont {Morino}}, \bibinfo {author} {\bibfnamefont
  {P.}~\bibnamefont {Springer}}, \bibinfo {author} {\bibfnamefont
  {S.}~\bibnamefont {Stanwyck}}, \bibinfo {author} {\bibfnamefont
  {I.}~\bibnamefont {Terentyev}}, \bibinfo {author} {\bibfnamefont
  {S.}~\bibnamefont {Varadhan}}, \bibinfo {author} {\bibfnamefont
  {J.}~\bibnamefont {Wong}},\ and\ \bibinfo {author} {\bibfnamefont
  {T.}~\bibnamefont {Yamaguchi}},\ }\href
  {https://doi.org/10.1109/QCE57702.2023.00119} {\bibinfo {title} {{cuQuantum
  SDK: A High-Performance Library for Accelerating Quantum Science}}} (\bibinfo
  {year} {2023})\BibitemShut {NoStop}%
\bibitem [{\citenamefont {Gray}(2018)}]{gray2018quimb}%
  \BibitemOpen
  \bibfield  {author} {\bibinfo {author} {\bibfnamefont {J.}~\bibnamefont
  {Gray}},\ }\bibfield  {title} {\bibinfo {title} {{quimb: A python package for
  quantum information and many-body calculations}},\ }\href@noop {} {\bibfield
  {journal} {\bibinfo  {journal} {Journal of Open Source Software}\ }\textbf
  {\bibinfo {volume} {3}},\ \bibinfo {pages} {819} (\bibinfo {year}
  {2018})}\BibitemShut {NoStop}%
\bibitem [{\citenamefont {Fishman}\ \emph {et~al.}(2022)\citenamefont
  {Fishman}, \citenamefont {White},\ and\ \citenamefont
  {Stoudenmire}}]{fishman2022itensor}%
  \BibitemOpen
  \bibfield  {author} {\bibinfo {author} {\bibfnamefont {M.}~\bibnamefont
  {Fishman}}, \bibinfo {author} {\bibfnamefont {S.}~\bibnamefont {White}},\
  and\ \bibinfo {author} {\bibfnamefont {E.}~\bibnamefont {Stoudenmire}},\
  }\bibfield  {title} {\bibinfo {title} {{The ITensor software library for
  tensor network calculations}},\ }\href@noop {} {\bibfield  {journal}
  {\bibinfo  {journal} {SciPost Physics Codebases}\ ,\ \bibinfo {pages} {004}}
  (\bibinfo {year} {2022})}\BibitemShut {NoStop}%
\bibitem [{\citenamefont {Gray}\ and\ \citenamefont
  {Kourtis}(2021)}]{gray2021hyper}%
  \BibitemOpen
  \bibfield  {author} {\bibinfo {author} {\bibfnamefont {J.}~\bibnamefont
  {Gray}}\ and\ \bibinfo {author} {\bibfnamefont {S.}~\bibnamefont {Kourtis}},\
  }\bibfield  {title} {\bibinfo {title} {Hyper-optimized tensor network
  contraction},\ }\href@noop {} {\bibfield  {journal} {\bibinfo  {journal}
  {Quantum}\ }\textbf {\bibinfo {volume} {5}},\ \bibinfo {pages} {410}
  (\bibinfo {year} {2021})}\BibitemShut {NoStop}%
\bibitem [{\citenamefont {Javadi-Abhari}\ \emph {et~al.}(2024)\citenamefont
  {Javadi-Abhari}, \citenamefont {Treinish}, \citenamefont {Krsulich},
  \citenamefont {Wood}, \citenamefont {Lishman}, \citenamefont {Gacon},
  \citenamefont {Martiel}, \citenamefont {Nation}, \citenamefont {Bishop},
  \citenamefont {Cross} \emph {et~al.}}]{javadi2024quantum}%
  \BibitemOpen
  \bibfield  {author} {\bibinfo {author} {\bibfnamefont {A.}~\bibnamefont
  {Javadi-Abhari}}, \bibinfo {author} {\bibfnamefont {M.}~\bibnamefont
  {Treinish}}, \bibinfo {author} {\bibfnamefont {K.}~\bibnamefont {Krsulich}},
  \bibinfo {author} {\bibfnamefont {C.~J.}\ \bibnamefont {Wood}}, \bibinfo
  {author} {\bibfnamefont {J.}~\bibnamefont {Lishman}}, \bibinfo {author}
  {\bibfnamefont {J.}~\bibnamefont {Gacon}}, \bibinfo {author} {\bibfnamefont
  {S.}~\bibnamefont {Martiel}}, \bibinfo {author} {\bibfnamefont {P.~D.}\
  \bibnamefont {Nation}}, \bibinfo {author} {\bibfnamefont {L.~S.}\
  \bibnamefont {Bishop}}, \bibinfo {author} {\bibfnamefont {A.~W.}\
  \bibnamefont {Cross}}, \emph {et~al.},\ }\bibfield  {title} {\bibinfo {title}
  {Quantum computing with qiskit},\ }\href@noop {} {\bibfield  {journal}
  {\bibinfo  {journal} {arXiv preprint arXiv:2405.08810}\ } (\bibinfo {year}
  {2024})}\BibitemShut {NoStop}%
\bibitem [{\citenamefont {Torlai}\ and\ \citenamefont
  {Fishman}(2020)}]{pastaq}%
  \BibitemOpen
  \bibfield  {author} {\bibinfo {author} {\bibfnamefont {G.}~\bibnamefont
  {Torlai}}\ and\ \bibinfo {author} {\bibfnamefont {M.}~\bibnamefont
  {Fishman}},\ }\href@noop {} {\bibinfo {title} {\mbox{PastaQ}: A package for
  simulation, tomography and analysis of quantum computers}} (\bibinfo {year}
  {2020})\BibitemShut {NoStop}%
\bibitem [{\citenamefont {Hauschild}\ and\ \citenamefont
  {Pollmann}(2018)}]{hauschild2018efficient}%
  \BibitemOpen
  \bibfield  {author} {\bibinfo {author} {\bibfnamefont {J.}~\bibnamefont
  {Hauschild}}\ and\ \bibinfo {author} {\bibfnamefont {F.}~\bibnamefont
  {Pollmann}},\ }\bibfield  {title} {\bibinfo {title} {{Efficient numerical
  simulations with tensor networks: Tensor Network Python (TeNPy)}},\
  }\href@noop {} {\bibfield  {journal} {\bibinfo  {journal} {SciPost Physics
  Lecture Notes}\ ,\ \bibinfo {pages} {005}} (\bibinfo {year}
  {2018})}\BibitemShut {NoStop}%
\bibitem [{\citenamefont {Alvarez}(2009)}]{alvarez2009density}%
  \BibitemOpen
  \bibfield  {author} {\bibinfo {author} {\bibfnamefont {G.}~\bibnamefont
  {Alvarez}},\ }\bibfield  {title} {\bibinfo {title} {The density matrix
  renormalization group for strongly correlated electron systems: A generic
  implementation},\ }\href@noop {} {\bibfield  {journal} {\bibinfo  {journal}
  {Computer Physics Communications}\ }\textbf {\bibinfo {volume} {180}},\
  \bibinfo {pages} {1572} (\bibinfo {year} {2009})}\BibitemShut {NoStop}%
\bibitem [{\citenamefont {Rams}\ \emph {et~al.}(2024)\citenamefont {Rams},
  \citenamefont {W{\'o}jtowicz}, \citenamefont {Sinha},\ and\ \citenamefont
  {Hasik}}]{rams2024yastn}%
  \BibitemOpen
  \bibfield  {author} {\bibinfo {author} {\bibfnamefont {M.~M.}\ \bibnamefont
  {Rams}}, \bibinfo {author} {\bibfnamefont {G.}~\bibnamefont {W{\'o}jtowicz}},
  \bibinfo {author} {\bibfnamefont {A.}~\bibnamefont {Sinha}},\ and\ \bibinfo
  {author} {\bibfnamefont {J.}~\bibnamefont {Hasik}},\ }\bibfield  {title}
  {\bibinfo {title} {Yastn: Yet another symmetric tensor networks; a python
  library for abelian symmetric tensor network calculations},\ }\href@noop {}
  {\bibfield  {journal} {\bibinfo  {journal} {arXiv preprint arXiv:2405.12196}\
  } (\bibinfo {year} {2024})}\BibitemShut {NoStop}%
\bibitem [{\citenamefont {Luo}\ \emph {et~al.}(2020)\citenamefont {Luo},
  \citenamefont {Liu}, \citenamefont {Zhang},\ and\ \citenamefont
  {Wang}}]{luo2020yao}%
  \BibitemOpen
  \bibfield  {author} {\bibinfo {author} {\bibfnamefont {X.-Z.}\ \bibnamefont
  {Luo}}, \bibinfo {author} {\bibfnamefont {J.-G.}\ \bibnamefont {Liu}},
  \bibinfo {author} {\bibfnamefont {P.}~\bibnamefont {Zhang}},\ and\ \bibinfo
  {author} {\bibfnamefont {L.}~\bibnamefont {Wang}},\ }\bibfield  {title}
  {\bibinfo {title} {Yao. jl: Extensible, efficient framework for quantum
  algorithm design},\ }\href@noop {} {\bibfield  {journal} {\bibinfo  {journal}
  {Quantum}\ }\textbf {\bibinfo {volume} {4}},\ \bibinfo {pages} {341}
  (\bibinfo {year} {2020})}\BibitemShut {NoStop}%
\bibitem [{\citenamefont {Brennan}\ \emph {et~al.}(2022)\citenamefont
  {Brennan}, \citenamefont {O’Riordan}, \citenamefont {Hanley}, \citenamefont
  {Doyle}, \citenamefont {Allalen}, \citenamefont {Brayford}, \citenamefont
  {Iapichino},\ and\ \citenamefont {Moran}}]{brennan2022qxtools}%
  \BibitemOpen
  \bibfield  {author} {\bibinfo {author} {\bibfnamefont {J.}~\bibnamefont
  {Brennan}}, \bibinfo {author} {\bibfnamefont {L.}~\bibnamefont
  {O’Riordan}}, \bibinfo {author} {\bibfnamefont {K.}~\bibnamefont {Hanley}},
  \bibinfo {author} {\bibfnamefont {M.}~\bibnamefont {Doyle}}, \bibinfo
  {author} {\bibfnamefont {M.}~\bibnamefont {Allalen}}, \bibinfo {author}
  {\bibfnamefont {D.}~\bibnamefont {Brayford}}, \bibinfo {author}
  {\bibfnamefont {L.}~\bibnamefont {Iapichino}},\ and\ \bibinfo {author}
  {\bibfnamefont {N.}~\bibnamefont {Moran}},\ }\bibfield  {title} {\bibinfo
  {title} {{QXTools: A Julia framework for distributed quantum circuit
  simulation}},\ }\href@noop {} {\bibfield  {journal} {\bibinfo  {journal}
  {Journal of Open Source Software}\ }\textbf {\bibinfo {volume} {7}},\
  \bibinfo {pages} {3711} (\bibinfo {year} {2022})}\BibitemShut {NoStop}%
\bibitem [{\citenamefont {Strano}\ and\ \citenamefont
  {Bollay}(2024)}]{strano2024qrack}%
  \BibitemOpen
  \bibfield  {author} {\bibinfo {author} {\bibfnamefont {D.}~\bibnamefont
  {Strano}}\ and\ \bibinfo {author} {\bibfnamefont {B.}~\bibnamefont
  {Bollay}},\ }\href@noop {} {\bibinfo {title} {{unitaryfund/qrack}}} (\bibinfo
  {year} {2017--2024})\BibitemShut {NoStop}%
\bibitem [{\citenamefont {Zhang}\ \emph {et~al.}(2023)\citenamefont {Zhang},
  \citenamefont {Allcock}, \citenamefont {Wan}, \citenamefont {Liu},
  \citenamefont {Sun}, \citenamefont {Yu}, \citenamefont {Yang}, \citenamefont
  {Qiu}, \citenamefont {Ye}, \citenamefont {Chen} \emph
  {et~al.}}]{zhang2023tensorcircuit}%
  \BibitemOpen
  \bibfield  {author} {\bibinfo {author} {\bibfnamefont {S.-X.}\ \bibnamefont
  {Zhang}}, \bibinfo {author} {\bibfnamefont {J.}~\bibnamefont {Allcock}},
  \bibinfo {author} {\bibfnamefont {Z.-Q.}\ \bibnamefont {Wan}}, \bibinfo
  {author} {\bibfnamefont {S.}~\bibnamefont {Liu}}, \bibinfo {author}
  {\bibfnamefont {J.}~\bibnamefont {Sun}}, \bibinfo {author} {\bibfnamefont
  {H.}~\bibnamefont {Yu}}, \bibinfo {author} {\bibfnamefont {X.-H.}\
  \bibnamefont {Yang}}, \bibinfo {author} {\bibfnamefont {J.}~\bibnamefont
  {Qiu}}, \bibinfo {author} {\bibfnamefont {Z.}~\bibnamefont {Ye}}, \bibinfo
  {author} {\bibfnamefont {Y.-Q.}\ \bibnamefont {Chen}}, \emph {et~al.},\
  }\bibfield  {title} {\bibinfo {title} {{TensorCircuit: a Quantum Software
  Framework for the NISQ Era}},\ }\href@noop {} {\bibfield  {journal} {\bibinfo
   {journal} {Quantum}\ }\textbf {\bibinfo {volume} {7}},\ \bibinfo {pages}
  {912} (\bibinfo {year} {2023})}\BibitemShut {NoStop}%
\bibitem [{\citenamefont {Lykov}\ \emph {et~al.}(2021)\citenamefont {Lykov},
  \citenamefont {Chen}, \citenamefont {Chen}, \citenamefont {Keipert},
  \citenamefont {Zhang}, \citenamefont {Gibbs},\ and\ \citenamefont
  {Alexeev}}]{lykov2021qtensor}%
  \BibitemOpen
  \bibfield  {author} {\bibinfo {author} {\bibfnamefont {D.}~\bibnamefont
  {Lykov}}, \bibinfo {author} {\bibfnamefont {A.}~\bibnamefont {Chen}},
  \bibinfo {author} {\bibfnamefont {H.}~\bibnamefont {Chen}}, \bibinfo {author}
  {\bibfnamefont {K.}~\bibnamefont {Keipert}}, \bibinfo {author} {\bibfnamefont
  {Z.}~\bibnamefont {Zhang}}, \bibinfo {author} {\bibfnamefont
  {T.}~\bibnamefont {Gibbs}},\ and\ \bibinfo {author} {\bibfnamefont
  {Y.}~\bibnamefont {Alexeev}},\ }\href
  {https://doi.org/10.1109/QCS54837.2021.00007} {\bibinfo {title} {{Performance
  Evaluation and Acceleration of the QTensor Quantum Circuit Simulator on
  GPUs}}} (\bibinfo {year} {2021})\BibitemShut {NoStop}%
\bibitem [{\citenamefont {Zhang}\ \emph {et~al.}(2019)\citenamefont {Zhang},
  \citenamefont {Huang}, \citenamefont {Newman}, \citenamefont {Cai},
  \citenamefont {Yu}, \citenamefont {Tian}, \citenamefont {Yuan}, \citenamefont
  {Xu}, \citenamefont {Wu}, \citenamefont {Gao} \emph
  {et~al.}}]{zhang2019alibaba}%
  \BibitemOpen
  \bibfield  {author} {\bibinfo {author} {\bibfnamefont {F.}~\bibnamefont
  {Zhang}}, \bibinfo {author} {\bibfnamefont {C.}~\bibnamefont {Huang}},
  \bibinfo {author} {\bibfnamefont {M.}~\bibnamefont {Newman}}, \bibinfo
  {author} {\bibfnamefont {J.}~\bibnamefont {Cai}}, \bibinfo {author}
  {\bibfnamefont {H.}~\bibnamefont {Yu}}, \bibinfo {author} {\bibfnamefont
  {Z.}~\bibnamefont {Tian}}, \bibinfo {author} {\bibfnamefont {B.}~\bibnamefont
  {Yuan}}, \bibinfo {author} {\bibfnamefont {H.}~\bibnamefont {Xu}}, \bibinfo
  {author} {\bibfnamefont {J.}~\bibnamefont {Wu}}, \bibinfo {author}
  {\bibfnamefont {X.}~\bibnamefont {Gao}}, \emph {et~al.},\ }\bibfield  {title}
  {\bibinfo {title} {{Alibaba cloud quantum development platform: Large-scale
  classical simulation of quantum circuits}},\ }\href@noop {} {\bibfield
  {journal} {\bibinfo  {journal} {arXiv preprint arXiv:1907.11217}\ } (\bibinfo
  {year} {2019})}\BibitemShut {NoStop}%
\bibitem [{\citenamefont {Villalonga}\ \emph {et~al.}(2020)\citenamefont
  {Villalonga}, \citenamefont {Lyakh}, \citenamefont {Boixo}, \citenamefont
  {Neven}, \citenamefont {Humble}, \citenamefont {Biswas}, \citenamefont
  {Rieffel}, \citenamefont {Ho},\ and\ \citenamefont
  {Mandr{\`a}}}]{villalonga2020establishing}%
  \BibitemOpen
  \bibfield  {author} {\bibinfo {author} {\bibfnamefont {B.}~\bibnamefont
  {Villalonga}}, \bibinfo {author} {\bibfnamefont {D.}~\bibnamefont {Lyakh}},
  \bibinfo {author} {\bibfnamefont {S.}~\bibnamefont {Boixo}}, \bibinfo
  {author} {\bibfnamefont {H.}~\bibnamefont {Neven}}, \bibinfo {author}
  {\bibfnamefont {T.~S.}\ \bibnamefont {Humble}}, \bibinfo {author}
  {\bibfnamefont {R.}~\bibnamefont {Biswas}}, \bibinfo {author} {\bibfnamefont
  {E.~G.}\ \bibnamefont {Rieffel}}, \bibinfo {author} {\bibfnamefont
  {A.}~\bibnamefont {Ho}},\ and\ \bibinfo {author} {\bibfnamefont
  {S.}~\bibnamefont {Mandr{\`a}}},\ }\bibfield  {title} {\bibinfo {title}
  {{Establishing the quantum supremacy frontier with a 281 Pflop/s
  simulation}},\ }\href@noop {} {\bibfield  {journal} {\bibinfo  {journal}
  {Quantum Science and Technology}\ }\textbf {\bibinfo {volume} {5}},\ \bibinfo
  {pages} {034003} (\bibinfo {year} {2020})}\BibitemShut {NoStop}%
\bibitem [{\citenamefont {Mandrà}\ \emph {et~al.}(2021)\citenamefont
  {Mandrà}, \citenamefont {Marshall}, \citenamefont {Rieffel},\ and\
  \citenamefont {Biswas}}]{salvatore2021hybridq}%
  \BibitemOpen
  \bibfield  {author} {\bibinfo {author} {\bibfnamefont {S.}~\bibnamefont
  {Mandrà}}, \bibinfo {author} {\bibfnamefont {J.}~\bibnamefont {Marshall}},
  \bibinfo {author} {\bibfnamefont {E.~G.}\ \bibnamefont {Rieffel}},\ and\
  \bibinfo {author} {\bibfnamefont {R.}~\bibnamefont {Biswas}},\ }\href
  {https://doi.org/10.1109/QCS54837.2021.00015} {\bibinfo {title} {{HybridQ: A
  Hybrid Simulator for Quantum Circuits}}} (\bibinfo {year} {2021})\BibitemShut
  {NoStop}%
\bibitem [{\citenamefont {Lyakh}\ \emph {et~al.}(2022)\citenamefont {Lyakh},
  \citenamefont {Nguyen}, \citenamefont {Claudino}, \citenamefont
  {Dumitrescu},\ and\ \citenamefont {McCaskey}}]{lyakh2022exatn}%
  \BibitemOpen
  \bibfield  {author} {\bibinfo {author} {\bibfnamefont {D.~I.}\ \bibnamefont
  {Lyakh}}, \bibinfo {author} {\bibfnamefont {T.}~\bibnamefont {Nguyen}},
  \bibinfo {author} {\bibfnamefont {D.}~\bibnamefont {Claudino}}, \bibinfo
  {author} {\bibfnamefont {E.}~\bibnamefont {Dumitrescu}},\ and\ \bibinfo
  {author} {\bibfnamefont {A.~J.}\ \bibnamefont {McCaskey}},\ }\bibfield
  {title} {\bibinfo {title} {{ExaTN: Scalable GPU-Accelerated High-Performance
  Processing of General Tensor Networks at Exascale}},\ }\href@noop {}
  {\bibfield  {journal} {\bibinfo  {journal} {Frontiers in Applied Mathematics
  and Statistics}\ }\textbf {\bibinfo {volume} {8}},\ \bibinfo {pages} {838601}
  (\bibinfo {year} {2022})}\BibitemShut {NoStop}%
\bibitem [{\citenamefont {Zhai}\ \emph {et~al.}(2023)\citenamefont {Zhai} \emph
  {et~al.}}]{zhai2023block2}%
  \BibitemOpen
  \bibfield  {author} {\bibinfo {author} {\bibfnamefont {H.}~\bibnamefont
  {Zhai}} \emph {et~al.},\ }\bibfield  {title} {\bibinfo {title} {Block2: A
  comprehensive open source framework to develop and apply state-of-the-art
  dmrg algorithms in electronic structure and beyond},\ }\href
  {https://doi.org/10.1063/5.0180424} {\bibfield  {journal} {\bibinfo
  {journal} {The Journal of Chemical Physics}\ }\textbf {\bibinfo {volume}
  {159}},\ \bibinfo {pages} {234801} (\bibinfo {year} {2023})}\BibitemShut
  {NoStop}%
\bibitem [{\citenamefont {White}(1992)}]{white1992density}%
  \BibitemOpen
  \bibfield  {author} {\bibinfo {author} {\bibfnamefont {S.~R.}\ \bibnamefont
  {White}},\ }\bibfield  {title} {\bibinfo {title} {Density matrix formulation
  for quantum renormalization groups},\ }\href@noop {} {\bibfield  {journal}
  {\bibinfo  {journal} {Physical review letters}\ }\textbf {\bibinfo {volume}
  {69}},\ \bibinfo {pages} {2863} (\bibinfo {year} {1992})}\BibitemShut
  {NoStop}%
\bibitem [{\citenamefont {Wang}\ \emph {et~al.}(2023)\citenamefont {Wang},
  \citenamefont {Pan}, \citenamefont {Xu}, \citenamefont {Yang}, \citenamefont
  {Li},\ and\ \citenamefont {Cichocki}}]{wang2023tensor}%
  \BibitemOpen
  \bibfield  {author} {\bibinfo {author} {\bibfnamefont {M.}~\bibnamefont
  {Wang}}, \bibinfo {author} {\bibfnamefont {Y.}~\bibnamefont {Pan}}, \bibinfo
  {author} {\bibfnamefont {Z.}~\bibnamefont {Xu}}, \bibinfo {author}
  {\bibfnamefont {X.}~\bibnamefont {Yang}}, \bibinfo {author} {\bibfnamefont
  {G.}~\bibnamefont {Li}},\ and\ \bibinfo {author} {\bibfnamefont
  {A.}~\bibnamefont {Cichocki}},\ }\bibfield  {title} {\bibinfo {title} {Tensor
  networks meet neural networks: A survey and future perspectives},\
  }\href@noop {} {\bibfield  {journal} {\bibinfo  {journal} {arXiv preprint
  arXiv:2302.09019}\ } (\bibinfo {year} {2023})}\BibitemShut {NoStop}%
\bibitem [{\citenamefont {Liao}\ \emph {et~al.}(2019)\citenamefont {Liao},
  \citenamefont {Liu}, \citenamefont {Wang},\ and\ \citenamefont
  {Xiang}}]{liao2019differentiable}%
  \BibitemOpen
  \bibfield  {author} {\bibinfo {author} {\bibfnamefont {H.-J.}\ \bibnamefont
  {Liao}}, \bibinfo {author} {\bibfnamefont {J.-G.}\ \bibnamefont {Liu}},
  \bibinfo {author} {\bibfnamefont {L.}~\bibnamefont {Wang}},\ and\ \bibinfo
  {author} {\bibfnamefont {T.}~\bibnamefont {Xiang}},\ }\bibfield  {title}
  {\bibinfo {title} {Differentiable programming tensor networks},\ }\href@noop
  {} {\bibfield  {journal} {\bibinfo  {journal} {Physical Review X}\ }\textbf
  {\bibinfo {volume} {9}},\ \bibinfo {pages} {031041} (\bibinfo {year}
  {2019})}\BibitemShut {NoStop}%
\bibitem [{\citenamefont {Hauru}\ \emph {et~al.}(2021)\citenamefont {Hauru},
  \citenamefont {Van~Damme},\ and\ \citenamefont
  {Haegeman}}]{hauru2021riemannian}%
  \BibitemOpen
  \bibfield  {author} {\bibinfo {author} {\bibfnamefont {M.}~\bibnamefont
  {Hauru}}, \bibinfo {author} {\bibfnamefont {M.}~\bibnamefont {Van~Damme}},\
  and\ \bibinfo {author} {\bibfnamefont {J.}~\bibnamefont {Haegeman}},\
  }\bibfield  {title} {\bibinfo {title} {Riemannian optimization of isometric
  tensor networks},\ }\href@noop {} {\bibfield  {journal} {\bibinfo  {journal}
  {SciPost Phys}\ }\textbf {\bibinfo {volume} {10}},\ \bibinfo {pages} {040}
  (\bibinfo {year} {2021})}\BibitemShut {NoStop}%
\bibitem [{\citenamefont {Luchnikov}\ \emph
  {et~al.}(2021{\natexlab{a}})\citenamefont {Luchnikov}, \citenamefont
  {Ryzhov}, \citenamefont {Filippov},\ and\ \citenamefont
  {Ouerdane}}]{luchnikov2021qgopt}%
  \BibitemOpen
  \bibfield  {author} {\bibinfo {author} {\bibfnamefont {I.}~\bibnamefont
  {Luchnikov}}, \bibinfo {author} {\bibfnamefont {A.}~\bibnamefont {Ryzhov}},
  \bibinfo {author} {\bibfnamefont {S.}~\bibnamefont {Filippov}},\ and\
  \bibinfo {author} {\bibfnamefont {H.}~\bibnamefont {Ouerdane}},\ }\bibfield
  {title} {\bibinfo {title} {{QGOpt: Riemannian optimization for quantum
  technologies}},\ }\href@noop {} {\bibfield  {journal} {\bibinfo  {journal}
  {SciPost Physics}\ }\textbf {\bibinfo {volume} {10}},\ \bibinfo {pages} {079}
  (\bibinfo {year} {2021}{\natexlab{a}})}\BibitemShut {NoStop}%
\bibitem [{\citenamefont {Berezutskii}\ \emph {et~al.}(2025)\citenamefont
  {Berezutskii}, \citenamefont {Luchnikov},\ and\ \citenamefont
  {Fedorov}}]{berezutskii2025simulating}%
  \BibitemOpen
  \bibfield  {author} {\bibinfo {author} {\bibfnamefont {A.}~\bibnamefont
  {Berezutskii}}, \bibinfo {author} {\bibfnamefont {I.}~\bibnamefont
  {Luchnikov}},\ and\ \bibinfo {author} {\bibfnamefont {A.}~\bibnamefont
  {Fedorov}},\ }\bibfield  {title} {\bibinfo {title} {Simulating quantum
  circuits using the multi-scale entanglement renormalization ansatz},\
  }\href@noop {} {\bibfield  {journal} {\bibinfo  {journal} {Physical Review
  Research}\ }\textbf {\bibinfo {volume} {7}},\ \bibinfo {pages} {013063}
  (\bibinfo {year} {2025})}\BibitemShut {NoStop}%
\bibitem [{\citenamefont {Luchnikov}\ \emph
  {et~al.}(2021{\natexlab{b}})\citenamefont {Luchnikov}, \citenamefont
  {Krechetov},\ and\ \citenamefont {Filippov}}]{luchnikov2021riemannian}%
  \BibitemOpen
  \bibfield  {author} {\bibinfo {author} {\bibfnamefont {I.}~\bibnamefont
  {Luchnikov}}, \bibinfo {author} {\bibfnamefont {M.}~\bibnamefont
  {Krechetov}},\ and\ \bibinfo {author} {\bibfnamefont {S.}~\bibnamefont
  {Filippov}},\ }\bibfield  {title} {\bibinfo {title} {Riemannian geometry and
  automatic differentiation for optimization problems of quantum physics and
  quantum technologies},\ }\href@noop {} {\bibfield  {journal} {\bibinfo
  {journal} {New Journal of Physics}\ } (\bibinfo {year}
  {2021}{\natexlab{b}})}\BibitemShut {NoStop}%
\bibitem [{\citenamefont {Sandvik}\ and\ \citenamefont
  {Vidal}(2007)}]{sandvik2007variational}%
  \BibitemOpen
  \bibfield  {author} {\bibinfo {author} {\bibfnamefont {A.~W.}\ \bibnamefont
  {Sandvik}}\ and\ \bibinfo {author} {\bibfnamefont {G.}~\bibnamefont
  {Vidal}},\ }\bibfield  {title} {\bibinfo {title} {{Variational quantum Monte
  Carlo simulations with tensor-network states}},\ }\href@noop {} {\bibfield
  {journal} {\bibinfo  {journal} {Physical review letters}\ }\textbf {\bibinfo
  {volume} {99}},\ \bibinfo {pages} {220602} (\bibinfo {year}
  {2007})}\BibitemShut {NoStop}%
\bibitem [{\citenamefont {Wang}\ \emph {et~al.}(2011)\citenamefont {Wang},
  \citenamefont {Pi{\v{z}}orn},\ and\ \citenamefont
  {Verstraete}}]{wang2011monte}%
  \BibitemOpen
  \bibfield  {author} {\bibinfo {author} {\bibfnamefont {L.}~\bibnamefont
  {Wang}}, \bibinfo {author} {\bibfnamefont {I.}~\bibnamefont {Pi{\v{z}}orn}},\
  and\ \bibinfo {author} {\bibfnamefont {F.}~\bibnamefont {Verstraete}},\
  }\bibfield  {title} {\bibinfo {title} {{Monte Carlo simulation with tensor
  network states}},\ }\href@noop {} {\bibfield  {journal} {\bibinfo  {journal}
  {Physical Review B—Condensed Matter and Materials Physics}\ }\textbf
  {\bibinfo {volume} {83}},\ \bibinfo {pages} {134421} (\bibinfo {year}
  {2011})}\BibitemShut {NoStop}%
\bibitem [{\citenamefont {Paeckel}\ \emph {et~al.}(2019)\citenamefont
  {Paeckel}, \citenamefont {K{\"o}hler}, \citenamefont {Swoboda}, \citenamefont
  {Manmana}, \citenamefont {Schollw{\"o}ck},\ and\ \citenamefont
  {Hubig}}]{paeckel2019time}%
  \BibitemOpen
  \bibfield  {author} {\bibinfo {author} {\bibfnamefont {S.}~\bibnamefont
  {Paeckel}}, \bibinfo {author} {\bibfnamefont {T.}~\bibnamefont {K{\"o}hler}},
  \bibinfo {author} {\bibfnamefont {A.}~\bibnamefont {Swoboda}}, \bibinfo
  {author} {\bibfnamefont {S.~R.}\ \bibnamefont {Manmana}}, \bibinfo {author}
  {\bibfnamefont {U.}~\bibnamefont {Schollw{\"o}ck}},\ and\ \bibinfo {author}
  {\bibfnamefont {C.}~\bibnamefont {Hubig}},\ }\bibfield  {title} {\bibinfo
  {title} {Time-evolution methods for matrix-product states},\ }\href@noop {}
  {\bibfield  {journal} {\bibinfo  {journal} {Annals of Physics}\ }\textbf
  {\bibinfo {volume} {411}},\ \bibinfo {pages} {167998} (\bibinfo {year}
  {2019})}\BibitemShut {NoStop}%
\bibitem [{\citenamefont {Feiguin}\ and\ \citenamefont
  {White}(2005)}]{feiguin2005time}%
  \BibitemOpen
  \bibfield  {author} {\bibinfo {author} {\bibfnamefont {A.~E.}\ \bibnamefont
  {Feiguin}}\ and\ \bibinfo {author} {\bibfnamefont {S.~R.}\ \bibnamefont
  {White}},\ }\bibfield  {title} {\bibinfo {title} {Time-step targeting methods
  for real-time dynamics using the density matrix renormalization group},\
  }\href@noop {} {\bibfield  {journal} {\bibinfo  {journal} {Physical Review
  B—Condensed Matter and Materials Physics}\ }\textbf {\bibinfo {volume}
  {72}},\ \bibinfo {pages} {020404} (\bibinfo {year} {2005})}\BibitemShut
  {NoStop}%
\bibitem [{\citenamefont {Alvarez}\ \emph {et~al.}(2011)\citenamefont
  {Alvarez}, \citenamefont {Dias~da Silva}, \citenamefont {Ponce},\ and\
  \citenamefont {Dagotto}}]{alvarez2011time}%
  \BibitemOpen
  \bibfield  {author} {\bibinfo {author} {\bibfnamefont {G.}~\bibnamefont
  {Alvarez}}, \bibinfo {author} {\bibfnamefont {L.~G.}\ \bibnamefont {Dias~da
  Silva}}, \bibinfo {author} {\bibfnamefont {E.}~\bibnamefont {Ponce}},\ and\
  \bibinfo {author} {\bibfnamefont {E.}~\bibnamefont {Dagotto}},\ }\bibfield
  {title} {\bibinfo {title} {Time evolution with the density-matrix
  renormalization-group algorithm: A generic implementation for strongly
  correlated electronic systems},\ }\href@noop {} {\bibfield  {journal}
  {\bibinfo  {journal} {Physical Review E—Statistical, Nonlinear, and Soft
  Matter Physics}\ }\textbf {\bibinfo {volume} {84}},\ \bibinfo {pages}
  {056706} (\bibinfo {year} {2011})}\BibitemShut {NoStop}%
\bibitem [{\citenamefont {Haegeman}\ \emph {et~al.}(2011)\citenamefont
  {Haegeman}, \citenamefont {Cirac}, \citenamefont {Osborne}, \citenamefont
  {Pi{\v{z}}orn}, \citenamefont {Verschelde},\ and\ \citenamefont
  {Verstraete}}]{haegeman2011time}%
  \BibitemOpen
  \bibfield  {author} {\bibinfo {author} {\bibfnamefont {J.}~\bibnamefont
  {Haegeman}}, \bibinfo {author} {\bibfnamefont {J.~I.}\ \bibnamefont {Cirac}},
  \bibinfo {author} {\bibfnamefont {T.~J.}\ \bibnamefont {Osborne}}, \bibinfo
  {author} {\bibfnamefont {I.}~\bibnamefont {Pi{\v{z}}orn}}, \bibinfo {author}
  {\bibfnamefont {H.}~\bibnamefont {Verschelde}},\ and\ \bibinfo {author}
  {\bibfnamefont {F.}~\bibnamefont {Verstraete}},\ }\bibfield  {title}
  {\bibinfo {title} {Time-dependent variational principle for quantum
  lattices},\ }\href@noop {} {\bibfield  {journal} {\bibinfo  {journal}
  {Physical review letters}\ }\textbf {\bibinfo {volume} {107}},\ \bibinfo
  {pages} {070601} (\bibinfo {year} {2011})}\BibitemShut {NoStop}%
\bibitem [{\citenamefont {Haegeman}\ \emph {et~al.}(2016)\citenamefont
  {Haegeman}, \citenamefont {Lubich}, \citenamefont {Oseledets}, \citenamefont
  {Vandereycken},\ and\ \citenamefont {Verstraete}}]{haegeman2016unifying}%
  \BibitemOpen
  \bibfield  {author} {\bibinfo {author} {\bibfnamefont {J.}~\bibnamefont
  {Haegeman}}, \bibinfo {author} {\bibfnamefont {C.}~\bibnamefont {Lubich}},
  \bibinfo {author} {\bibfnamefont {I.}~\bibnamefont {Oseledets}}, \bibinfo
  {author} {\bibfnamefont {B.}~\bibnamefont {Vandereycken}},\ and\ \bibinfo
  {author} {\bibfnamefont {F.}~\bibnamefont {Verstraete}},\ }\bibfield  {title}
  {\bibinfo {title} {Unifying time evolution and optimization with matrix
  product states},\ }\href@noop {} {\bibfield  {journal} {\bibinfo  {journal}
  {Physical Review B}\ }\textbf {\bibinfo {volume} {94}},\ \bibinfo {pages}
  {165116} (\bibinfo {year} {2016})}\BibitemShut {NoStop}%
\bibitem [{\citenamefont {Ran}\ \emph {et~al.}(2020)\citenamefont {Ran},
  \citenamefont {Sun}, \citenamefont {Fei}, \citenamefont {Su},\ and\
  \citenamefont {Lewenstein}}]{ran2020tensor}%
  \BibitemOpen
  \bibfield  {author} {\bibinfo {author} {\bibfnamefont {S.-J.}\ \bibnamefont
  {Ran}}, \bibinfo {author} {\bibfnamefont {Z.-Z.}\ \bibnamefont {Sun}},
  \bibinfo {author} {\bibfnamefont {S.-M.}\ \bibnamefont {Fei}}, \bibinfo
  {author} {\bibfnamefont {G.}~\bibnamefont {Su}},\ and\ \bibinfo {author}
  {\bibfnamefont {M.}~\bibnamefont {Lewenstein}},\ }\bibfield  {title}
  {\bibinfo {title} {Tensor network compressed sensing with unsupervised
  machine learning},\ }\href@noop {} {\bibfield  {journal} {\bibinfo  {journal}
  {Physical Review Research}\ }\textbf {\bibinfo {volume} {2}},\ \bibinfo
  {pages} {033293} (\bibinfo {year} {2020})}\BibitemShut {NoStop}%
\bibitem [{\citenamefont {Arnborg}\ \emph {et~al.}(1987)\citenamefont
  {Arnborg}, \citenamefont {Corneil},\ and\ \citenamefont
  {Proskurowski}}]{arnborg1987complexity}%
  \BibitemOpen
  \bibfield  {author} {\bibinfo {author} {\bibfnamefont {S.}~\bibnamefont
  {Arnborg}}, \bibinfo {author} {\bibfnamefont {D.~G.}\ \bibnamefont
  {Corneil}},\ and\ \bibinfo {author} {\bibfnamefont {A.}~\bibnamefont
  {Proskurowski}},\ }\bibfield  {title} {\bibinfo {title} {Complexity of
  finding embeddings in ak-tree},\ }\href@noop {} {\bibfield  {journal}
  {\bibinfo  {journal} {SIAM Journal on Algebraic Discrete Methods}\ }\textbf
  {\bibinfo {volume} {8}},\ \bibinfo {pages} {277} (\bibinfo {year}
  {1987})}\BibitemShut {NoStop}%
\bibitem [{\citenamefont {Pfeifer}\ \emph {et~al.}(2014)\citenamefont
  {Pfeifer}, \citenamefont {Haegeman},\ and\ \citenamefont
  {Verstraete}}]{pfeifer2014faster}%
  \BibitemOpen
  \bibfield  {author} {\bibinfo {author} {\bibfnamefont {R.~N.}\ \bibnamefont
  {Pfeifer}}, \bibinfo {author} {\bibfnamefont {J.}~\bibnamefont {Haegeman}},\
  and\ \bibinfo {author} {\bibfnamefont {F.}~\bibnamefont {Verstraete}},\
  }\bibfield  {title} {\bibinfo {title} {Faster identification of optimal
  contraction sequences for tensor networks},\ }\href@noop {} {\bibfield
  {journal} {\bibinfo  {journal} {Physical Review E}\ }\textbf {\bibinfo
  {volume} {90}},\ \bibinfo {pages} {033315} (\bibinfo {year}
  {2014})}\BibitemShut {NoStop}%
\bibitem [{\citenamefont {Kourtis}\ \emph {et~al.}(2019)\citenamefont
  {Kourtis}, \citenamefont {Chamon}, \citenamefont {Mucciolo},\ and\
  \citenamefont {Ruckenstein}}]{kourtis2019fast}%
  \BibitemOpen
  \bibfield  {author} {\bibinfo {author} {\bibfnamefont {S.}~\bibnamefont
  {Kourtis}}, \bibinfo {author} {\bibfnamefont {C.}~\bibnamefont {Chamon}},
  \bibinfo {author} {\bibfnamefont {E.}~\bibnamefont {Mucciolo}},\ and\
  \bibinfo {author} {\bibfnamefont {A.}~\bibnamefont {Ruckenstein}},\
  }\bibfield  {title} {\bibinfo {title} {Fast counting with tensor networks},\
  }\href@noop {} {\bibfield  {journal} {\bibinfo  {journal} {SciPost Physics}\
  }\textbf {\bibinfo {volume} {7}},\ \bibinfo {pages} {060} (\bibinfo {year}
  {2019})}\BibitemShut {NoStop}%
\bibitem [{\citenamefont {Pan}\ and\ \citenamefont
  {Zhang}(2021)}]{pan2021simulating}%
  \BibitemOpen
  \bibfield  {author} {\bibinfo {author} {\bibfnamefont {F.}~\bibnamefont
  {Pan}}\ and\ \bibinfo {author} {\bibfnamefont {P.}~\bibnamefont {Zhang}},\
  }\bibfield  {title} {\bibinfo {title} {{Simulating the Sycamore quantum
  supremacy circuits}},\ }\href@noop {} {\bibfield  {journal} {\bibinfo
  {journal} {arXiv preprint arXiv:2103.03074}\ } (\bibinfo {year}
  {2021})}\BibitemShut {NoStop}%
\bibitem [{\citenamefont {Kalachev}\ \emph
  {et~al.}(2021{\natexlab{a}})\citenamefont {Kalachev}, \citenamefont
  {Panteleev}, \citenamefont {Zhou},\ and\ \citenamefont
  {Yung}}]{kalachev2021classical}%
  \BibitemOpen
  \bibfield  {author} {\bibinfo {author} {\bibfnamefont {G.}~\bibnamefont
  {Kalachev}}, \bibinfo {author} {\bibfnamefont {P.}~\bibnamefont {Panteleev}},
  \bibinfo {author} {\bibfnamefont {P.}~\bibnamefont {Zhou}},\ and\ \bibinfo
  {author} {\bibfnamefont {M.-H.}\ \bibnamefont {Yung}},\ }\bibfield  {title}
  {\bibinfo {title} {Classical sampling of random quantum circuits with bounded
  fidelity},\ }\href@noop {} {\bibfield  {journal} {\bibinfo  {journal} {arXiv
  preprint arXiv:2112.15083}\ } (\bibinfo {year}
  {2021}{\natexlab{a}})}\BibitemShut {NoStop}%
\bibitem [{\citenamefont {Morvan}\ \emph {et~al.}(2024)\citenamefont {Morvan},
  \citenamefont {Villalonga}, \citenamefont {Mi}, \citenamefont {Mandra},
  \citenamefont {Bengtsson}, \citenamefont {Klimov}, \citenamefont {Chen},
  \citenamefont {Hong}, \citenamefont {Erickson}, \citenamefont {Drozdov} \emph
  {et~al.}}]{morvan2023phase}%
  \BibitemOpen
  \bibfield  {author} {\bibinfo {author} {\bibfnamefont {A.}~\bibnamefont
  {Morvan}}, \bibinfo {author} {\bibfnamefont {B.}~\bibnamefont {Villalonga}},
  \bibinfo {author} {\bibfnamefont {X.}~\bibnamefont {Mi}}, \bibinfo {author}
  {\bibfnamefont {S.}~\bibnamefont {Mandra}}, \bibinfo {author} {\bibfnamefont
  {A.}~\bibnamefont {Bengtsson}}, \bibinfo {author} {\bibfnamefont
  {P.}~\bibnamefont {Klimov}}, \bibinfo {author} {\bibfnamefont
  {Z.}~\bibnamefont {Chen}}, \bibinfo {author} {\bibfnamefont {S.}~\bibnamefont
  {Hong}}, \bibinfo {author} {\bibfnamefont {C.}~\bibnamefont {Erickson}},
  \bibinfo {author} {\bibfnamefont {I.}~\bibnamefont {Drozdov}}, \emph
  {et~al.},\ }\bibfield  {title} {\bibinfo {title} {Phase transitions in random
  circuit sampling},\ }\href@noop {} {\bibfield  {journal} {\bibinfo  {journal}
  {Nature}\ }\textbf {\bibinfo {volume} {634}},\ \bibinfo {pages} {328}
  (\bibinfo {year} {2024})}\BibitemShut {NoStop}%
\bibitem [{\citenamefont {Meirom}\ \emph {et~al.}(2022)\citenamefont {Meirom},
  \citenamefont {Maron}, \citenamefont {Mannor},\ and\ \citenamefont
  {Chechik}}]{meirom2022optimizing}%
  \BibitemOpen
  \bibfield  {author} {\bibinfo {author} {\bibfnamefont {E.}~\bibnamefont
  {Meirom}}, \bibinfo {author} {\bibfnamefont {H.}~\bibnamefont {Maron}},
  \bibinfo {author} {\bibfnamefont {S.}~\bibnamefont {Mannor}},\ and\ \bibinfo
  {author} {\bibfnamefont {G.}~\bibnamefont {Chechik}},\ }\href@noop {}
  {\bibinfo {title} {Optimizing tensor network contraction using reinforcement
  learning}} (\bibinfo {year} {2022})\BibitemShut {NoStop}%
\bibitem [{\citenamefont {Liu}\ and\ \citenamefont
  {Zhang}(2023)}]{liu2023classical}%
  \BibitemOpen
  \bibfield  {author} {\bibinfo {author} {\bibfnamefont {X.-Y.}\ \bibnamefont
  {Liu}}\ and\ \bibinfo {author} {\bibfnamefont {Z.}~\bibnamefont {Zhang}},\
  }\href@noop {} {\bibinfo {title} {Classical simulation of quantum circuits
  using reinforcement learning: Parallel environments and benchmark}} (\bibinfo
  {year} {2023})\BibitemShut {NoStop}%
\bibitem [{\citenamefont {Kalachev}\ \emph
  {et~al.}(2021{\natexlab{b}})\citenamefont {Kalachev}, \citenamefont
  {Panteleev},\ and\ \citenamefont {Yung}}]{kalachev2021multi}%
  \BibitemOpen
  \bibfield  {author} {\bibinfo {author} {\bibfnamefont {G.}~\bibnamefont
  {Kalachev}}, \bibinfo {author} {\bibfnamefont {P.}~\bibnamefont
  {Panteleev}},\ and\ \bibinfo {author} {\bibfnamefont {M.-H.}\ \bibnamefont
  {Yung}},\ }\bibfield  {title} {\bibinfo {title} {Multi-tensor contraction for
  xeb verification of quantum circuits},\ }\href@noop {} {\bibfield  {journal}
  {\bibinfo  {journal} {arXiv preprint arXiv:2108.05665}\ } (\bibinfo {year}
  {2021}{\natexlab{b}})}\BibitemShut {NoStop}%
\bibitem [{\citenamefont {Huang}\ \emph {et~al.}(2020)\citenamefont {Huang},
  \citenamefont {Zhang}, \citenamefont {Newman}, \citenamefont {Cai},
  \citenamefont {Gao}, \citenamefont {Tian}, \citenamefont {Wu}, \citenamefont
  {Xu}, \citenamefont {Yu}, \citenamefont {Yuan} \emph
  {et~al.}}]{huang2020classical}%
  \BibitemOpen
  \bibfield  {author} {\bibinfo {author} {\bibfnamefont {C.}~\bibnamefont
  {Huang}}, \bibinfo {author} {\bibfnamefont {F.}~\bibnamefont {Zhang}},
  \bibinfo {author} {\bibfnamefont {M.}~\bibnamefont {Newman}}, \bibinfo
  {author} {\bibfnamefont {J.}~\bibnamefont {Cai}}, \bibinfo {author}
  {\bibfnamefont {X.}~\bibnamefont {Gao}}, \bibinfo {author} {\bibfnamefont
  {Z.}~\bibnamefont {Tian}}, \bibinfo {author} {\bibfnamefont {J.}~\bibnamefont
  {Wu}}, \bibinfo {author} {\bibfnamefont {H.}~\bibnamefont {Xu}}, \bibinfo
  {author} {\bibfnamefont {H.}~\bibnamefont {Yu}}, \bibinfo {author}
  {\bibfnamefont {B.}~\bibnamefont {Yuan}}, \emph {et~al.},\ }\bibfield
  {title} {\bibinfo {title} {Classical simulation of quantum supremacy
  circuits},\ }\href@noop {} {\bibfield  {journal} {\bibinfo  {journal} {arXiv
  preprint arXiv:2005.06787}\ } (\bibinfo {year} {2020})}\BibitemShut {NoStop}%
\bibitem [{\citenamefont {Aaronson}\ and\ \citenamefont
  {Chen}(2016)}]{aaronson2016complexity}%
  \BibitemOpen
  \bibfield  {author} {\bibinfo {author} {\bibfnamefont {S.}~\bibnamefont
  {Aaronson}}\ and\ \bibinfo {author} {\bibfnamefont {L.}~\bibnamefont
  {Chen}},\ }\bibfield  {title} {\bibinfo {title} {Complexity-theoretic
  foundations of quantum supremacy experiments},\ }\href@noop {} {\bibfield
  {journal} {\bibinfo  {journal} {arXiv preprint arXiv:1612.05903}\ } (\bibinfo
  {year} {2016})}\BibitemShut {NoStop}%
\bibitem [{\citenamefont {Chen}\ \emph
  {et~al.}(2018{\natexlab{a}})\citenamefont {Chen}, \citenamefont {Zhang},
  \citenamefont {Huang}, \citenamefont {Newman},\ and\ \citenamefont
  {Shi}}]{chen2018classical}%
  \BibitemOpen
  \bibfield  {author} {\bibinfo {author} {\bibfnamefont {J.}~\bibnamefont
  {Chen}}, \bibinfo {author} {\bibfnamefont {F.}~\bibnamefont {Zhang}},
  \bibinfo {author} {\bibfnamefont {C.}~\bibnamefont {Huang}}, \bibinfo
  {author} {\bibfnamefont {M.}~\bibnamefont {Newman}},\ and\ \bibinfo {author}
  {\bibfnamefont {Y.}~\bibnamefont {Shi}},\ }\bibfield  {title} {\bibinfo
  {title} {Classical simulation of intermediate-size quantum circuits},\
  }\href@noop {} {\bibfield  {journal} {\bibinfo  {journal} {arXiv preprint
  arXiv:1805.01450}\ } (\bibinfo {year} {2018}{\natexlab{a}})}\BibitemShut
  {NoStop}%
\bibitem [{\citenamefont {Markov}\ \emph {et~al.}(2018)\citenamefont {Markov},
  \citenamefont {Fatima}, \citenamefont {Isakov},\ and\ \citenamefont
  {Boixo}}]{markov2018quantum}%
  \BibitemOpen
  \bibfield  {author} {\bibinfo {author} {\bibfnamefont {I.~L.}\ \bibnamefont
  {Markov}}, \bibinfo {author} {\bibfnamefont {A.}~\bibnamefont {Fatima}},
  \bibinfo {author} {\bibfnamefont {S.~V.}\ \bibnamefont {Isakov}},\ and\
  \bibinfo {author} {\bibfnamefont {S.}~\bibnamefont {Boixo}},\ }\bibfield
  {title} {\bibinfo {title} {Quantum supremacy is both closer and farther than
  it appears},\ }\href@noop {} {\bibfield  {journal} {\bibinfo  {journal}
  {arXiv preprint arXiv:1807.10749}\ } (\bibinfo {year} {2018})}\BibitemShut
  {NoStop}%
\bibitem [{\citenamefont {Villalonga}\ \emph {et~al.}(2019)\citenamefont
  {Villalonga}, \citenamefont {Boixo}, \citenamefont {Nelson}, \citenamefont
  {Henze}, \citenamefont {Rieffel}, \citenamefont {Biswas},\ and\ \citenamefont
  {Mandr{\`a}}}]{villalonga2019flexible}%
  \BibitemOpen
  \bibfield  {author} {\bibinfo {author} {\bibfnamefont {B.}~\bibnamefont
  {Villalonga}}, \bibinfo {author} {\bibfnamefont {S.}~\bibnamefont {Boixo}},
  \bibinfo {author} {\bibfnamefont {B.}~\bibnamefont {Nelson}}, \bibinfo
  {author} {\bibfnamefont {C.}~\bibnamefont {Henze}}, \bibinfo {author}
  {\bibfnamefont {E.}~\bibnamefont {Rieffel}}, \bibinfo {author} {\bibfnamefont
  {R.}~\bibnamefont {Biswas}},\ and\ \bibinfo {author} {\bibfnamefont
  {S.}~\bibnamefont {Mandr{\`a}}},\ }\bibfield  {title} {\bibinfo {title} {A
  flexible high-performance simulator for verifying and benchmarking quantum
  circuits implemented on real hardware},\ }\href@noop {} {\bibfield  {journal}
  {\bibinfo  {journal} {npj Quantum Information}\ }\textbf {\bibinfo {volume}
  {5}},\ \bibinfo {pages} {86} (\bibinfo {year} {2019})}\BibitemShut {NoStop}%
\bibitem [{\citenamefont {Pednault}\ \emph {et~al.}(2017)\citenamefont
  {Pednault}, \citenamefont {Gunnels}, \citenamefont {Nannicini}, \citenamefont
  {Horesh}, \citenamefont {Magerlein}, \citenamefont {Solomonik}, \citenamefont
  {Draeger}, \citenamefont {Holland},\ and\ \citenamefont
  {Wisnieff}}]{pednault2017pareto}%
  \BibitemOpen
  \bibfield  {author} {\bibinfo {author} {\bibfnamefont {E.}~\bibnamefont
  {Pednault}}, \bibinfo {author} {\bibfnamefont {J.~A.}\ \bibnamefont
  {Gunnels}}, \bibinfo {author} {\bibfnamefont {G.}~\bibnamefont {Nannicini}},
  \bibinfo {author} {\bibfnamefont {L.}~\bibnamefont {Horesh}}, \bibinfo
  {author} {\bibfnamefont {T.}~\bibnamefont {Magerlein}}, \bibinfo {author}
  {\bibfnamefont {E.}~\bibnamefont {Solomonik}}, \bibinfo {author}
  {\bibfnamefont {E.~W.}\ \bibnamefont {Draeger}}, \bibinfo {author}
  {\bibfnamefont {E.~T.}\ \bibnamefont {Holland}},\ and\ \bibinfo {author}
  {\bibfnamefont {R.}~\bibnamefont {Wisnieff}},\ }\bibfield  {title} {\bibinfo
  {title} {Pareto-efficient quantum circuit simulation using tensor contraction
  deferral},\ }\href@noop {} {\bibfield  {journal} {\bibinfo  {journal} {arXiv
  preprint arXiv:1710.05867}\ } (\bibinfo {year} {2017})}\BibitemShut {NoStop}%
\bibitem [{\citenamefont {Huang}\ \emph {et~al.}(2021)\citenamefont {Huang},
  \citenamefont {Zhang}, \citenamefont {Newman}, \citenamefont {Ni},
  \citenamefont {Ding}, \citenamefont {Cai}, \citenamefont {Gao}, \citenamefont
  {Wang}, \citenamefont {Wu}, \citenamefont {Zhang} \emph
  {et~al.}}]{huang2021efficient}%
  \BibitemOpen
  \bibfield  {author} {\bibinfo {author} {\bibfnamefont {C.}~\bibnamefont
  {Huang}}, \bibinfo {author} {\bibfnamefont {F.}~\bibnamefont {Zhang}},
  \bibinfo {author} {\bibfnamefont {M.}~\bibnamefont {Newman}}, \bibinfo
  {author} {\bibfnamefont {X.}~\bibnamefont {Ni}}, \bibinfo {author}
  {\bibfnamefont {D.}~\bibnamefont {Ding}}, \bibinfo {author} {\bibfnamefont
  {J.}~\bibnamefont {Cai}}, \bibinfo {author} {\bibfnamefont {X.}~\bibnamefont
  {Gao}}, \bibinfo {author} {\bibfnamefont {T.}~\bibnamefont {Wang}}, \bibinfo
  {author} {\bibfnamefont {F.}~\bibnamefont {Wu}}, \bibinfo {author}
  {\bibfnamefont {G.}~\bibnamefont {Zhang}}, \emph {et~al.},\ }\bibfield
  {title} {\bibinfo {title} {Efficient parallelization of tensor network
  contraction for simulating quantum computation},\ }\href@noop {} {\bibfield
  {journal} {\bibinfo  {journal} {Nature Computational Science}\ }\textbf
  {\bibinfo {volume} {1}},\ \bibinfo {pages} {578} (\bibinfo {year}
  {2021})}\BibitemShut {NoStop}%
\bibitem [{\citenamefont {Pan}\ \emph {et~al.}(2022)\citenamefont {Pan},
  \citenamefont {Chen},\ and\ \citenamefont {Zhang}}]{pan2022solving}%
  \BibitemOpen
  \bibfield  {author} {\bibinfo {author} {\bibfnamefont {F.}~\bibnamefont
  {Pan}}, \bibinfo {author} {\bibfnamefont {K.}~\bibnamefont {Chen}},\ and\
  \bibinfo {author} {\bibfnamefont {P.}~\bibnamefont {Zhang}},\ }\bibfield
  {title} {\bibinfo {title} {{Solving the Sampling Problem of the Sycamore
  Quantum Circuits}},\ }\href@noop {} {\bibfield  {journal} {\bibinfo
  {journal} {Physical Review Letters}\ }\textbf {\bibinfo {volume} {129}},\
  \bibinfo {pages} {090502} (\bibinfo {year} {2022})}\BibitemShut {NoStop}%
\bibitem [{\citenamefont {Liu}\ \emph {et~al.}(2024)\citenamefont {Liu},
  \citenamefont {Chen}, \citenamefont {Guo}, \citenamefont {Song},
  \citenamefont {Shi}, \citenamefont {Gan}, \citenamefont {Wu}, \citenamefont
  {Wu}, \citenamefont {Fu}, \citenamefont {Liu} \emph
  {et~al.}}]{liu2024verifying}%
  \BibitemOpen
  \bibfield  {author} {\bibinfo {author} {\bibfnamefont {Y.}~\bibnamefont
  {Liu}}, \bibinfo {author} {\bibfnamefont {Y.}~\bibnamefont {Chen}}, \bibinfo
  {author} {\bibfnamefont {C.}~\bibnamefont {Guo}}, \bibinfo {author}
  {\bibfnamefont {J.}~\bibnamefont {Song}}, \bibinfo {author} {\bibfnamefont
  {X.}~\bibnamefont {Shi}}, \bibinfo {author} {\bibfnamefont {L.}~\bibnamefont
  {Gan}}, \bibinfo {author} {\bibfnamefont {W.}~\bibnamefont {Wu}}, \bibinfo
  {author} {\bibfnamefont {W.}~\bibnamefont {Wu}}, \bibinfo {author}
  {\bibfnamefont {H.}~\bibnamefont {Fu}}, \bibinfo {author} {\bibfnamefont
  {X.}~\bibnamefont {Liu}}, \emph {et~al.},\ }\bibfield  {title} {\bibinfo
  {title} {{Verifying Quantum Advantage Experiments with Multiple Amplitude
  Tensor Network Contraction}},\ }\href@noop {} {\bibfield  {journal} {\bibinfo
   {journal} {Physical Review Letters}\ }\textbf {\bibinfo {volume} {132}},\
  \bibinfo {pages} {030601} (\bibinfo {year} {2024})}\BibitemShut {NoStop}%
\bibitem [{\citenamefont {Zhao}\ \emph {et~al.}(2025)\citenamefont {Zhao},
  \citenamefont {Zhong}, \citenamefont {Pan}, \citenamefont {Chen},
  \citenamefont {Fu}, \citenamefont {Su}, \citenamefont {Xie}, \citenamefont
  {Zhao}, \citenamefont {Zhang}, \citenamefont {Ouyang} \emph
  {et~al.}}]{zhao2024leapfrogging}%
  \BibitemOpen
  \bibfield  {author} {\bibinfo {author} {\bibfnamefont {X.-H.}\ \bibnamefont
  {Zhao}}, \bibinfo {author} {\bibfnamefont {H.-S.}\ \bibnamefont {Zhong}},
  \bibinfo {author} {\bibfnamefont {F.}~\bibnamefont {Pan}}, \bibinfo {author}
  {\bibfnamefont {Z.-H.}\ \bibnamefont {Chen}}, \bibinfo {author}
  {\bibfnamefont {R.}~\bibnamefont {Fu}}, \bibinfo {author} {\bibfnamefont
  {Z.}~\bibnamefont {Su}}, \bibinfo {author} {\bibfnamefont {X.}~\bibnamefont
  {Xie}}, \bibinfo {author} {\bibfnamefont {C.}~\bibnamefont {Zhao}}, \bibinfo
  {author} {\bibfnamefont {P.}~\bibnamefont {Zhang}}, \bibinfo {author}
  {\bibfnamefont {W.}~\bibnamefont {Ouyang}}, \emph {et~al.},\ }\bibfield
  {title} {\bibinfo {title} {Leapfrogging sycamore: harnessing 1432 gpus for
  7$\times$ faster quantum random circuit sampling},\ }\href@noop {} {\bibfield
   {journal} {\bibinfo  {journal} {National Science Review}\ }\textbf {\bibinfo
  {volume} {12}},\ \bibinfo {pages} {nwae317} (\bibinfo {year}
  {2025})}\BibitemShut {NoStop}%
\bibitem [{\citenamefont {Roth}(1996)}]{roth1996hardness}%
  \BibitemOpen
  \bibfield  {author} {\bibinfo {author} {\bibfnamefont {D.}~\bibnamefont
  {Roth}},\ }\bibfield  {title} {\bibinfo {title} {On the hardness of
  approximate reasoning},\ }\href@noop {} {\bibfield  {journal} {\bibinfo
  {journal} {Artificial Intelligence}\ }\textbf {\bibinfo {volume} {82}},\
  \bibinfo {pages} {273} (\bibinfo {year} {1996})}\BibitemShut {NoStop}%
\bibitem [{\citenamefont {Nishino}\ and\ \citenamefont
  {Okunishi}(1996)}]{nishino1996corner}%
  \BibitemOpen
  \bibfield  {author} {\bibinfo {author} {\bibfnamefont {T.}~\bibnamefont
  {Nishino}}\ and\ \bibinfo {author} {\bibfnamefont {K.}~\bibnamefont
  {Okunishi}},\ }\bibfield  {title} {\bibinfo {title} {Corner transfer matrix
  renormalization group method},\ }\href@noop {} {\bibfield  {journal}
  {\bibinfo  {journal} {Journal of the Physical Society of Japan}\ }\textbf
  {\bibinfo {volume} {65}},\ \bibinfo {pages} {891} (\bibinfo {year}
  {1996})}\BibitemShut {NoStop}%
\bibitem [{\citenamefont {Levin}\ and\ \citenamefont
  {Nave}(2007)}]{levin2007tensor}%
  \BibitemOpen
  \bibfield  {author} {\bibinfo {author} {\bibfnamefont {M.}~\bibnamefont
  {Levin}}\ and\ \bibinfo {author} {\bibfnamefont {C.~P.}\ \bibnamefont
  {Nave}},\ }\bibfield  {title} {\bibinfo {title} {Tensor renormalization group
  approach to two-dimensional classical lattice models},\ }\href@noop {}
  {\bibfield  {journal} {\bibinfo  {journal} {Physical review letters}\
  }\textbf {\bibinfo {volume} {99}},\ \bibinfo {pages} {120601} (\bibinfo
  {year} {2007})}\BibitemShut {NoStop}%
\bibitem [{\citenamefont {Xie}\ \emph {et~al.}(2012)\citenamefont {Xie},
  \citenamefont {Chen}, \citenamefont {Qin}, \citenamefont {Zhu}, \citenamefont
  {Yang},\ and\ \citenamefont {Xiang}}]{xie2012coarse}%
  \BibitemOpen
  \bibfield  {author} {\bibinfo {author} {\bibfnamefont {Z.-Y.}\ \bibnamefont
  {Xie}}, \bibinfo {author} {\bibfnamefont {J.}~\bibnamefont {Chen}}, \bibinfo
  {author} {\bibfnamefont {M.-P.}\ \bibnamefont {Qin}}, \bibinfo {author}
  {\bibfnamefont {J.~W.}\ \bibnamefont {Zhu}}, \bibinfo {author} {\bibfnamefont
  {L.-P.}\ \bibnamefont {Yang}},\ and\ \bibinfo {author} {\bibfnamefont
  {T.}~\bibnamefont {Xiang}},\ }\bibfield  {title} {\bibinfo {title}
  {Coarse-graining renormalization by higher-order singular value
  decomposition},\ }\href@noop {} {\bibfield  {journal} {\bibinfo  {journal}
  {Physical Review B—Condensed Matter and Materials Physics}\ }\textbf
  {\bibinfo {volume} {86}},\ \bibinfo {pages} {045139} (\bibinfo {year}
  {2012})}\BibitemShut {NoStop}%
\bibitem [{\citenamefont {Evenbly}\ and\ \citenamefont
  {Vidal}(2015)}]{evenbly2015tensor}%
  \BibitemOpen
  \bibfield  {author} {\bibinfo {author} {\bibfnamefont {G.}~\bibnamefont
  {Evenbly}}\ and\ \bibinfo {author} {\bibfnamefont {G.}~\bibnamefont
  {Vidal}},\ }\bibfield  {title} {\bibinfo {title} {Tensor network
  renormalization},\ }\href@noop {} {\bibfield  {journal} {\bibinfo  {journal}
  {Physical review letters}\ }\textbf {\bibinfo {volume} {115}},\ \bibinfo
  {pages} {180405} (\bibinfo {year} {2015})}\BibitemShut {NoStop}%
\bibitem [{\citenamefont {Chen}\ \emph {et~al.}(2024)\citenamefont {Chen},
  \citenamefont {Jiang}, \citenamefont {Hangleiter},\ and\ \citenamefont
  {Schuch}}]{chen2024sign}%
  \BibitemOpen
  \bibfield  {author} {\bibinfo {author} {\bibfnamefont {J.}~\bibnamefont
  {Chen}}, \bibinfo {author} {\bibfnamefont {J.}~\bibnamefont {Jiang}},
  \bibinfo {author} {\bibfnamefont {D.}~\bibnamefont {Hangleiter}},\ and\
  \bibinfo {author} {\bibfnamefont {N.}~\bibnamefont {Schuch}},\ }\bibfield
  {title} {\bibinfo {title} {Sign problem in tensor network contraction},\
  }\href@noop {} {\bibfield  {journal} {\bibinfo  {journal} {arXiv preprint
  arXiv:2404.19023}\ } (\bibinfo {year} {2024})}\BibitemShut {NoStop}%
\bibitem [{\citenamefont {Jiang}\ \emph {et~al.}(2024)\citenamefont {Jiang},
  \citenamefont {Chen}, \citenamefont {Schuch},\ and\ \citenamefont
  {Hangleiter}}]{jiang2024positive}%
  \BibitemOpen
  \bibfield  {author} {\bibinfo {author} {\bibfnamefont {J.}~\bibnamefont
  {Jiang}}, \bibinfo {author} {\bibfnamefont {J.}~\bibnamefont {Chen}},
  \bibinfo {author} {\bibfnamefont {N.}~\bibnamefont {Schuch}},\ and\ \bibinfo
  {author} {\bibfnamefont {D.}~\bibnamefont {Hangleiter}},\ }\bibfield  {title}
  {\bibinfo {title} {Positive bias makes tensor-network contraction
  tractable},\ }\href@noop {} {\bibfield  {journal} {\bibinfo  {journal} {arXiv
  preprint arXiv:2410.05414}\ } (\bibinfo {year} {2024})}\BibitemShut {NoStop}%
\bibitem [{\citenamefont {Vidal}(2008)}]{vidal2008class}%
  \BibitemOpen
  \bibfield  {author} {\bibinfo {author} {\bibfnamefont {G.}~\bibnamefont
  {Vidal}},\ }\bibfield  {title} {\bibinfo {title} {Class of quantum many-body
  states that can be efficiently simulated},\ }\href@noop {} {\bibfield
  {journal} {\bibinfo  {journal} {Physical review letters}\ }\textbf {\bibinfo
  {volume} {101}},\ \bibinfo {pages} {110501} (\bibinfo {year}
  {2008})}\BibitemShut {NoStop}%
\bibitem [{\citenamefont {Verstraete}\ and\ \citenamefont
  {Cirac}(2004)}]{verstraete2004renormalization}%
  \BibitemOpen
  \bibfield  {author} {\bibinfo {author} {\bibfnamefont {F.}~\bibnamefont
  {Verstraete}}\ and\ \bibinfo {author} {\bibfnamefont {J.~I.}\ \bibnamefont
  {Cirac}},\ }\bibfield  {title} {\bibinfo {title} {Renormalization algorithms
  for quantum-many body systems in two and higher dimensions},\ }\href@noop {}
  {\bibfield  {journal} {\bibinfo  {journal} {arXiv preprint cond-mat/0407066}\
  } (\bibinfo {year} {2004})}\BibitemShut {NoStop}%
\bibitem [{\citenamefont {Jiang}\ \emph {et~al.}(2008)\citenamefont {Jiang},
  \citenamefont {Weng},\ and\ \citenamefont {Xiang}}]{jiang2008accurate}%
  \BibitemOpen
  \bibfield  {author} {\bibinfo {author} {\bibfnamefont {H.-C.}\ \bibnamefont
  {Jiang}}, \bibinfo {author} {\bibfnamefont {Z.-Y.}\ \bibnamefont {Weng}},\
  and\ \bibinfo {author} {\bibfnamefont {T.}~\bibnamefont {Xiang}},\ }\bibfield
   {title} {\bibinfo {title} {Accurate determination of tensor network state of
  quantum lattice models in two dimensions},\ }\href@noop {} {\bibfield
  {journal} {\bibinfo  {journal} {Physical review letters}\ }\textbf {\bibinfo
  {volume} {101}},\ \bibinfo {pages} {090603} (\bibinfo {year}
  {2008})}\BibitemShut {NoStop}%
\bibitem [{\citenamefont {Corboz}\ \emph {et~al.}(2010)\citenamefont {Corboz},
  \citenamefont {Jordan},\ and\ \citenamefont {Vidal}}]{corboz2010simulation}%
  \BibitemOpen
  \bibfield  {author} {\bibinfo {author} {\bibfnamefont {P.}~\bibnamefont
  {Corboz}}, \bibinfo {author} {\bibfnamefont {J.}~\bibnamefont {Jordan}},\
  and\ \bibinfo {author} {\bibfnamefont {G.}~\bibnamefont {Vidal}},\ }\bibfield
   {title} {\bibinfo {title} {Simulation of fermionic lattice models in two
  dimensions with projected entangled-pair states: Next-nearest neighbor
  hamiltonians},\ }\href@noop {} {\bibfield  {journal} {\bibinfo  {journal}
  {Physical Review B—Condensed Matter and Materials Physics}\ }\textbf
  {\bibinfo {volume} {82}},\ \bibinfo {pages} {245119} (\bibinfo {year}
  {2010})}\BibitemShut {NoStop}%
\bibitem [{\citenamefont {Lubasch}\ \emph {et~al.}(2014)\citenamefont
  {Lubasch}, \citenamefont {Cirac},\ and\ \citenamefont
  {Banuls}}]{lubasch2014unifying}%
  \BibitemOpen
  \bibfield  {author} {\bibinfo {author} {\bibfnamefont {M.}~\bibnamefont
  {Lubasch}}, \bibinfo {author} {\bibfnamefont {J.~I.}\ \bibnamefont {Cirac}},\
  and\ \bibinfo {author} {\bibfnamefont {M.-C.}\ \bibnamefont {Banuls}},\
  }\bibfield  {title} {\bibinfo {title} {Unifying projected entangled pair
  state contractions},\ }\href@noop {} {\bibfield  {journal} {\bibinfo
  {journal} {New Journal of Physics}\ }\textbf {\bibinfo {volume} {16}},\
  \bibinfo {pages} {033014} (\bibinfo {year} {2014})}\BibitemShut {NoStop}%
\bibitem [{\citenamefont {Pan}\ \emph {et~al.}(2020)\citenamefont {Pan},
  \citenamefont {Zhou}, \citenamefont {Li},\ and\ \citenamefont
  {Zhang}}]{pan2020contracting}%
  \BibitemOpen
  \bibfield  {author} {\bibinfo {author} {\bibfnamefont {F.}~\bibnamefont
  {Pan}}, \bibinfo {author} {\bibfnamefont {P.}~\bibnamefont {Zhou}}, \bibinfo
  {author} {\bibfnamefont {S.}~\bibnamefont {Li}},\ and\ \bibinfo {author}
  {\bibfnamefont {P.}~\bibnamefont {Zhang}},\ }\bibfield  {title} {\bibinfo
  {title} {Contracting arbitrary tensor networks: general approximate algorithm
  and applications in graphical models and quantum circuit simulations},\
  }\href@noop {} {\bibfield  {journal} {\bibinfo  {journal} {Physical Review
  Letters}\ }\textbf {\bibinfo {volume} {125}},\ \bibinfo {pages} {060503}
  (\bibinfo {year} {2020})}\BibitemShut {NoStop}%
\bibitem [{\citenamefont {Ma}\ \emph {et~al.}(2024)\citenamefont {Ma},
  \citenamefont {Fishman}, \citenamefont {Stoudenmire},\ and\ \citenamefont
  {Solomonik}}]{ma2024approximate}%
  \BibitemOpen
  \bibfield  {author} {\bibinfo {author} {\bibfnamefont {L.}~\bibnamefont
  {Ma}}, \bibinfo {author} {\bibfnamefont {M.}~\bibnamefont {Fishman}},
  \bibinfo {author} {\bibfnamefont {M.}~\bibnamefont {Stoudenmire}},\ and\
  \bibinfo {author} {\bibfnamefont {E.}~\bibnamefont {Solomonik}},\ }\bibfield
  {title} {\bibinfo {title} {Approximate contraction of arbitrary tensor
  networks with a flexible and efficient density matrix algorithm},\
  }\href@noop {} {\bibfield  {journal} {\bibinfo  {journal} {arXiv preprint
  arXiv:2406.09769}\ } (\bibinfo {year} {2024})}\BibitemShut {NoStop}%
\bibitem [{\citenamefont {Gray}\ and\ \citenamefont
  {Chan}(2024)}]{gray2024hyperoptimized}%
  \BibitemOpen
  \bibfield  {author} {\bibinfo {author} {\bibfnamefont {J.}~\bibnamefont
  {Gray}}\ and\ \bibinfo {author} {\bibfnamefont {G.~K.-L.}\ \bibnamefont
  {Chan}},\ }\bibfield  {title} {\bibinfo {title} {Hyperoptimized approximate
  contraction of tensor networks with arbitrary geometry},\ }\href@noop {}
  {\bibfield  {journal} {\bibinfo  {journal} {Physical Review X}\ }\textbf
  {\bibinfo {volume} {14}},\ \bibinfo {pages} {011009} (\bibinfo {year}
  {2024})}\BibitemShut {NoStop}%
\bibitem [{\citenamefont {Zhou}\ \emph {et~al.}(2020)\citenamefont {Zhou},
  \citenamefont {Stoudenmire},\ and\ \citenamefont {Waintal}}]{zhou2020limits}%
  \BibitemOpen
  \bibfield  {author} {\bibinfo {author} {\bibfnamefont {Y.}~\bibnamefont
  {Zhou}}, \bibinfo {author} {\bibfnamefont {E.~M.}\ \bibnamefont
  {Stoudenmire}},\ and\ \bibinfo {author} {\bibfnamefont {X.}~\bibnamefont
  {Waintal}},\ }\bibfield  {title} {\bibinfo {title} {What limits the
  simulation of quantum computers?},\ }\href@noop {} {\bibfield  {journal}
  {\bibinfo  {journal} {Physical Review X}\ }\textbf {\bibinfo {volume} {10}},\
  \bibinfo {pages} {041038} (\bibinfo {year} {2020})}\BibitemShut {NoStop}%
\bibitem [{\citenamefont {Liu}\ \emph {et~al.}(2022)\citenamefont {Liu},
  \citenamefont {Liu}, \citenamefont {Alexeev},\ and\ \citenamefont
  {Jiang}}]{liu2022estimating}%
  \BibitemOpen
  \bibfield  {author} {\bibinfo {author} {\bibfnamefont {M.}~\bibnamefont
  {Liu}}, \bibinfo {author} {\bibfnamefont {J.}~\bibnamefont {Liu}}, \bibinfo
  {author} {\bibfnamefont {Y.}~\bibnamefont {Alexeev}},\ and\ \bibinfo {author}
  {\bibfnamefont {L.}~\bibnamefont {Jiang}},\ }\bibfield  {title} {\bibinfo
  {title} {Estimating the randomness of quantum circuit ensembles up to 50
  qubits},\ }\href@noop {} {\bibfield  {journal} {\bibinfo  {journal} {npj
  Quantum Information}\ }\textbf {\bibinfo {volume} {8}},\ \bibinfo {pages}
  {137} (\bibinfo {year} {2022})}\BibitemShut {NoStop}%
\bibitem [{\citenamefont {Kim}\ \emph {et~al.}(2023)\citenamefont {Kim},
  \citenamefont {Eddins}, \citenamefont {Anand}, \citenamefont {Wei},
  \citenamefont {Van Den~Berg}, \citenamefont {Rosenblatt}, \citenamefont
  {Nayfeh}, \citenamefont {Wu}, \citenamefont {Zaletel}, \citenamefont {Temme}
  \emph {et~al.}}]{kim2023evidence}%
  \BibitemOpen
  \bibfield  {author} {\bibinfo {author} {\bibfnamefont {Y.}~\bibnamefont
  {Kim}}, \bibinfo {author} {\bibfnamefont {A.}~\bibnamefont {Eddins}},
  \bibinfo {author} {\bibfnamefont {S.}~\bibnamefont {Anand}}, \bibinfo
  {author} {\bibfnamefont {K.~X.}\ \bibnamefont {Wei}}, \bibinfo {author}
  {\bibfnamefont {E.}~\bibnamefont {Van Den~Berg}}, \bibinfo {author}
  {\bibfnamefont {S.}~\bibnamefont {Rosenblatt}}, \bibinfo {author}
  {\bibfnamefont {H.}~\bibnamefont {Nayfeh}}, \bibinfo {author} {\bibfnamefont
  {Y.}~\bibnamefont {Wu}}, \bibinfo {author} {\bibfnamefont {M.}~\bibnamefont
  {Zaletel}}, \bibinfo {author} {\bibfnamefont {K.}~\bibnamefont {Temme}},
  \emph {et~al.},\ }\bibfield  {title} {\bibinfo {title} {Evidence for the
  utility of quantum computing before fault tolerance},\ }\href@noop {}
  {\bibfield  {journal} {\bibinfo  {journal} {Nature}\ }\textbf {\bibinfo
  {volume} {618}},\ \bibinfo {pages} {500} (\bibinfo {year}
  {2023})}\BibitemShut {NoStop}%
\bibitem [{\citenamefont {Zaletel}\ and\ \citenamefont
  {Pollmann}(2020)}]{zaletel2020isometric}%
  \BibitemOpen
  \bibfield  {author} {\bibinfo {author} {\bibfnamefont {M.~P.}\ \bibnamefont
  {Zaletel}}\ and\ \bibinfo {author} {\bibfnamefont {F.}~\bibnamefont
  {Pollmann}},\ }\bibfield  {title} {\bibinfo {title} {Isometric tensor network
  states in two dimensions},\ }\href@noop {} {\bibfield  {journal} {\bibinfo
  {journal} {Physical review letters}\ }\textbf {\bibinfo {volume} {124}},\
  \bibinfo {pages} {037201} (\bibinfo {year} {2020})}\BibitemShut {NoStop}%
\bibitem [{\citenamefont {Tindall}\ and\ \citenamefont
  {Fishman}(2023)}]{tindall2023gauging}%
  \BibitemOpen
  \bibfield  {author} {\bibinfo {author} {\bibfnamefont {J.}~\bibnamefont
  {Tindall}}\ and\ \bibinfo {author} {\bibfnamefont {M.}~\bibnamefont
  {Fishman}},\ }\bibfield  {title} {\bibinfo {title} {Gauging tensor networks
  with belief propagation},\ }\href@noop {} {\bibfield  {journal} {\bibinfo
  {journal} {SciPost Physics}\ }\textbf {\bibinfo {volume} {15}},\ \bibinfo
  {pages} {222} (\bibinfo {year} {2023})}\BibitemShut {NoStop}%
\bibitem [{\citenamefont {Anand}\ \emph
  {et~al.}(2023{\natexlab{a}})\citenamefont {Anand}, \citenamefont {Temme},
  \citenamefont {Kandala},\ and\ \citenamefont {Zaletel}}]{anand2023classical}%
  \BibitemOpen
  \bibfield  {author} {\bibinfo {author} {\bibfnamefont {S.}~\bibnamefont
  {Anand}}, \bibinfo {author} {\bibfnamefont {K.}~\bibnamefont {Temme}},
  \bibinfo {author} {\bibfnamefont {A.}~\bibnamefont {Kandala}},\ and\ \bibinfo
  {author} {\bibfnamefont {M.}~\bibnamefont {Zaletel}},\ }\bibfield  {title}
  {\bibinfo {title} {Classical benchmarking of zero noise extrapolation beyond
  the exactly-verifiable regime},\ }\href@noop {} {\bibfield  {journal}
  {\bibinfo  {journal} {arXiv preprint arXiv:2306.17839}\ } (\bibinfo {year}
  {2023}{\natexlab{a}})}\BibitemShut {NoStop}%
\bibitem [{\citenamefont {Patra}\ \emph {et~al.}(2024)\citenamefont {Patra},
  \citenamefont {Jahromi}, \citenamefont {Singh},\ and\ \citenamefont
  {Or{\'u}s}}]{patra2024efficient}%
  \BibitemOpen
  \bibfield  {author} {\bibinfo {author} {\bibfnamefont {S.}~\bibnamefont
  {Patra}}, \bibinfo {author} {\bibfnamefont {S.~S.}\ \bibnamefont {Jahromi}},
  \bibinfo {author} {\bibfnamefont {S.}~\bibnamefont {Singh}},\ and\ \bibinfo
  {author} {\bibfnamefont {R.}~\bibnamefont {Or{\'u}s}},\ }\bibfield  {title}
  {\bibinfo {title} {Efficient tensor network simulation of ibm's largest
  quantum processors},\ }\href@noop {} {\bibfield  {journal} {\bibinfo
  {journal} {Physical Review Research}\ }\textbf {\bibinfo {volume} {6}},\
  \bibinfo {pages} {013326} (\bibinfo {year} {2024})}\BibitemShut {NoStop}%
\bibitem [{\citenamefont {Rudolph}\ \emph
  {et~al.}(2023{\natexlab{a}})\citenamefont {Rudolph}, \citenamefont {Fontana},
  \citenamefont {Holmes},\ and\ \citenamefont {Cincio}}]{rudolph2023classical}%
  \BibitemOpen
  \bibfield  {author} {\bibinfo {author} {\bibfnamefont {M.~S.}\ \bibnamefont
  {Rudolph}}, \bibinfo {author} {\bibfnamefont {E.}~\bibnamefont {Fontana}},
  \bibinfo {author} {\bibfnamefont {Z.}~\bibnamefont {Holmes}},\ and\ \bibinfo
  {author} {\bibfnamefont {L.}~\bibnamefont {Cincio}},\ }\bibfield  {title}
  {\bibinfo {title} {Classical surrogate simulation of quantum systems with
  lowesa},\ }\href@noop {} {\bibfield  {journal} {\bibinfo  {journal} {arXiv
  preprint arXiv:2308.09109}\ } (\bibinfo {year}
  {2023}{\natexlab{a}})}\BibitemShut {NoStop}%
\bibitem [{\citenamefont {Bouland}\ \emph {et~al.}(2022)\citenamefont
  {Bouland}, \citenamefont {Fefferman}, \citenamefont {Landau},\ and\
  \citenamefont {Liu}}]{bouland2022noise}%
  \BibitemOpen
  \bibfield  {author} {\bibinfo {author} {\bibfnamefont {A.}~\bibnamefont
  {Bouland}}, \bibinfo {author} {\bibfnamefont {B.}~\bibnamefont {Fefferman}},
  \bibinfo {author} {\bibfnamefont {Z.}~\bibnamefont {Landau}},\ and\ \bibinfo
  {author} {\bibfnamefont {Y.}~\bibnamefont {Liu}},\ }\href@noop {} {\bibinfo
  {title} {Noise and the frontier of quantum supremacy}} (\bibinfo {year}
  {2022})\BibitemShut {NoStop}%
\bibitem [{\citenamefont {Krovi}(2022)}]{krovi2022average}%
  \BibitemOpen
  \bibfield  {author} {\bibinfo {author} {\bibfnamefont {H.}~\bibnamefont
  {Krovi}},\ }\bibfield  {title} {\bibinfo {title} {Average-case hardness of
  estimating probabilities of random quantum circuits with a linear scaling in
  the error exponent},\ }\href@noop {} {\bibfield  {journal} {\bibinfo
  {journal} {arXiv preprint arXiv:2206.05642}\ } (\bibinfo {year}
  {2022})}\BibitemShut {NoStop}%
\bibitem [{\citenamefont {Arute}\ \emph {et~al.}(2019)\citenamefont {Arute},
  \citenamefont {Arya}, \citenamefont {Babbush}, \citenamefont {Bacon},
  \citenamefont {Bardin}, \citenamefont {Barends}, \citenamefont {Biswas},
  \citenamefont {Boixo}, \citenamefont {Brandao}, \citenamefont {Buell} \emph
  {et~al.}}]{arute2019quantum}%
  \BibitemOpen
  \bibfield  {author} {\bibinfo {author} {\bibfnamefont {F.}~\bibnamefont
  {Arute}}, \bibinfo {author} {\bibfnamefont {K.}~\bibnamefont {Arya}},
  \bibinfo {author} {\bibfnamefont {R.}~\bibnamefont {Babbush}}, \bibinfo
  {author} {\bibfnamefont {D.}~\bibnamefont {Bacon}}, \bibinfo {author}
  {\bibfnamefont {J.~C.}\ \bibnamefont {Bardin}}, \bibinfo {author}
  {\bibfnamefont {R.}~\bibnamefont {Barends}}, \bibinfo {author} {\bibfnamefont
  {R.}~\bibnamefont {Biswas}}, \bibinfo {author} {\bibfnamefont
  {S.}~\bibnamefont {Boixo}}, \bibinfo {author} {\bibfnamefont {F.~G.}\
  \bibnamefont {Brandao}}, \bibinfo {author} {\bibfnamefont {D.~A.}\
  \bibnamefont {Buell}}, \emph {et~al.},\ }\bibfield  {title} {\bibinfo {title}
  {Quantum supremacy using a programmable superconducting processor},\
  }\href@noop {} {\bibfield  {journal} {\bibinfo  {journal} {Nature}\ }\textbf
  {\bibinfo {volume} {574}},\ \bibinfo {pages} {505} (\bibinfo {year}
  {2019})}\BibitemShut {NoStop}%
\bibitem [{\citenamefont {Wu}\ \emph {et~al.}(2021)\citenamefont {Wu},
  \citenamefont {Bao}, \citenamefont {Cao}, \citenamefont {Chen}, \citenamefont
  {Chen}, \citenamefont {Chen}, \citenamefont {Chung}, \citenamefont {Deng},
  \citenamefont {Du}, \citenamefont {Fan} \emph {et~al.}}]{wu2021strong}%
  \BibitemOpen
  \bibfield  {author} {\bibinfo {author} {\bibfnamefont {Y.}~\bibnamefont
  {Wu}}, \bibinfo {author} {\bibfnamefont {W.-S.}\ \bibnamefont {Bao}},
  \bibinfo {author} {\bibfnamefont {S.}~\bibnamefont {Cao}}, \bibinfo {author}
  {\bibfnamefont {F.}~\bibnamefont {Chen}}, \bibinfo {author} {\bibfnamefont
  {M.-C.}\ \bibnamefont {Chen}}, \bibinfo {author} {\bibfnamefont
  {X.}~\bibnamefont {Chen}}, \bibinfo {author} {\bibfnamefont {T.-H.}\
  \bibnamefont {Chung}}, \bibinfo {author} {\bibfnamefont {H.}~\bibnamefont
  {Deng}}, \bibinfo {author} {\bibfnamefont {Y.}~\bibnamefont {Du}}, \bibinfo
  {author} {\bibfnamefont {D.}~\bibnamefont {Fan}}, \emph {et~al.},\ }\bibfield
   {title} {\bibinfo {title} {Strong quantum computational advantage using a
  superconducting quantum processor},\ }\href@noop {} {\bibfield  {journal}
  {\bibinfo  {journal} {Physical review letters}\ }\textbf {\bibinfo {volume}
  {127}},\ \bibinfo {pages} {180501} (\bibinfo {year} {2021})}\BibitemShut
  {NoStop}%
\bibitem [{\citenamefont {Zhu}\ \emph {et~al.}(2022)\citenamefont {Zhu},
  \citenamefont {Cao}, \citenamefont {Chen}, \citenamefont {Chen},
  \citenamefont {Chen}, \citenamefont {Chung}, \citenamefont {Deng},
  \citenamefont {Du}, \citenamefont {Fan}, \citenamefont {Gong} \emph
  {et~al.}}]{zhu2022quantum}%
  \BibitemOpen
  \bibfield  {author} {\bibinfo {author} {\bibfnamefont {Q.}~\bibnamefont
  {Zhu}}, \bibinfo {author} {\bibfnamefont {S.}~\bibnamefont {Cao}}, \bibinfo
  {author} {\bibfnamefont {F.}~\bibnamefont {Chen}}, \bibinfo {author}
  {\bibfnamefont {M.-C.}\ \bibnamefont {Chen}}, \bibinfo {author}
  {\bibfnamefont {X.}~\bibnamefont {Chen}}, \bibinfo {author} {\bibfnamefont
  {T.-H.}\ \bibnamefont {Chung}}, \bibinfo {author} {\bibfnamefont
  {H.}~\bibnamefont {Deng}}, \bibinfo {author} {\bibfnamefont {Y.}~\bibnamefont
  {Du}}, \bibinfo {author} {\bibfnamefont {D.}~\bibnamefont {Fan}}, \bibinfo
  {author} {\bibfnamefont {M.}~\bibnamefont {Gong}}, \emph {et~al.},\
  }\bibfield  {title} {\bibinfo {title} {Quantum computational advantage via
  60-qubit 24-cycle random circuit sampling},\ }\href@noop {} {\bibfield
  {journal} {\bibinfo  {journal} {Science bulletin}\ }\textbf {\bibinfo
  {volume} {67}},\ \bibinfo {pages} {240} (\bibinfo {year} {2022})}\BibitemShut
  {NoStop}%
\bibitem [{\citenamefont {Aaronson}\ and\ \citenamefont
  {Hung}(2023)}]{aaronson2023certified}%
  \BibitemOpen
  \bibfield  {author} {\bibinfo {author} {\bibfnamefont {S.}~\bibnamefont
  {Aaronson}}\ and\ \bibinfo {author} {\bibfnamefont {S.-H.}\ \bibnamefont
  {Hung}},\ }\href@noop {} {\bibinfo {title} {Certified randomness from quantum
  supremacy}} (\bibinfo {year} {2023})\BibitemShut {NoStop}%
\bibitem [{\citenamefont {Liu}\ \emph {et~al.}(2025)\citenamefont {Liu} \emph
  {et~al.}}]{liu2025certified}%
  \BibitemOpen
  \bibfield  {author} {\bibinfo {author} {\bibfnamefont {M.}~\bibnamefont
  {Liu}} \emph {et~al.},\ }\bibfield  {title} {\bibinfo {title} {Certified
  randomness with a trapped-ion quantum processor},\ }\href
  {https://doi.org/10.1038/s41586-025-08737-1} {\bibfield  {journal} {\bibinfo
  {journal} {Nature}\ }\textbf {\bibinfo {volume} {640}},\ \bibinfo {pages}
  {343–348} (\bibinfo {year} {2025})}\BibitemShut {NoStop}%
\bibitem [{\citenamefont {Amer}\ \emph {et~al.}(2024)\citenamefont {Amer},
  \citenamefont {Chakraborty}, \citenamefont {Cui}, \citenamefont {Kaleoglu},
  \citenamefont {Lim}, \citenamefont {Liu},\ and\ \citenamefont
  {Pistoia}}]{amer2024certified}%
  \BibitemOpen
  \bibfield  {author} {\bibinfo {author} {\bibfnamefont {O.}~\bibnamefont
  {Amer}}, \bibinfo {author} {\bibfnamefont {K.}~\bibnamefont {Chakraborty}},
  \bibinfo {author} {\bibfnamefont {D.}~\bibnamefont {Cui}}, \bibinfo {author}
  {\bibfnamefont {F.}~\bibnamefont {Kaleoglu}}, \bibinfo {author}
  {\bibfnamefont {C.}~\bibnamefont {Lim}}, \bibinfo {author} {\bibfnamefont
  {M.}~\bibnamefont {Liu}},\ and\ \bibinfo {author} {\bibfnamefont
  {M.}~\bibnamefont {Pistoia}},\ }\bibfield  {title} {\bibinfo {title}
  {Certified randomness implies secure classical position-verification},\
  }\href@noop {} {\bibfield  {journal} {\bibinfo  {journal} {arXiv preprint
  arXiv:2410.03982}\ } (\bibinfo {year} {2024})}\BibitemShut {NoStop}%
\bibitem [{\citenamefont {Amer}\ \emph {et~al.}(2025)\citenamefont {Amer},
  \citenamefont {Chakrabarti}, \citenamefont {Chakraborty}, \citenamefont
  {Eloul}, \citenamefont {Kumar}, \citenamefont {Lim}, \citenamefont {Liu},
  \citenamefont {Niroula}, \citenamefont {Satsangi}, \citenamefont
  {Shaydulin},\ and\ \citenamefont {Pistoia}}]{amer2025applications}%
  \BibitemOpen
  \bibfield  {author} {\bibinfo {author} {\bibfnamefont {O.}~\bibnamefont
  {Amer}}, \bibinfo {author} {\bibfnamefont {S.}~\bibnamefont {Chakrabarti}},
  \bibinfo {author} {\bibfnamefont {K.}~\bibnamefont {Chakraborty}}, \bibinfo
  {author} {\bibfnamefont {S.}~\bibnamefont {Eloul}}, \bibinfo {author}
  {\bibfnamefont {N.}~\bibnamefont {Kumar}}, \bibinfo {author} {\bibfnamefont
  {C.}~\bibnamefont {Lim}}, \bibinfo {author} {\bibfnamefont {M.}~\bibnamefont
  {Liu}}, \bibinfo {author} {\bibfnamefont {P.}~\bibnamefont {Niroula}},
  \bibinfo {author} {\bibfnamefont {Y.}~\bibnamefont {Satsangi}}, \bibinfo
  {author} {\bibfnamefont {R.}~\bibnamefont {Shaydulin}},\ and\ \bibinfo
  {author} {\bibfnamefont {M.}~\bibnamefont {Pistoia}},\ }\bibfield  {title}
  {\bibinfo {title} {Applications of certified randomness},\ }\href@noop {}
  {\bibfield  {journal} {\bibinfo  {journal} {arXiv preprint arXiv:2503.19759}\
  } (\bibinfo {year} {2025})}\BibitemShut {NoStop}%
\bibitem [{\citenamefont {Pednault}\ \emph {et~al.}(2019)\citenamefont
  {Pednault}, \citenamefont {Gunnels}, \citenamefont {Maslov},\ and\
  \citenamefont {Gambetta}}]{pednault2019quantum}%
  \BibitemOpen
  \bibfield  {author} {\bibinfo {author} {\bibfnamefont {E.}~\bibnamefont
  {Pednault}}, \bibinfo {author} {\bibfnamefont {J.}~\bibnamefont {Gunnels}},
  \bibinfo {author} {\bibfnamefont {D.}~\bibnamefont {Maslov}},\ and\ \bibinfo
  {author} {\bibfnamefont {J.}~\bibnamefont {Gambetta}},\ }\bibfield  {title}
  {\bibinfo {title} {On “quantum supremacy”},\ }\href@noop {} {\bibfield
  {journal} {\bibinfo  {journal} {IBM Research Blog}\ }\textbf {\bibinfo
  {volume} {21}} (\bibinfo {year} {2019})}\BibitemShut {NoStop}%
\bibitem [{\citenamefont {Schutski}\ \emph {et~al.}(2020)\citenamefont
  {Schutski}, \citenamefont {Lykov},\ and\ \citenamefont
  {Oseledets}}]{schutski2020adaptive}%
  \BibitemOpen
  \bibfield  {author} {\bibinfo {author} {\bibfnamefont {R.}~\bibnamefont
  {Schutski}}, \bibinfo {author} {\bibfnamefont {D.}~\bibnamefont {Lykov}},\
  and\ \bibinfo {author} {\bibfnamefont {I.}~\bibnamefont {Oseledets}},\
  }\bibfield  {title} {\bibinfo {title} {Adaptive algorithm for quantum circuit
  simulation},\ }\href@noop {} {\bibfield  {journal} {\bibinfo  {journal}
  {Physical Review A}\ }\textbf {\bibinfo {volume} {101}},\ \bibinfo {pages}
  {042335} (\bibinfo {year} {2020})}\BibitemShut {NoStop}%
\bibitem [{\citenamefont {Ayral}\ \emph {et~al.}(2023)\citenamefont {Ayral},
  \citenamefont {Louvet}, \citenamefont {Zhou}, \citenamefont {Lambert},
  \citenamefont {Stoudenmire},\ and\ \citenamefont
  {Waintal}}]{ayral2023density}%
  \BibitemOpen
  \bibfield  {author} {\bibinfo {author} {\bibfnamefont {T.}~\bibnamefont
  {Ayral}}, \bibinfo {author} {\bibfnamefont {T.}~\bibnamefont {Louvet}},
  \bibinfo {author} {\bibfnamefont {Y.}~\bibnamefont {Zhou}}, \bibinfo {author}
  {\bibfnamefont {C.}~\bibnamefont {Lambert}}, \bibinfo {author} {\bibfnamefont
  {E.~M.}\ \bibnamefont {Stoudenmire}},\ and\ \bibinfo {author} {\bibfnamefont
  {X.}~\bibnamefont {Waintal}},\ }\bibfield  {title} {\bibinfo {title}
  {Density-matrix renormalization group algorithm for simulating quantum
  circuits with a finite fidelity},\ }\href@noop {} {\bibfield  {journal}
  {\bibinfo  {journal} {PRX Quantum}\ }\textbf {\bibinfo {volume} {4}},\
  \bibinfo {pages} {020304} (\bibinfo {year} {2023})}\BibitemShut {NoStop}%
\bibitem [{\citenamefont {Haghshenas}\ \emph {et~al.}(2025)\citenamefont
  {Haghshenas}, \citenamefont {Chertkov}, \citenamefont {Mills}, \citenamefont
  {Kadow}, \citenamefont {Lin}, \citenamefont {Chen}, \citenamefont {Cade},
  \citenamefont {Niesen}, \citenamefont {Begu{\v{s}}i{\'c}}, \citenamefont
  {Rudolph} \emph {et~al.}}]{haghshenas2025digital}%
  \BibitemOpen
  \bibfield  {author} {\bibinfo {author} {\bibfnamefont {R.}~\bibnamefont
  {Haghshenas}}, \bibinfo {author} {\bibfnamefont {E.}~\bibnamefont
  {Chertkov}}, \bibinfo {author} {\bibfnamefont {M.}~\bibnamefont {Mills}},
  \bibinfo {author} {\bibfnamefont {W.}~\bibnamefont {Kadow}}, \bibinfo
  {author} {\bibfnamefont {S.-H.}\ \bibnamefont {Lin}}, \bibinfo {author}
  {\bibfnamefont {Y.-H.}\ \bibnamefont {Chen}}, \bibinfo {author}
  {\bibfnamefont {C.}~\bibnamefont {Cade}}, \bibinfo {author} {\bibfnamefont
  {I.}~\bibnamefont {Niesen}}, \bibinfo {author} {\bibfnamefont
  {T.}~\bibnamefont {Begu{\v{s}}i{\'c}}}, \bibinfo {author} {\bibfnamefont
  {M.~S.}\ \bibnamefont {Rudolph}}, \emph {et~al.},\ }\bibfield  {title}
  {\bibinfo {title} {Digital quantum magnetism at the frontier of classical
  simulations},\ }\href@noop {} {\bibfield  {journal} {\bibinfo  {journal}
  {arXiv preprint arXiv:2503.20870}\ } (\bibinfo {year} {2025})}\BibitemShut
  {NoStop}%
\bibitem [{\citenamefont {Noh}\ \emph {et~al.}(2020)\citenamefont {Noh},
  \citenamefont {Jiang},\ and\ \citenamefont {Fefferman}}]{noh2020efficient}%
  \BibitemOpen
  \bibfield  {author} {\bibinfo {author} {\bibfnamefont {K.}~\bibnamefont
  {Noh}}, \bibinfo {author} {\bibfnamefont {L.}~\bibnamefont {Jiang}},\ and\
  \bibinfo {author} {\bibfnamefont {B.}~\bibnamefont {Fefferman}},\ }\bibfield
  {title} {\bibinfo {title} {Efficient classical simulation of noisy random
  quantum circuits in one dimension},\ }\href@noop {} {\bibfield  {journal}
  {\bibinfo  {journal} {Quantum}\ }\textbf {\bibinfo {volume} {4}},\ \bibinfo
  {pages} {318} (\bibinfo {year} {2020})}\BibitemShut {NoStop}%
\bibitem [{\citenamefont {Cheng}\ \emph {et~al.}(2021)\citenamefont {Cheng},
  \citenamefont {Cao}, \citenamefont {Zhang}, \citenamefont {Liu},
  \citenamefont {Hou}, \citenamefont {Xu},\ and\ \citenamefont
  {Zeng}}]{cheng2021simulating}%
  \BibitemOpen
  \bibfield  {author} {\bibinfo {author} {\bibfnamefont {S.}~\bibnamefont
  {Cheng}}, \bibinfo {author} {\bibfnamefont {C.}~\bibnamefont {Cao}}, \bibinfo
  {author} {\bibfnamefont {C.}~\bibnamefont {Zhang}}, \bibinfo {author}
  {\bibfnamefont {Y.}~\bibnamefont {Liu}}, \bibinfo {author} {\bibfnamefont
  {S.-Y.}\ \bibnamefont {Hou}}, \bibinfo {author} {\bibfnamefont
  {P.}~\bibnamefont {Xu}},\ and\ \bibinfo {author} {\bibfnamefont
  {B.}~\bibnamefont {Zeng}},\ }\bibfield  {title} {\bibinfo {title} {Simulating
  noisy quantum circuits with matrix product density operators},\ }\href@noop
  {} {\bibfield  {journal} {\bibinfo  {journal} {Physical review research}\
  }\textbf {\bibinfo {volume} {3}},\ \bibinfo {pages} {023005} (\bibinfo {year}
  {2021})}\BibitemShut {NoStop}%
\bibitem [{\citenamefont {Guo}\ \emph {et~al.}(2019)\citenamefont {Guo},
  \citenamefont {Liu}, \citenamefont {Xiong}, \citenamefont {Xue},
  \citenamefont {Fu}, \citenamefont {Huang}, \citenamefont {Qiang},
  \citenamefont {Xu}, \citenamefont {Liu}, \citenamefont {Zheng} \emph
  {et~al.}}]{guo2019general}%
  \BibitemOpen
  \bibfield  {author} {\bibinfo {author} {\bibfnamefont {C.}~\bibnamefont
  {Guo}}, \bibinfo {author} {\bibfnamefont {Y.}~\bibnamefont {Liu}}, \bibinfo
  {author} {\bibfnamefont {M.}~\bibnamefont {Xiong}}, \bibinfo {author}
  {\bibfnamefont {S.}~\bibnamefont {Xue}}, \bibinfo {author} {\bibfnamefont
  {X.}~\bibnamefont {Fu}}, \bibinfo {author} {\bibfnamefont {A.}~\bibnamefont
  {Huang}}, \bibinfo {author} {\bibfnamefont {X.}~\bibnamefont {Qiang}},
  \bibinfo {author} {\bibfnamefont {P.}~\bibnamefont {Xu}}, \bibinfo {author}
  {\bibfnamefont {J.}~\bibnamefont {Liu}}, \bibinfo {author} {\bibfnamefont
  {S.}~\bibnamefont {Zheng}}, \emph {et~al.},\ }\bibfield  {title} {\bibinfo
  {title} {General-purpose quantum circuit simulator with projected
  entangled-pair states and the quantum supremacy frontier},\ }\href@noop {}
  {\bibfield  {journal} {\bibinfo  {journal} {Physical review letters}\
  }\textbf {\bibinfo {volume} {123}},\ \bibinfo {pages} {190501} (\bibinfo
  {year} {2019})}\BibitemShut {NoStop}%
\bibitem [{\citenamefont {Ellerbrock}\ and\ \citenamefont
  {Martinez}(2020)}]{ellerbrock2020multilayer}%
  \BibitemOpen
  \bibfield  {author} {\bibinfo {author} {\bibfnamefont {R.}~\bibnamefont
  {Ellerbrock}}\ and\ \bibinfo {author} {\bibfnamefont {T.~J.}\ \bibnamefont
  {Martinez}},\ }\bibfield  {title} {\bibinfo {title} {A multilayer
  multi-configurational approach to efficiently simulate large-scale
  circuit-based quantum computers on classical machines},\ }\href@noop {}
  {\bibfield  {journal} {\bibinfo  {journal} {The Journal of Chemical Physics}\
  }\textbf {\bibinfo {volume} {153}} (\bibinfo {year} {2020})}\BibitemShut
  {NoStop}%
\bibitem [{\citenamefont {Dumitrescu}(2017)}]{dumitrescu2017tree}%
  \BibitemOpen
  \bibfield  {author} {\bibinfo {author} {\bibfnamefont {E.}~\bibnamefont
  {Dumitrescu}},\ }\bibfield  {title} {\bibinfo {title} {Tree tensor network
  approach to simulating shor's algorithm},\ }\href@noop {} {\bibfield
  {journal} {\bibinfo  {journal} {Physical Review A}\ }\textbf {\bibinfo
  {volume} {96}},\ \bibinfo {pages} {062322} (\bibinfo {year}
  {2017})}\BibitemShut {NoStop}%
\bibitem [{\citenamefont {Gao}\ \emph {et~al.}(2025)\citenamefont {Gao} \emph
  {et~al.}}]{gao2025establishing}%
  \BibitemOpen
  \bibfield  {author} {\bibinfo {author} {\bibfnamefont {D.}~\bibnamefont
  {Gao}} \emph {et~al.},\ }\bibfield  {title} {\bibinfo {title} {Establishing a
  new benchmark in quantum computational advantage with 105-qubit zuchongzhi
  3.0 processor},\ }\href {https://doi.org/10.1103/PhysRevLett.134.090601}
  {\bibfield  {journal} {\bibinfo  {journal} {Phys. Rev. Lett.}\ }\textbf
  {\bibinfo {volume} {134}},\ \bibinfo {pages} {090601} (\bibinfo {year}
  {2025})}\BibitemShut {NoStop}%
\bibitem [{\citenamefont {Gross}\ and\ \citenamefont
  {Bloch}(2017)}]{gross2017quantum}%
  \BibitemOpen
  \bibfield  {author} {\bibinfo {author} {\bibfnamefont {C.}~\bibnamefont
  {Gross}}\ and\ \bibinfo {author} {\bibfnamefont {I.}~\bibnamefont {Bloch}},\
  }\bibfield  {title} {\bibinfo {title} {Quantum simulations with ultracold
  atoms in optical lattices},\ }\href@noop {} {\bibfield  {journal} {\bibinfo
  {journal} {Science}\ }\textbf {\bibinfo {volume} {357}},\ \bibinfo {pages}
  {995} (\bibinfo {year} {2017})}\BibitemShut {NoStop}%
\bibitem [{\citenamefont {Blatt}\ and\ \citenamefont
  {Roos}(2012)}]{blatt2012quantum}%
  \BibitemOpen
  \bibfield  {author} {\bibinfo {author} {\bibfnamefont {R.}~\bibnamefont
  {Blatt}}\ and\ \bibinfo {author} {\bibfnamefont {C.~F.}\ \bibnamefont
  {Roos}},\ }\bibfield  {title} {\bibinfo {title} {Quantum simulations with
  trapped ions},\ }\href@noop {} {\bibfield  {journal} {\bibinfo  {journal}
  {Nature Physics}\ }\textbf {\bibinfo {volume} {8}},\ \bibinfo {pages} {277}
  (\bibinfo {year} {2012})}\BibitemShut {NoStop}%
\bibitem [{\citenamefont {Browaeys}\ and\ \citenamefont
  {Lahaye}(2020)}]{browaeys2020many}%
  \BibitemOpen
  \bibfield  {author} {\bibinfo {author} {\bibfnamefont {A.}~\bibnamefont
  {Browaeys}}\ and\ \bibinfo {author} {\bibfnamefont {T.}~\bibnamefont
  {Lahaye}},\ }\bibfield  {title} {\bibinfo {title} {Many-body physics with
  individually controlled rydberg atoms},\ }\href@noop {} {\bibfield  {journal}
  {\bibinfo  {journal} {Nature Physics}\ }\textbf {\bibinfo {volume} {16}},\
  \bibinfo {pages} {132} (\bibinfo {year} {2020})}\BibitemShut {NoStop}%
\bibitem [{\citenamefont {Aspuru-Guzik}\ and\ \citenamefont
  {Walther}(2012)}]{aspuru2012photonic}%
  \BibitemOpen
  \bibfield  {author} {\bibinfo {author} {\bibfnamefont {A.}~\bibnamefont
  {Aspuru-Guzik}}\ and\ \bibinfo {author} {\bibfnamefont {P.}~\bibnamefont
  {Walther}},\ }\bibfield  {title} {\bibinfo {title} {Photonic quantum
  simulators},\ }\href@noop {} {\bibfield  {journal} {\bibinfo  {journal}
  {Nature physics}\ }\textbf {\bibinfo {volume} {8}},\ \bibinfo {pages} {285}
  (\bibinfo {year} {2012})}\BibitemShut {NoStop}%
\bibitem [{\citenamefont {Houck}\ \emph {et~al.}(2012)\citenamefont {Houck},
  \citenamefont {T{\"u}reci},\ and\ \citenamefont {Koch}}]{houck2012chip}%
  \BibitemOpen
  \bibfield  {author} {\bibinfo {author} {\bibfnamefont {A.~A.}\ \bibnamefont
  {Houck}}, \bibinfo {author} {\bibfnamefont {H.~E.}\ \bibnamefont
  {T{\"u}reci}},\ and\ \bibinfo {author} {\bibfnamefont {J.}~\bibnamefont
  {Koch}},\ }\bibfield  {title} {\bibinfo {title} {On-chip quantum simulation
  with superconducting circuits},\ }\href@noop {} {\bibfield  {journal}
  {\bibinfo  {journal} {Nature Physics}\ }\textbf {\bibinfo {volume} {8}},\
  \bibinfo {pages} {292} (\bibinfo {year} {2012})}\BibitemShut {NoStop}%
\bibitem [{\citenamefont {Kadowaki}\ and\ \citenamefont
  {Nishimori}(1998)}]{kadowaki1998quantum}%
  \BibitemOpen
  \bibfield  {author} {\bibinfo {author} {\bibfnamefont {T.}~\bibnamefont
  {Kadowaki}}\ and\ \bibinfo {author} {\bibfnamefont {H.}~\bibnamefont
  {Nishimori}},\ }\bibfield  {title} {\bibinfo {title} {Quantum annealing in
  the transverse ising model},\ }\href@noop {} {\bibfield  {journal} {\bibinfo
  {journal} {Physical Review E}\ }\textbf {\bibinfo {volume} {58}},\ \bibinfo
  {pages} {5355} (\bibinfo {year} {1998})}\BibitemShut {NoStop}%
\bibitem [{\citenamefont {King}\ \emph {et~al.}(2024)\citenamefont {King},
  \citenamefont {Nocera}, \citenamefont {Rams}, \citenamefont {Dziarmaga},
  \citenamefont {Wiersema}, \citenamefont {Bernoudy}, \citenamefont {Raymond},
  \citenamefont {Kaushal}, \citenamefont {Heinsdorf}, \citenamefont {Harris}
  \emph {et~al.}}]{king2024computational}%
  \BibitemOpen
  \bibfield  {author} {\bibinfo {author} {\bibfnamefont {A.~D.}\ \bibnamefont
  {King}}, \bibinfo {author} {\bibfnamefont {A.}~\bibnamefont {Nocera}},
  \bibinfo {author} {\bibfnamefont {M.~M.}\ \bibnamefont {Rams}}, \bibinfo
  {author} {\bibfnamefont {J.}~\bibnamefont {Dziarmaga}}, \bibinfo {author}
  {\bibfnamefont {R.}~\bibnamefont {Wiersema}}, \bibinfo {author}
  {\bibfnamefont {W.}~\bibnamefont {Bernoudy}}, \bibinfo {author}
  {\bibfnamefont {J.}~\bibnamefont {Raymond}}, \bibinfo {author} {\bibfnamefont
  {N.}~\bibnamefont {Kaushal}}, \bibinfo {author} {\bibfnamefont
  {N.}~\bibnamefont {Heinsdorf}}, \bibinfo {author} {\bibfnamefont
  {R.}~\bibnamefont {Harris}}, \emph {et~al.},\ }\bibfield  {title} {\bibinfo
  {title} {Computational supremacy in quantum simulation},\ }\href@noop {}
  {\bibfield  {journal} {\bibinfo  {journal} {arXiv preprint arXiv:2403.00910}\
  } (\bibinfo {year} {2024})}\BibitemShut {NoStop}%
\bibitem [{\citenamefont {Shaw}\ \emph {et~al.}(2024)\citenamefont {Shaw},
  \citenamefont {Chen}, \citenamefont {Choi}, \citenamefont {Mark},
  \citenamefont {Scholl}, \citenamefont {Finkelstein}, \citenamefont {Elben},
  \citenamefont {Choi},\ and\ \citenamefont {Endres}}]{shaw2024benchmarking}%
  \BibitemOpen
  \bibfield  {author} {\bibinfo {author} {\bibfnamefont {A.~L.}\ \bibnamefont
  {Shaw}}, \bibinfo {author} {\bibfnamefont {Z.}~\bibnamefont {Chen}}, \bibinfo
  {author} {\bibfnamefont {J.}~\bibnamefont {Choi}}, \bibinfo {author}
  {\bibfnamefont {D.~K.}\ \bibnamefont {Mark}}, \bibinfo {author}
  {\bibfnamefont {P.}~\bibnamefont {Scholl}}, \bibinfo {author} {\bibfnamefont
  {R.}~\bibnamefont {Finkelstein}}, \bibinfo {author} {\bibfnamefont
  {A.}~\bibnamefont {Elben}}, \bibinfo {author} {\bibfnamefont
  {S.}~\bibnamefont {Choi}},\ and\ \bibinfo {author} {\bibfnamefont
  {M.}~\bibnamefont {Endres}},\ }\bibfield  {title} {\bibinfo {title}
  {Benchmarking highly entangled states on a 60-atom analogue quantum
  simulator},\ }\href@noop {} {\bibfield  {journal} {\bibinfo  {journal}
  {Nature}\ }\textbf {\bibinfo {volume} {628}},\ \bibinfo {pages} {71}
  (\bibinfo {year} {2024})}\BibitemShut {NoStop}%
\bibitem [{\citenamefont {Tindall}\ \emph {et~al.}(2025)\citenamefont
  {Tindall}, \citenamefont {Mello}, \citenamefont {Fishman}, \citenamefont
  {Stoudenmire},\ and\ \citenamefont {Sels}}]{tindall2025dynamics}%
  \BibitemOpen
  \bibfield  {author} {\bibinfo {author} {\bibfnamefont {J.}~\bibnamefont
  {Tindall}}, \bibinfo {author} {\bibfnamefont {A.~F.}\ \bibnamefont {Mello}},
  \bibinfo {author} {\bibfnamefont {M.}~\bibnamefont {Fishman}}, \bibinfo
  {author} {\bibfnamefont {E.~M.}\ \bibnamefont {Stoudenmire}},\ and\ \bibinfo
  {author} {\bibfnamefont {D.}~\bibnamefont {Sels}},\ }\bibfield  {title}
  {\bibinfo {title} {Dynamics of disordered quantum systems with two- and
  three-dimensional tensor networks},\ }\href@noop {} {\bibfield  {journal}
  {\bibinfo  {journal} {arXiv preprint arXiv:2503.05693}\ } (\bibinfo {year}
  {2025})}\BibitemShut {NoStop}%
\bibitem [{\citenamefont {King}\ \emph {et~al.}(2025)\citenamefont {King},
  \citenamefont {Nocera}, \citenamefont {Rams}, \citenamefont {Dziarmaga},
  \citenamefont {Raymond}, \citenamefont {Kaushal}, \citenamefont {Sandvik},
  \citenamefont {Alvarez}, \citenamefont {Carrasquilla}, \citenamefont {Franz}
  \emph {et~al.}}]{king2025comment}%
  \BibitemOpen
  \bibfield  {author} {\bibinfo {author} {\bibfnamefont {A.~D.}\ \bibnamefont
  {King}}, \bibinfo {author} {\bibfnamefont {A.}~\bibnamefont {Nocera}},
  \bibinfo {author} {\bibfnamefont {M.~M.}\ \bibnamefont {Rams}}, \bibinfo
  {author} {\bibfnamefont {J.}~\bibnamefont {Dziarmaga}}, \bibinfo {author}
  {\bibfnamefont {J.}~\bibnamefont {Raymond}}, \bibinfo {author} {\bibfnamefont
  {N.}~\bibnamefont {Kaushal}}, \bibinfo {author} {\bibfnamefont {A.~W.}\
  \bibnamefont {Sandvik}}, \bibinfo {author} {\bibfnamefont {G.}~\bibnamefont
  {Alvarez}}, \bibinfo {author} {\bibfnamefont {J.}~\bibnamefont
  {Carrasquilla}}, \bibinfo {author} {\bibfnamefont {M.}~\bibnamefont {Franz}},
  \emph {et~al.},\ }\bibfield  {title} {\bibinfo {title} {Comment on:" dynamics
  of disordered quantum systems with two-and three-dimensional tensor networks"
  arxiv: 2503.05693},\ }\href@noop {} {\bibfield  {journal} {\bibinfo
  {journal} {arXiv preprint arXiv:2504.06283}\ } (\bibinfo {year}
  {2025})}\BibitemShut {NoStop}%
\bibitem [{\citenamefont {Aaronson}\ and\ \citenamefont
  {Arkhipov}(2011)}]{aaronson2011computational}%
  \BibitemOpen
  \bibfield  {author} {\bibinfo {author} {\bibfnamefont {S.}~\bibnamefont
  {Aaronson}}\ and\ \bibinfo {author} {\bibfnamefont {A.}~\bibnamefont
  {Arkhipov}},\ }\href@noop {} {\bibinfo {title} {The computational complexity
  of linear optics}} (\bibinfo {year} {2011})\BibitemShut {NoStop}%
\bibitem [{\citenamefont {Zhong}\ \emph {et~al.}(2020)\citenamefont {Zhong},
  \citenamefont {Wang}, \citenamefont {Deng}, \citenamefont {Chen},
  \citenamefont {Peng}, \citenamefont {Luo}, \citenamefont {Qin}, \citenamefont
  {Wu}, \citenamefont {Ding}, \citenamefont {Hu} \emph
  {et~al.}}]{zhong2020quantum}%
  \BibitemOpen
  \bibfield  {author} {\bibinfo {author} {\bibfnamefont {H.-S.}\ \bibnamefont
  {Zhong}}, \bibinfo {author} {\bibfnamefont {H.}~\bibnamefont {Wang}},
  \bibinfo {author} {\bibfnamefont {Y.-H.}\ \bibnamefont {Deng}}, \bibinfo
  {author} {\bibfnamefont {M.-C.}\ \bibnamefont {Chen}}, \bibinfo {author}
  {\bibfnamefont {L.-C.}\ \bibnamefont {Peng}}, \bibinfo {author}
  {\bibfnamefont {Y.-H.}\ \bibnamefont {Luo}}, \bibinfo {author} {\bibfnamefont
  {J.}~\bibnamefont {Qin}}, \bibinfo {author} {\bibfnamefont {D.}~\bibnamefont
  {Wu}}, \bibinfo {author} {\bibfnamefont {X.}~\bibnamefont {Ding}}, \bibinfo
  {author} {\bibfnamefont {Y.}~\bibnamefont {Hu}}, \emph {et~al.},\ }\bibfield
  {title} {\bibinfo {title} {Quantum computational advantage using photons},\
  }\href@noop {} {\bibfield  {journal} {\bibinfo  {journal} {Science}\ }\textbf
  {\bibinfo {volume} {370}},\ \bibinfo {pages} {1460} (\bibinfo {year}
  {2020})}\BibitemShut {NoStop}%
\bibitem [{\citenamefont {Zhong}\ \emph {et~al.}(2021)\citenamefont {Zhong},
  \citenamefont {Deng}, \citenamefont {Qin}, \citenamefont {Wang},
  \citenamefont {Chen}, \citenamefont {Peng}, \citenamefont {Luo},
  \citenamefont {Wu}, \citenamefont {Gong}, \citenamefont {Su} \emph
  {et~al.}}]{zhong2021phase}%
  \BibitemOpen
  \bibfield  {author} {\bibinfo {author} {\bibfnamefont {H.-S.}\ \bibnamefont
  {Zhong}}, \bibinfo {author} {\bibfnamefont {Y.-H.}\ \bibnamefont {Deng}},
  \bibinfo {author} {\bibfnamefont {J.}~\bibnamefont {Qin}}, \bibinfo {author}
  {\bibfnamefont {H.}~\bibnamefont {Wang}}, \bibinfo {author} {\bibfnamefont
  {M.-C.}\ \bibnamefont {Chen}}, \bibinfo {author} {\bibfnamefont {L.-C.}\
  \bibnamefont {Peng}}, \bibinfo {author} {\bibfnamefont {Y.-H.}\ \bibnamefont
  {Luo}}, \bibinfo {author} {\bibfnamefont {D.}~\bibnamefont {Wu}}, \bibinfo
  {author} {\bibfnamefont {S.-Q.}\ \bibnamefont {Gong}}, \bibinfo {author}
  {\bibfnamefont {H.}~\bibnamefont {Su}}, \emph {et~al.},\ }\bibfield  {title}
  {\bibinfo {title} {{Phase-Programmable Gaussian Boson Sampling Using
  Stimulated Squeezed Light}},\ }\href@noop {} {\bibfield  {journal} {\bibinfo
  {journal} {Physical review letters}\ }\textbf {\bibinfo {volume} {127}},\
  \bibinfo {pages} {180502} (\bibinfo {year} {2021})}\BibitemShut {NoStop}%
\bibitem [{\citenamefont {Madsen}\ \emph {et~al.}(2022)\citenamefont {Madsen},
  \citenamefont {Laudenbach}, \citenamefont {Askarani}, \citenamefont
  {Rortais}, \citenamefont {Vincent}, \citenamefont {Bulmer}, \citenamefont
  {Miatto}, \citenamefont {Neuhaus}, \citenamefont {Helt}, \citenamefont
  {Collins} \emph {et~al.}}]{madsen2022quantum}%
  \BibitemOpen
  \bibfield  {author} {\bibinfo {author} {\bibfnamefont {L.~S.}\ \bibnamefont
  {Madsen}}, \bibinfo {author} {\bibfnamefont {F.}~\bibnamefont {Laudenbach}},
  \bibinfo {author} {\bibfnamefont {M.~F.}\ \bibnamefont {Askarani}}, \bibinfo
  {author} {\bibfnamefont {F.}~\bibnamefont {Rortais}}, \bibinfo {author}
  {\bibfnamefont {T.}~\bibnamefont {Vincent}}, \bibinfo {author} {\bibfnamefont
  {J.~F.}\ \bibnamefont {Bulmer}}, \bibinfo {author} {\bibfnamefont {F.~M.}\
  \bibnamefont {Miatto}}, \bibinfo {author} {\bibfnamefont {L.}~\bibnamefont
  {Neuhaus}}, \bibinfo {author} {\bibfnamefont {L.~G.}\ \bibnamefont {Helt}},
  \bibinfo {author} {\bibfnamefont {M.~J.}\ \bibnamefont {Collins}}, \emph
  {et~al.},\ }\bibfield  {title} {\bibinfo {title} {Quantum computational
  advantage with a programmable photonic processor},\ }\href@noop {} {\bibfield
   {journal} {\bibinfo  {journal} {Nature}\ }\textbf {\bibinfo {volume}
  {606}},\ \bibinfo {pages} {75} (\bibinfo {year} {2022})}\BibitemShut
  {NoStop}%
\bibitem [{\citenamefont {Deng}\ \emph {et~al.}(2023)\citenamefont {Deng},
  \citenamefont {Gu}, \citenamefont {Liu}, \citenamefont {Gong}, \citenamefont
  {Su}, \citenamefont {Zhang}, \citenamefont {Tang}, \citenamefont {Jia},
  \citenamefont {Xu}, \citenamefont {Chen} \emph {et~al.}}]{deng2023gaussian}%
  \BibitemOpen
  \bibfield  {author} {\bibinfo {author} {\bibfnamefont {Y.-H.}\ \bibnamefont
  {Deng}}, \bibinfo {author} {\bibfnamefont {Y.-C.}\ \bibnamefont {Gu}},
  \bibinfo {author} {\bibfnamefont {H.-L.}\ \bibnamefont {Liu}}, \bibinfo
  {author} {\bibfnamefont {S.-Q.}\ \bibnamefont {Gong}}, \bibinfo {author}
  {\bibfnamefont {H.}~\bibnamefont {Su}}, \bibinfo {author} {\bibfnamefont
  {Z.-J.}\ \bibnamefont {Zhang}}, \bibinfo {author} {\bibfnamefont {H.-Y.}\
  \bibnamefont {Tang}}, \bibinfo {author} {\bibfnamefont {M.-H.}\ \bibnamefont
  {Jia}}, \bibinfo {author} {\bibfnamefont {J.-M.}\ \bibnamefont {Xu}},
  \bibinfo {author} {\bibfnamefont {M.-C.}\ \bibnamefont {Chen}}, \emph
  {et~al.},\ }\bibfield  {title} {\bibinfo {title} {Gaussian boson sampling
  with pseudo-photon-number-resolving detectors and quantum computational
  advantage},\ }\href@noop {} {\bibfield  {journal} {\bibinfo  {journal}
  {Physical review letters}\ }\textbf {\bibinfo {volume} {131}},\ \bibinfo
  {pages} {150601} (\bibinfo {year} {2023})}\BibitemShut {NoStop}%
\bibitem [{\citenamefont {Huang}\ \emph {et~al.}(2019)\citenamefont {Huang},
  \citenamefont {Bao},\ and\ \citenamefont {Guo}}]{huang2019simulating}%
  \BibitemOpen
  \bibfield  {author} {\bibinfo {author} {\bibfnamefont {H.-L.}\ \bibnamefont
  {Huang}}, \bibinfo {author} {\bibfnamefont {W.-S.}\ \bibnamefont {Bao}},\
  and\ \bibinfo {author} {\bibfnamefont {C.}~\bibnamefont {Guo}},\ }\bibfield
  {title} {\bibinfo {title} {Simulating the dynamics of single photons in boson
  sampling devices with matrix product states},\ }\href@noop {} {\bibfield
  {journal} {\bibinfo  {journal} {Physical Review A}\ }\textbf {\bibinfo
  {volume} {100}},\ \bibinfo {pages} {032305} (\bibinfo {year}
  {2019})}\BibitemShut {NoStop}%
\bibitem [{\citenamefont {Oh}\ \emph {et~al.}(2021)\citenamefont {Oh},
  \citenamefont {Noh}, \citenamefont {Fefferman},\ and\ \citenamefont
  {Jiang}}]{oh2021classical}%
  \BibitemOpen
  \bibfield  {author} {\bibinfo {author} {\bibfnamefont {C.}~\bibnamefont
  {Oh}}, \bibinfo {author} {\bibfnamefont {K.}~\bibnamefont {Noh}}, \bibinfo
  {author} {\bibfnamefont {B.}~\bibnamefont {Fefferman}},\ and\ \bibinfo
  {author} {\bibfnamefont {L.}~\bibnamefont {Jiang}},\ }\bibfield  {title}
  {\bibinfo {title} {Classical simulation of lossy boson sampling using matrix
  product operators},\ }\href@noop {} {\bibfield  {journal} {\bibinfo
  {journal} {Physical Review A}\ }\textbf {\bibinfo {volume} {104}},\ \bibinfo
  {pages} {022407} (\bibinfo {year} {2021})}\BibitemShut {NoStop}%
\bibitem [{\citenamefont {Liu}\ \emph {et~al.}(2023{\natexlab{a}})\citenamefont
  {Liu}, \citenamefont {Oh}, \citenamefont {Liu}, \citenamefont {Jiang},\ and\
  \citenamefont {Alexeev}}]{liu2023simulating}%
  \BibitemOpen
  \bibfield  {author} {\bibinfo {author} {\bibfnamefont {M.}~\bibnamefont
  {Liu}}, \bibinfo {author} {\bibfnamefont {C.}~\bibnamefont {Oh}}, \bibinfo
  {author} {\bibfnamefont {J.}~\bibnamefont {Liu}}, \bibinfo {author}
  {\bibfnamefont {L.}~\bibnamefont {Jiang}},\ and\ \bibinfo {author}
  {\bibfnamefont {Y.}~\bibnamefont {Alexeev}},\ }\bibfield  {title} {\bibinfo
  {title} {{Simulating lossy Gaussian boson sampling with matrix-product
  operators}},\ }\href@noop {} {\bibfield  {journal} {\bibinfo  {journal}
  {Physical Review A}\ }\textbf {\bibinfo {volume} {108}},\ \bibinfo {pages}
  {052604} (\bibinfo {year} {2023}{\natexlab{a}})}\BibitemShut {NoStop}%
\bibitem [{\citenamefont {Hamilton}\ \emph {et~al.}(2017)\citenamefont
  {Hamilton}, \citenamefont {Kruse}, \citenamefont {Sansoni}, \citenamefont
  {Barkhofen}, \citenamefont {Silberhorn},\ and\ \citenamefont
  {Jex}}]{hamilton2017gaussian}%
  \BibitemOpen
  \bibfield  {author} {\bibinfo {author} {\bibfnamefont {C.~S.}\ \bibnamefont
  {Hamilton}}, \bibinfo {author} {\bibfnamefont {R.}~\bibnamefont {Kruse}},
  \bibinfo {author} {\bibfnamefont {L.}~\bibnamefont {Sansoni}}, \bibinfo
  {author} {\bibfnamefont {S.}~\bibnamefont {Barkhofen}}, \bibinfo {author}
  {\bibfnamefont {C.}~\bibnamefont {Silberhorn}},\ and\ \bibinfo {author}
  {\bibfnamefont {I.}~\bibnamefont {Jex}},\ }\bibfield  {title} {\bibinfo
  {title} {Gaussian boson sampling},\ }\href@noop {} {\bibfield  {journal}
  {\bibinfo  {journal} {Physical review letters}\ }\textbf {\bibinfo {volume}
  {119}},\ \bibinfo {pages} {170501} (\bibinfo {year} {2017})}\BibitemShut
  {NoStop}%
\bibitem [{\citenamefont {Quesada}\ \emph {et~al.}(2022)\citenamefont
  {Quesada}, \citenamefont {Chadwick}, \citenamefont {Bell}, \citenamefont
  {Arrazola}, \citenamefont {Vincent}, \citenamefont {Qi},\ and\ \citenamefont
  {Garc{\'\i}a-Patr{\'o}n}}]{quesada2022quadratic}%
  \BibitemOpen
  \bibfield  {author} {\bibinfo {author} {\bibfnamefont {N.}~\bibnamefont
  {Quesada}}, \bibinfo {author} {\bibfnamefont {R.~S.}\ \bibnamefont
  {Chadwick}}, \bibinfo {author} {\bibfnamefont {B.~A.}\ \bibnamefont {Bell}},
  \bibinfo {author} {\bibfnamefont {J.~M.}\ \bibnamefont {Arrazola}}, \bibinfo
  {author} {\bibfnamefont {T.}~\bibnamefont {Vincent}}, \bibinfo {author}
  {\bibfnamefont {H.}~\bibnamefont {Qi}},\ and\ \bibinfo {author}
  {\bibfnamefont {R.}~\bibnamefont {Garc{\'\i}a-Patr{\'o}n}},\ }\bibfield
  {title} {\bibinfo {title} {Quadratic speed-up for simulating gaussian boson
  sampling},\ }\href@noop {} {\bibfield  {journal} {\bibinfo  {journal} {PRX
  Quantum}\ }\textbf {\bibinfo {volume} {3}},\ \bibinfo {pages} {010306}
  (\bibinfo {year} {2022})}\BibitemShut {NoStop}%
\bibitem [{\citenamefont {Cilluffo}\ \emph {et~al.}(2023)\citenamefont
  {Cilluffo}, \citenamefont {Lorenzoni},\ and\ \citenamefont
  {Plenio}}]{cilluffo2023simulating}%
  \BibitemOpen
  \bibfield  {author} {\bibinfo {author} {\bibfnamefont {D.}~\bibnamefont
  {Cilluffo}}, \bibinfo {author} {\bibfnamefont {N.}~\bibnamefont
  {Lorenzoni}},\ and\ \bibinfo {author} {\bibfnamefont {M.~B.}\ \bibnamefont
  {Plenio}},\ }\bibfield  {title} {\bibinfo {title} {{Simulating Gaussian Boson
  Sampling with Tensor Networks in the Heisenberg picture}},\ }\href@noop {}
  {\bibfield  {journal} {\bibinfo  {journal} {arXiv preprint arXiv:2305.11215}\
  } (\bibinfo {year} {2023})}\BibitemShut {NoStop}%
\bibitem [{\citenamefont {Nielsen}\ and\ \citenamefont
  {Chuang}(2010)}]{nielsen2010quantum}%
  \BibitemOpen
  \bibfield  {author} {\bibinfo {author} {\bibfnamefont {M.~A.}\ \bibnamefont
  {Nielsen}}\ and\ \bibinfo {author} {\bibfnamefont {I.~L.}\ \bibnamefont
  {Chuang}},\ }\href@noop {} {\emph {\bibinfo {title} {Quantum computation and
  quantum information}}}\ (\bibinfo  {publisher} {Cambridge university press},\
  \bibinfo {year} {2010})\BibitemShut {NoStop}%
\bibitem [{\citenamefont {Biamonte}\ and\ \citenamefont
  {Bergholm}(2017)}]{biamonte2017tensor}%
  \BibitemOpen
  \bibfield  {author} {\bibinfo {author} {\bibfnamefont {J.}~\bibnamefont
  {Biamonte}}\ and\ \bibinfo {author} {\bibfnamefont {V.}~\bibnamefont
  {Bergholm}},\ }\bibfield  {title} {\bibinfo {title} {Tensor networks in a
  nutshell},\ }\href@noop {} {\bibfield  {journal} {\bibinfo  {journal} {arXiv
  preprint arXiv:1708.00006}\ } (\bibinfo {year} {2017})}\BibitemShut {NoStop}%
\bibitem [{\citenamefont {Vidal}(2003{\natexlab{b}})}]{vidal2003efficient}%
  \BibitemOpen
  \bibfield  {author} {\bibinfo {author} {\bibfnamefont {G.}~\bibnamefont
  {Vidal}},\ }\bibfield  {title} {\bibinfo {title} {Efficient classical
  simulation of slightly entangled quantum computations},\ }\href@noop {}
  {\bibfield  {journal} {\bibinfo  {journal} {Physical review letters}\
  }\textbf {\bibinfo {volume} {91}},\ \bibinfo {pages} {147902} (\bibinfo
  {year} {2003}{\natexlab{b}})}\BibitemShut {NoStop}%
\bibitem [{\citenamefont {Sch{\"o}n}\ \emph {et~al.}(2005)\citenamefont
  {Sch{\"o}n}, \citenamefont {Solano}, \citenamefont {Verstraete},
  \citenamefont {Cirac},\ and\ \citenamefont {Wolf}}]{schon2005sequential}%
  \BibitemOpen
  \bibfield  {author} {\bibinfo {author} {\bibfnamefont {C.}~\bibnamefont
  {Sch{\"o}n}}, \bibinfo {author} {\bibfnamefont {E.}~\bibnamefont {Solano}},
  \bibinfo {author} {\bibfnamefont {F.}~\bibnamefont {Verstraete}}, \bibinfo
  {author} {\bibfnamefont {J.~I.}\ \bibnamefont {Cirac}},\ and\ \bibinfo
  {author} {\bibfnamefont {M.~M.}\ \bibnamefont {Wolf}},\ }\bibfield  {title}
  {\bibinfo {title} {Sequential generation of entangled multiqubit states},\
  }\href@noop {} {\bibfield  {journal} {\bibinfo  {journal} {Physical review
  letters}\ }\textbf {\bibinfo {volume} {95}},\ \bibinfo {pages} {110503}
  (\bibinfo {year} {2005})}\BibitemShut {NoStop}%
\bibitem [{\citenamefont {Perez-Garcia}\ \emph {et~al.}(2006)\citenamefont
  {Perez-Garcia}, \citenamefont {Verstraete}, \citenamefont {Wolf},\ and\
  \citenamefont {Cirac}}]{perez2006matrix}%
  \BibitemOpen
  \bibfield  {author} {\bibinfo {author} {\bibfnamefont {D.}~\bibnamefont
  {Perez-Garcia}}, \bibinfo {author} {\bibfnamefont {F.}~\bibnamefont
  {Verstraete}}, \bibinfo {author} {\bibfnamefont {M.~M.}\ \bibnamefont
  {Wolf}},\ and\ \bibinfo {author} {\bibfnamefont {J.~I.}\ \bibnamefont
  {Cirac}},\ }\bibfield  {title} {\bibinfo {title} {Matrix product state
  representations},\ }\href@noop {} {\bibfield  {journal} {\bibinfo  {journal}
  {arXiv preprint quant-ph/0608197}\ } (\bibinfo {year} {2006})}\BibitemShut
  {NoStop}%
\bibitem [{\citenamefont {Ran}(2020)}]{ran2020encoding}%
  \BibitemOpen
  \bibfield  {author} {\bibinfo {author} {\bibfnamefont {S.-J.}\ \bibnamefont
  {Ran}},\ }\bibfield  {title} {\bibinfo {title} {Encoding of matrix product
  states into quantum circuits of one-and two-qubit gates},\ }\href@noop {}
  {\bibfield  {journal} {\bibinfo  {journal} {Physical Review A}\ }\textbf
  {\bibinfo {volume} {101}},\ \bibinfo {pages} {032310} (\bibinfo {year}
  {2020})}\BibitemShut {NoStop}%
\bibitem [{\citenamefont {Wei}\ \emph {et~al.}(2023)\citenamefont {Wei},
  \citenamefont {Malz},\ and\ \citenamefont {Cirac}}]{wei2023efficient}%
  \BibitemOpen
  \bibfield  {author} {\bibinfo {author} {\bibfnamefont {Z.-Y.}\ \bibnamefont
  {Wei}}, \bibinfo {author} {\bibfnamefont {D.}~\bibnamefont {Malz}},\ and\
  \bibinfo {author} {\bibfnamefont {J.~I.}\ \bibnamefont {Cirac}},\ }\bibfield
  {title} {\bibinfo {title} {Efficient adiabatic preparation of tensor network
  states},\ }\href@noop {} {\bibfield  {journal} {\bibinfo  {journal} {Physical
  Review Research}\ }\textbf {\bibinfo {volume} {5}},\ \bibinfo {pages}
  {L022037} (\bibinfo {year} {2023})}\BibitemShut {NoStop}%
\bibitem [{\citenamefont {Malz}\ \emph {et~al.}(2024)\citenamefont {Malz},
  \citenamefont {Styliaris}, \citenamefont {Wei},\ and\ \citenamefont
  {Cirac}}]{malz2024preparation}%
  \BibitemOpen
  \bibfield  {author} {\bibinfo {author} {\bibfnamefont {D.}~\bibnamefont
  {Malz}}, \bibinfo {author} {\bibfnamefont {G.}~\bibnamefont {Styliaris}},
  \bibinfo {author} {\bibfnamefont {Z.-Y.}\ \bibnamefont {Wei}},\ and\ \bibinfo
  {author} {\bibfnamefont {J.~I.}\ \bibnamefont {Cirac}},\ }\bibfield  {title}
  {\bibinfo {title} {Preparation of matrix product states with log-depth
  quantum circuits},\ }\href@noop {} {\bibfield  {journal} {\bibinfo  {journal}
  {Physical Review Letters}\ }\textbf {\bibinfo {volume} {132}},\ \bibinfo
  {pages} {040404} (\bibinfo {year} {2024})}\BibitemShut {NoStop}%
\bibitem [{\citenamefont {Haghshenas}\ \emph {et~al.}(2022)\citenamefont
  {Haghshenas}, \citenamefont {Gray}, \citenamefont {Potter},\ and\
  \citenamefont {Chan}}]{haghshenas2022variational}%
  \BibitemOpen
  \bibfield  {author} {\bibinfo {author} {\bibfnamefont {R.}~\bibnamefont
  {Haghshenas}}, \bibinfo {author} {\bibfnamefont {J.}~\bibnamefont {Gray}},
  \bibinfo {author} {\bibfnamefont {A.~C.}\ \bibnamefont {Potter}},\ and\
  \bibinfo {author} {\bibfnamefont {G.~K.-L.}\ \bibnamefont {Chan}},\
  }\bibfield  {title} {\bibinfo {title} {Variational power of quantum circuit
  tensor networks},\ }\href@noop {} {\bibfield  {journal} {\bibinfo  {journal}
  {Physical Review X}\ }\textbf {\bibinfo {volume} {12}},\ \bibinfo {pages}
  {011047} (\bibinfo {year} {2022})}\BibitemShut {NoStop}%
\bibitem [{\citenamefont {Foss-Feig}\ \emph {et~al.}(2021)\citenamefont
  {Foss-Feig}, \citenamefont {Hayes}, \citenamefont {Dreiling}, \citenamefont
  {Figgatt}, \citenamefont {Gaebler}, \citenamefont {Moses}, \citenamefont
  {Pino},\ and\ \citenamefont {Potter}}]{foss2021holographic}%
  \BibitemOpen
  \bibfield  {author} {\bibinfo {author} {\bibfnamefont {M.}~\bibnamefont
  {Foss-Feig}}, \bibinfo {author} {\bibfnamefont {D.}~\bibnamefont {Hayes}},
  \bibinfo {author} {\bibfnamefont {J.~M.}\ \bibnamefont {Dreiling}}, \bibinfo
  {author} {\bibfnamefont {C.}~\bibnamefont {Figgatt}}, \bibinfo {author}
  {\bibfnamefont {J.~P.}\ \bibnamefont {Gaebler}}, \bibinfo {author}
  {\bibfnamefont {S.~A.}\ \bibnamefont {Moses}}, \bibinfo {author}
  {\bibfnamefont {J.~M.}\ \bibnamefont {Pino}},\ and\ \bibinfo {author}
  {\bibfnamefont {A.~C.}\ \bibnamefont {Potter}},\ }\bibfield  {title}
  {\bibinfo {title} {Holographic quantum algorithms for simulating correlated
  spin systems},\ }\href@noop {} {\bibfield  {journal} {\bibinfo  {journal}
  {Physical Review Research}\ }\textbf {\bibinfo {volume} {3}},\ \bibinfo
  {pages} {033002} (\bibinfo {year} {2021})}\BibitemShut {NoStop}%
\bibitem [{\citenamefont {Barratt}\ \emph {et~al.}(2021)\citenamefont
  {Barratt}, \citenamefont {Dborin}, \citenamefont {Bal}, \citenamefont
  {Stojevic}, \citenamefont {Pollmann},\ and\ \citenamefont
  {Green}}]{barratt2021parallel}%
  \BibitemOpen
  \bibfield  {author} {\bibinfo {author} {\bibfnamefont {F.}~\bibnamefont
  {Barratt}}, \bibinfo {author} {\bibfnamefont {J.}~\bibnamefont {Dborin}},
  \bibinfo {author} {\bibfnamefont {M.}~\bibnamefont {Bal}}, \bibinfo {author}
  {\bibfnamefont {V.}~\bibnamefont {Stojevic}}, \bibinfo {author}
  {\bibfnamefont {F.}~\bibnamefont {Pollmann}},\ and\ \bibinfo {author}
  {\bibfnamefont {A.~G.}\ \bibnamefont {Green}},\ }\bibfield  {title} {\bibinfo
  {title} {Parallel quantum simulation of large systems on small nisq
  computers},\ }\href@noop {} {\bibfield  {journal} {\bibinfo  {journal} {npj
  Quantum Information}\ }\textbf {\bibinfo {volume} {7}},\ \bibinfo {pages}
  {79} (\bibinfo {year} {2021})}\BibitemShut {NoStop}%
\bibitem [{\citenamefont {Smith}\ \emph {et~al.}(2022)\citenamefont {Smith},
  \citenamefont {Jobst}, \citenamefont {Green},\ and\ \citenamefont
  {Pollmann}}]{smith2022crossing}%
  \BibitemOpen
  \bibfield  {author} {\bibinfo {author} {\bibfnamefont {A.}~\bibnamefont
  {Smith}}, \bibinfo {author} {\bibfnamefont {B.}~\bibnamefont {Jobst}},
  \bibinfo {author} {\bibfnamefont {A.~G.}\ \bibnamefont {Green}},\ and\
  \bibinfo {author} {\bibfnamefont {F.}~\bibnamefont {Pollmann}},\ }\bibfield
  {title} {\bibinfo {title} {Crossing a topological phase transition with a
  quantum computer},\ }\href@noop {} {\bibfield  {journal} {\bibinfo  {journal}
  {Physical Review Research}\ }\textbf {\bibinfo {volume} {4}},\ \bibinfo
  {pages} {L022020} (\bibinfo {year} {2022})}\BibitemShut {NoStop}%
\bibitem [{\citenamefont {Meth}\ \emph {et~al.}(2022)\citenamefont {Meth},
  \citenamefont {Kuzmin}, \citenamefont {van Bijnen}, \citenamefont {Postler},
  \citenamefont {Stricker}, \citenamefont {Blatt}, \citenamefont {Ringbauer},
  \citenamefont {Monz}, \citenamefont {Silvi},\ and\ \citenamefont
  {Schindler}}]{meth2022probing}%
  \BibitemOpen
  \bibfield  {author} {\bibinfo {author} {\bibfnamefont {M.}~\bibnamefont
  {Meth}}, \bibinfo {author} {\bibfnamefont {V.}~\bibnamefont {Kuzmin}},
  \bibinfo {author} {\bibfnamefont {R.}~\bibnamefont {van Bijnen}}, \bibinfo
  {author} {\bibfnamefont {L.}~\bibnamefont {Postler}}, \bibinfo {author}
  {\bibfnamefont {R.}~\bibnamefont {Stricker}}, \bibinfo {author}
  {\bibfnamefont {R.}~\bibnamefont {Blatt}}, \bibinfo {author} {\bibfnamefont
  {M.}~\bibnamefont {Ringbauer}}, \bibinfo {author} {\bibfnamefont
  {T.}~\bibnamefont {Monz}}, \bibinfo {author} {\bibfnamefont {P.}~\bibnamefont
  {Silvi}},\ and\ \bibinfo {author} {\bibfnamefont {P.}~\bibnamefont
  {Schindler}},\ }\bibfield  {title} {\bibinfo {title} {Probing phases of
  quantum matter with an ion-trap tensor-network quantum eigensolver},\
  }\href@noop {} {\bibfield  {journal} {\bibinfo  {journal} {Physical Review
  X}\ }\textbf {\bibinfo {volume} {12}},\ \bibinfo {pages} {041035} (\bibinfo
  {year} {2022})}\BibitemShut {NoStop}%
\bibitem [{\citenamefont {Lin}\ \emph {et~al.}(2021)\citenamefont {Lin},
  \citenamefont {Dilip}, \citenamefont {Green}, \citenamefont {Smith},\ and\
  \citenamefont {Pollmann}}]{lin2021real}%
  \BibitemOpen
  \bibfield  {author} {\bibinfo {author} {\bibfnamefont {S.-H.}\ \bibnamefont
  {Lin}}, \bibinfo {author} {\bibfnamefont {R.}~\bibnamefont {Dilip}}, \bibinfo
  {author} {\bibfnamefont {A.~G.}\ \bibnamefont {Green}}, \bibinfo {author}
  {\bibfnamefont {A.}~\bibnamefont {Smith}},\ and\ \bibinfo {author}
  {\bibfnamefont {F.}~\bibnamefont {Pollmann}},\ }\bibfield  {title} {\bibinfo
  {title} {Real-and imaginary-time evolution with compressed quantum
  circuits},\ }\href@noop {} {\bibfield  {journal} {\bibinfo  {journal} {PRX
  Quantum}\ }\textbf {\bibinfo {volume} {2}},\ \bibinfo {pages} {010342}
  (\bibinfo {year} {2021})}\BibitemShut {NoStop}%
\bibitem [{\citenamefont {Haghshenas}\ \emph {et~al.}(2019)\citenamefont
  {Haghshenas}, \citenamefont {O'Rourke},\ and\ \citenamefont
  {Chan}}]{haghshenas2019conversion}%
  \BibitemOpen
  \bibfield  {author} {\bibinfo {author} {\bibfnamefont {R.}~\bibnamefont
  {Haghshenas}}, \bibinfo {author} {\bibfnamefont {M.~J.}\ \bibnamefont
  {O'Rourke}},\ and\ \bibinfo {author} {\bibfnamefont {G.~K.-L.}\ \bibnamefont
  {Chan}},\ }\bibfield  {title} {\bibinfo {title} {Conversion of projected
  entangled pair states into a canonical form},\ }\href@noop {} {\bibfield
  {journal} {\bibinfo  {journal} {Physical Review B}\ }\textbf {\bibinfo
  {volume} {100}},\ \bibinfo {pages} {054404} (\bibinfo {year}
  {2019})}\BibitemShut {NoStop}%
\bibitem [{\citenamefont {Wei}\ \emph {et~al.}(2022)\citenamefont {Wei},
  \citenamefont {Malz},\ and\ \citenamefont {Cirac}}]{wei2022sequential}%
  \BibitemOpen
  \bibfield  {author} {\bibinfo {author} {\bibfnamefont {Z.-Y.}\ \bibnamefont
  {Wei}}, \bibinfo {author} {\bibfnamefont {D.}~\bibnamefont {Malz}},\ and\
  \bibinfo {author} {\bibfnamefont {J.~I.}\ \bibnamefont {Cirac}},\ }\bibfield
  {title} {\bibinfo {title} {Sequential generation of projected entangled-pair
  states},\ }\href@noop {} {\bibfield  {journal} {\bibinfo  {journal} {Physical
  Review Letters}\ }\textbf {\bibinfo {volume} {128}},\ \bibinfo {pages}
  {010607} (\bibinfo {year} {2022})}\BibitemShut {NoStop}%
\bibitem [{\citenamefont {Kim}\ and\ \citenamefont
  {Swingle}(2017)}]{kim2017robust}%
  \BibitemOpen
  \bibfield  {author} {\bibinfo {author} {\bibfnamefont {I.~H.}\ \bibnamefont
  {Kim}}\ and\ \bibinfo {author} {\bibfnamefont {B.}~\bibnamefont {Swingle}},\
  }\bibfield  {title} {\bibinfo {title} {Robust entanglement renormalization on
  a noisy quantum computer},\ }\href@noop {} {\bibfield  {journal} {\bibinfo
  {journal} {arXiv preprint arXiv:1711.07500}\ } (\bibinfo {year}
  {2017})}\BibitemShut {NoStop}%
\bibitem [{\citenamefont {Mart{\'\i}n}\ \emph {et~al.}(2023)\citenamefont
  {Mart{\'\i}n}, \citenamefont {Plekhanov},\ and\ \citenamefont
  {Lubasch}}]{martin2023barren}%
  \BibitemOpen
  \bibfield  {author} {\bibinfo {author} {\bibfnamefont {E.~C.}\ \bibnamefont
  {Mart{\'\i}n}}, \bibinfo {author} {\bibfnamefont {K.}~\bibnamefont
  {Plekhanov}},\ and\ \bibinfo {author} {\bibfnamefont {M.}~\bibnamefont
  {Lubasch}},\ }\bibfield  {title} {\bibinfo {title} {Barren plateaus in
  quantum tensor network optimization},\ }\href@noop {} {\bibfield  {journal}
  {\bibinfo  {journal} {Quantum}\ }\textbf {\bibinfo {volume} {7}},\ \bibinfo
  {pages} {974} (\bibinfo {year} {2023})}\BibitemShut {NoStop}%
\bibitem [{\citenamefont {Barthel}\ and\ \citenamefont
  {Miao}(2025)}]{barthel2025absence}%
  \BibitemOpen
  \bibfield  {author} {\bibinfo {author} {\bibfnamefont {T.}~\bibnamefont
  {Barthel}}\ and\ \bibinfo {author} {\bibfnamefont {Q.}~\bibnamefont {Miao}},\
  }\bibfield  {title} {\bibinfo {title} {Absence of barren plateaus and scaling
  of gradients in the energy optimization of isometric tensor network states},\
  }\href@noop {} {\bibfield  {journal} {\bibinfo  {journal} {Communications in
  Mathematical Physics}\ }\textbf {\bibinfo {volume} {406}},\ \bibinfo {pages}
  {86} (\bibinfo {year} {2025})}\BibitemShut {NoStop}%
\bibitem [{\citenamefont {Sewell}\ and\ \citenamefont
  {Jordan}(2021)}]{sewell2021preparing}%
  \BibitemOpen
  \bibfield  {author} {\bibinfo {author} {\bibfnamefont {T.~J.}\ \bibnamefont
  {Sewell}}\ and\ \bibinfo {author} {\bibfnamefont {S.~P.}\ \bibnamefont
  {Jordan}},\ }\bibfield  {title} {\bibinfo {title} {Preparing renormalization
  group fixed points on nisq hardware},\ }\href@noop {} {\bibfield  {journal}
  {\bibinfo  {journal} {arXiv preprint arXiv:2109.09787}\ } (\bibinfo {year}
  {2021})}\BibitemShut {NoStop}%
\bibitem [{\citenamefont {Anand}\ \emph
  {et~al.}(2023{\natexlab{b}})\citenamefont {Anand}, \citenamefont {Hauschild},
  \citenamefont {Zhang}, \citenamefont {Potter},\ and\ \citenamefont
  {Zaletel}}]{anand2023holographic}%
  \BibitemOpen
  \bibfield  {author} {\bibinfo {author} {\bibfnamefont {S.}~\bibnamefont
  {Anand}}, \bibinfo {author} {\bibfnamefont {J.}~\bibnamefont {Hauschild}},
  \bibinfo {author} {\bibfnamefont {Y.}~\bibnamefont {Zhang}}, \bibinfo
  {author} {\bibfnamefont {A.~C.}\ \bibnamefont {Potter}},\ and\ \bibinfo
  {author} {\bibfnamefont {M.~P.}\ \bibnamefont {Zaletel}},\ }\bibfield
  {title} {\bibinfo {title} {Holographic quantum simulation of entanglement
  renormalization circuits},\ }\href@noop {} {\bibfield  {journal} {\bibinfo
  {journal} {PRX Quantum}\ }\textbf {\bibinfo {volume} {4}},\ \bibinfo {pages}
  {030334} (\bibinfo {year} {2023}{\natexlab{b}})}\BibitemShut {NoStop}%
\bibitem [{\citenamefont {Haghshenas}\ \emph {et~al.}(2023)\citenamefont
  {Haghshenas}, \citenamefont {Chertkov}, \citenamefont {DeCross},
  \citenamefont {Gatterman}, \citenamefont {Gerber}, \citenamefont {Gilmore},
  \citenamefont {Gresh}, \citenamefont {Hewitt}, \citenamefont {Horst},
  \citenamefont {Matheny} \emph {et~al.}}]{haghshenas2023probing}%
  \BibitemOpen
  \bibfield  {author} {\bibinfo {author} {\bibfnamefont {R.}~\bibnamefont
  {Haghshenas}}, \bibinfo {author} {\bibfnamefont {E.}~\bibnamefont
  {Chertkov}}, \bibinfo {author} {\bibfnamefont {M.}~\bibnamefont {DeCross}},
  \bibinfo {author} {\bibfnamefont {T.~M.}\ \bibnamefont {Gatterman}}, \bibinfo
  {author} {\bibfnamefont {J.~A.}\ \bibnamefont {Gerber}}, \bibinfo {author}
  {\bibfnamefont {K.}~\bibnamefont {Gilmore}}, \bibinfo {author} {\bibfnamefont
  {D.}~\bibnamefont {Gresh}}, \bibinfo {author} {\bibfnamefont
  {N.}~\bibnamefont {Hewitt}}, \bibinfo {author} {\bibfnamefont {C.~V.}\
  \bibnamefont {Horst}}, \bibinfo {author} {\bibfnamefont {M.}~\bibnamefont
  {Matheny}}, \emph {et~al.},\ }\bibfield  {title} {\bibinfo {title} {Probing
  critical states of matter on a digital quantum computer},\ }\href@noop {}
  {\bibfield  {journal} {\bibinfo  {journal} {arXiv preprint arXiv:2305.01650}\
  } (\bibinfo {year} {2023})}\BibitemShut {NoStop}%
\bibitem [{\citenamefont {Miao}\ \emph {et~al.}(2024)\citenamefont {Miao},
  \citenamefont {Wang}, \citenamefont {Brown}, \citenamefont {Barthel},\ and\
  \citenamefont {Cetina}}]{miao2024probing}%
  \BibitemOpen
  \bibfield  {author} {\bibinfo {author} {\bibfnamefont {Q.}~\bibnamefont
  {Miao}}, \bibinfo {author} {\bibfnamefont {T.}~\bibnamefont {Wang}}, \bibinfo
  {author} {\bibfnamefont {K.~R.}\ \bibnamefont {Brown}}, \bibinfo {author}
  {\bibfnamefont {T.}~\bibnamefont {Barthel}},\ and\ \bibinfo {author}
  {\bibfnamefont {M.}~\bibnamefont {Cetina}},\ }\bibfield  {title} {\bibinfo
  {title} {Probing entanglement scaling across a quantum phase transition on a
  quantum computer},\ }\href@noop {} {\bibfield  {journal} {\bibinfo  {journal}
  {arXiv preprint arXiv:2412.18602}\ } (\bibinfo {year} {2024})}\BibitemShut
  {NoStop}%
\bibitem [{\citenamefont {Zhang}\ \emph {et~al.}(2024)\citenamefont {Zhang},
  \citenamefont {Gopalakrishnan},\ and\ \citenamefont
  {Styliaris}}]{zhang2024characterizing}%
  \BibitemOpen
  \bibfield  {author} {\bibinfo {author} {\bibfnamefont {Y.}~\bibnamefont
  {Zhang}}, \bibinfo {author} {\bibfnamefont {S.}~\bibnamefont
  {Gopalakrishnan}},\ and\ \bibinfo {author} {\bibfnamefont {G.}~\bibnamefont
  {Styliaris}},\ }\bibfield  {title} {\bibinfo {title} {Characterizing
  matrix-product states and projected entangled-pair states preparable via
  measurement and feedback},\ }\href@noop {} {\bibfield  {journal} {\bibinfo
  {journal} {PRX Quantum}\ }\textbf {\bibinfo {volume} {5}},\ \bibinfo {pages}
  {040304} (\bibinfo {year} {2024})}\BibitemShut {NoStop}%
\bibitem [{\citenamefont {Larsen}\ \emph {et~al.}(2024)\citenamefont {Larsen},
  \citenamefont {Grace}, \citenamefont {Baczewski},\ and\ \citenamefont
  {Magann}}]{larsen2024feedback}%
  \BibitemOpen
  \bibfield  {author} {\bibinfo {author} {\bibfnamefont {J.~B.}\ \bibnamefont
  {Larsen}}, \bibinfo {author} {\bibfnamefont {M.~D.}\ \bibnamefont {Grace}},
  \bibinfo {author} {\bibfnamefont {A.~D.}\ \bibnamefont {Baczewski}},\ and\
  \bibinfo {author} {\bibfnamefont {A.~B.}\ \bibnamefont {Magann}},\ }\bibfield
   {title} {\bibinfo {title} {Feedback-based quantum algorithms for ground
  state preparation},\ }\href@noop {} {\bibfield  {journal} {\bibinfo
  {journal} {Physical Review Research}\ }\textbf {\bibinfo {volume} {6}},\
  \bibinfo {pages} {033336} (\bibinfo {year} {2024})}\BibitemShut {NoStop}%
\bibitem [{\citenamefont {Stephen}\ and\ \citenamefont
  {Hart}(2024)}]{stephen2024preparing}%
  \BibitemOpen
  \bibfield  {author} {\bibinfo {author} {\bibfnamefont {D.~T.}\ \bibnamefont
  {Stephen}}\ and\ \bibinfo {author} {\bibfnamefont {O.}~\bibnamefont {Hart}},\
  }\bibfield  {title} {\bibinfo {title} {Preparing matrix product states via
  fusion: constraints and extensions},\ }\href@noop {} {\bibfield  {journal}
  {\bibinfo  {journal} {arXiv preprint arXiv:2404.16360}\ } (\bibinfo {year}
  {2024})}\BibitemShut {NoStop}%
\bibitem [{\citenamefont {Smith}\ \emph {et~al.}(2023)\citenamefont {Smith},
  \citenamefont {Crane}, \citenamefont {Wiebe},\ and\ \citenamefont
  {Girvin}}]{smith2023deterministic}%
  \BibitemOpen
  \bibfield  {author} {\bibinfo {author} {\bibfnamefont {K.~C.}\ \bibnamefont
  {Smith}}, \bibinfo {author} {\bibfnamefont {E.}~\bibnamefont {Crane}},
  \bibinfo {author} {\bibfnamefont {N.}~\bibnamefont {Wiebe}},\ and\ \bibinfo
  {author} {\bibfnamefont {S.}~\bibnamefont {Girvin}},\ }\bibfield  {title}
  {\bibinfo {title} {Deterministic constant-depth preparation of the aklt state
  on a quantum processor using fusion measurements},\ }\href@noop {} {\bibfield
   {journal} {\bibinfo  {journal} {PRX Quantum}\ }\textbf {\bibinfo {volume}
  {4}},\ \bibinfo {pages} {020315} (\bibinfo {year} {2023})}\BibitemShut
  {NoStop}%
\bibitem [{\citenamefont {Sahay}\ and\ \citenamefont
  {Verresen}(2024{\natexlab{a}})}]{sahay2024classifying}%
  \BibitemOpen
  \bibfield  {author} {\bibinfo {author} {\bibfnamefont {R.}~\bibnamefont
  {Sahay}}\ and\ \bibinfo {author} {\bibfnamefont {R.}~\bibnamefont
  {Verresen}},\ }\bibfield  {title} {\bibinfo {title} {Classifying
  one-dimensional quantum states prepared by a single round of measurements},\
  }\href@noop {} {\bibfield  {journal} {\bibinfo  {journal} {arXiv preprint
  arXiv:2404.16753}\ } (\bibinfo {year} {2024}{\natexlab{a}})}\BibitemShut
  {NoStop}%
\bibitem [{\citenamefont {Sahay}\ and\ \citenamefont
  {Verresen}(2024{\natexlab{b}})}]{sahay2024finite}%
  \BibitemOpen
  \bibfield  {author} {\bibinfo {author} {\bibfnamefont {R.}~\bibnamefont
  {Sahay}}\ and\ \bibinfo {author} {\bibfnamefont {R.}~\bibnamefont
  {Verresen}},\ }\bibfield  {title} {\bibinfo {title} {Finite-depth preparation
  of tensor network states from measurement},\ }\href@noop {} {\bibfield
  {journal} {\bibinfo  {journal} {arXiv preprint arXiv:2404.17087}\ } (\bibinfo
  {year} {2024}{\natexlab{b}})}\BibitemShut {NoStop}%
\bibitem [{\citenamefont {Smith}\ \emph {et~al.}(2024)\citenamefont {Smith},
  \citenamefont {Khan}, \citenamefont {Clark}, \citenamefont {Girvin},\ and\
  \citenamefont {Wei}}]{smith2024constant}%
  \BibitemOpen
  \bibfield  {author} {\bibinfo {author} {\bibfnamefont {K.~C.}\ \bibnamefont
  {Smith}}, \bibinfo {author} {\bibfnamefont {A.}~\bibnamefont {Khan}},
  \bibinfo {author} {\bibfnamefont {B.~K.}\ \bibnamefont {Clark}}, \bibinfo
  {author} {\bibfnamefont {S.}~\bibnamefont {Girvin}},\ and\ \bibinfo {author}
  {\bibfnamefont {T.-C.}\ \bibnamefont {Wei}},\ }\bibfield  {title} {\bibinfo
  {title} {Constant-depth preparation of matrix product states with adaptive
  quantum circuits},\ }\href@noop {} {\bibfield  {journal} {\bibinfo  {journal}
  {PRX Quantum}\ }\textbf {\bibinfo {volume} {5}},\ \bibinfo {pages} {030344}
  (\bibinfo {year} {2024})}\BibitemShut {NoStop}%
\bibitem [{\citenamefont {Ben-Dov}\ \emph {et~al.}(2024)\citenamefont
  {Ben-Dov}, \citenamefont {Shnaiderov}, \citenamefont {Makmal},\ and\
  \citenamefont {Dalla~Torre}}]{ben2024approximate}%
  \BibitemOpen
  \bibfield  {author} {\bibinfo {author} {\bibfnamefont {M.}~\bibnamefont
  {Ben-Dov}}, \bibinfo {author} {\bibfnamefont {D.}~\bibnamefont {Shnaiderov}},
  \bibinfo {author} {\bibfnamefont {A.}~\bibnamefont {Makmal}},\ and\ \bibinfo
  {author} {\bibfnamefont {E.~G.}\ \bibnamefont {Dalla~Torre}},\ }\bibfield
  {title} {\bibinfo {title} {Approximate encoding of quantum states using
  shallow circuits},\ }\href@noop {} {\bibfield  {journal} {\bibinfo  {journal}
  {npj Quantum Information}\ }\textbf {\bibinfo {volume} {10}},\ \bibinfo
  {pages} {65} (\bibinfo {year} {2024})}\BibitemShut {NoStop}%
\bibitem [{\citenamefont {Rudolph}\ \emph
  {et~al.}(2023{\natexlab{b}})\citenamefont {Rudolph}, \citenamefont {Chen},
  \citenamefont {Miller}, \citenamefont {Acharya},\ and\ \citenamefont
  {Perdomo-Ortiz}}]{rudolph2023decomposition}%
  \BibitemOpen
  \bibfield  {author} {\bibinfo {author} {\bibfnamefont {M.~S.}\ \bibnamefont
  {Rudolph}}, \bibinfo {author} {\bibfnamefont {J.}~\bibnamefont {Chen}},
  \bibinfo {author} {\bibfnamefont {J.}~\bibnamefont {Miller}}, \bibinfo
  {author} {\bibfnamefont {A.}~\bibnamefont {Acharya}},\ and\ \bibinfo {author}
  {\bibfnamefont {A.}~\bibnamefont {Perdomo-Ortiz}},\ }\bibfield  {title}
  {\bibinfo {title} {Decomposition of matrix product states into shallow
  quantum circuits},\ }\href@noop {} {\bibfield  {journal} {\bibinfo  {journal}
  {Quantum Science and Technology}\ }\textbf {\bibinfo {volume} {9}},\ \bibinfo
  {pages} {015012} (\bibinfo {year} {2023}{\natexlab{b}})}\BibitemShut
  {NoStop}%
\bibitem [{\citenamefont {Melnikov}\ \emph {et~al.}(2023)\citenamefont
  {Melnikov}, \citenamefont {Termanova}, \citenamefont {Dolgov}, \citenamefont
  {Neukart},\ and\ \citenamefont {Perelshtein}}]{melnikov2023quantum}%
  \BibitemOpen
  \bibfield  {author} {\bibinfo {author} {\bibfnamefont {A.~A.}\ \bibnamefont
  {Melnikov}}, \bibinfo {author} {\bibfnamefont {A.~A.}\ \bibnamefont
  {Termanova}}, \bibinfo {author} {\bibfnamefont {S.~V.}\ \bibnamefont
  {Dolgov}}, \bibinfo {author} {\bibfnamefont {F.}~\bibnamefont {Neukart}},\
  and\ \bibinfo {author} {\bibfnamefont {M.}~\bibnamefont {Perelshtein}},\
  }\bibfield  {title} {\bibinfo {title} {Quantum state preparation using tensor
  networks},\ }\href@noop {} {\bibfield  {journal} {\bibinfo  {journal}
  {Quantum Science and Technology}\ }\textbf {\bibinfo {volume} {8}},\ \bibinfo
  {pages} {035027} (\bibinfo {year} {2023})}\BibitemShut {NoStop}%
\bibitem [{\citenamefont {Termanova}\ \emph {et~al.}(2024)\citenamefont
  {Termanova}, \citenamefont {Melnikov}, \citenamefont {Mamenchikov},
  \citenamefont {Belokonev}, \citenamefont {Dolgov}, \citenamefont
  {Berezutskii}, \citenamefont {Ellerbrock}, \citenamefont {Mansell},\ and\
  \citenamefont {Perelshtein}}]{termanova2024tensor}%
  \BibitemOpen
  \bibfield  {author} {\bibinfo {author} {\bibfnamefont {A.}~\bibnamefont
  {Termanova}}, \bibinfo {author} {\bibfnamefont {A.}~\bibnamefont {Melnikov}},
  \bibinfo {author} {\bibfnamefont {E.}~\bibnamefont {Mamenchikov}}, \bibinfo
  {author} {\bibfnamefont {N.}~\bibnamefont {Belokonev}}, \bibinfo {author}
  {\bibfnamefont {S.}~\bibnamefont {Dolgov}}, \bibinfo {author} {\bibfnamefont
  {A.}~\bibnamefont {Berezutskii}}, \bibinfo {author} {\bibfnamefont
  {R.}~\bibnamefont {Ellerbrock}}, \bibinfo {author} {\bibfnamefont
  {C.}~\bibnamefont {Mansell}},\ and\ \bibinfo {author} {\bibfnamefont
  {M.}~\bibnamefont {Perelshtein}},\ }\bibfield  {title} {\bibinfo {title}
  {Tensor quantum programming},\ }\href@noop {} {\bibfield  {journal} {\bibinfo
   {journal} {New Journal of Physics}\ }\textbf {\bibinfo {volume} {26}},\
  \bibinfo {pages} {123019} (\bibinfo {year} {2024})}\BibitemShut {NoStop}%
\bibitem [{\citenamefont {Jaderberg}\ \emph {et~al.}(2025)\citenamefont
  {Jaderberg}, \citenamefont {Pennington}, \citenamefont {Marshall},
  \citenamefont {Anderson}, \citenamefont {Agarwal}, \citenamefont {Lindoy},
  \citenamefont {Rungger}, \citenamefont {Mensa},\ and\ \citenamefont
  {Crain}}]{jaderberg2025variational}%
  \BibitemOpen
  \bibfield  {author} {\bibinfo {author} {\bibfnamefont {B.}~\bibnamefont
  {Jaderberg}}, \bibinfo {author} {\bibfnamefont {G.}~\bibnamefont
  {Pennington}}, \bibinfo {author} {\bibfnamefont {K.~V.}\ \bibnamefont
  {Marshall}}, \bibinfo {author} {\bibfnamefont {L.~W.}\ \bibnamefont
  {Anderson}}, \bibinfo {author} {\bibfnamefont {A.}~\bibnamefont {Agarwal}},
  \bibinfo {author} {\bibfnamefont {L.~P.}\ \bibnamefont {Lindoy}}, \bibinfo
  {author} {\bibfnamefont {I.}~\bibnamefont {Rungger}}, \bibinfo {author}
  {\bibfnamefont {S.}~\bibnamefont {Mensa}},\ and\ \bibinfo {author}
  {\bibfnamefont {J.}~\bibnamefont {Crain}},\ }\bibfield  {title} {\bibinfo
  {title} {Variational preparation of normal matrix product states on quantum
  computers},\ }\href@noop {} {\bibfield  {journal} {\bibinfo  {journal} {arXiv
  preprint arXiv:2503.09683}\ } (\bibinfo {year} {2025})}\BibitemShut {NoStop}%
\bibitem [{\citenamefont {Kukliansky}\ \emph {et~al.}(2023)\citenamefont
  {Kukliansky}, \citenamefont {Younis}, \citenamefont {Cincio},\ and\
  \citenamefont {Iancu}}]{kukliansky2023qfactor}%
  \BibitemOpen
  \bibfield  {author} {\bibinfo {author} {\bibfnamefont {A.}~\bibnamefont
  {Kukliansky}}, \bibinfo {author} {\bibfnamefont {E.}~\bibnamefont {Younis}},
  \bibinfo {author} {\bibfnamefont {L.}~\bibnamefont {Cincio}},\ and\ \bibinfo
  {author} {\bibfnamefont {C.}~\bibnamefont {Iancu}},\ }\href@noop {} {\bibinfo
  {title} {{QFactor}: A domain-specific optimizer for quantum circuit
  instantiation}} (\bibinfo {year} {2023})\BibitemShut {NoStop}%
\bibitem [{\citenamefont {Lidar}\ and\ \citenamefont
  {Brun}(2013)}]{lidar2013quantum}%
  \BibitemOpen
  \bibfield  {author} {\bibinfo {author} {\bibfnamefont {D.~A.}\ \bibnamefont
  {Lidar}}\ and\ \bibinfo {author} {\bibfnamefont {T.~A.}\ \bibnamefont
  {Brun}},\ }\href@noop {} {\emph {\bibinfo {title} {Quantum error
  correction}}}\ (\bibinfo  {publisher} {Cambridge university press},\ \bibinfo
  {year} {2013})\BibitemShut {NoStop}%
\bibitem [{\citenamefont {Ferris}\ and\ \citenamefont
  {Poulin}(2014)}]{ferris2014tensor}%
  \BibitemOpen
  \bibfield  {author} {\bibinfo {author} {\bibfnamefont {A.~J.}\ \bibnamefont
  {Ferris}}\ and\ \bibinfo {author} {\bibfnamefont {D.}~\bibnamefont
  {Poulin}},\ }\bibfield  {title} {\bibinfo {title} {Tensor networks and
  quantum error correction},\ }\href@noop {} {\bibfield  {journal} {\bibinfo
  {journal} {Physical review letters}\ }\textbf {\bibinfo {volume} {113}},\
  \bibinfo {pages} {030501} (\bibinfo {year} {2014})}\BibitemShut {NoStop}%
\bibitem [{\citenamefont {Farrelly}\ \emph {et~al.}(2021)\citenamefont
  {Farrelly}, \citenamefont {Harris}, \citenamefont {McMahon},\ and\
  \citenamefont {Stace}}]{farrelly2021tensor}%
  \BibitemOpen
  \bibfield  {author} {\bibinfo {author} {\bibfnamefont {T.}~\bibnamefont
  {Farrelly}}, \bibinfo {author} {\bibfnamefont {R.~J.}\ \bibnamefont
  {Harris}}, \bibinfo {author} {\bibfnamefont {N.~A.}\ \bibnamefont
  {McMahon}},\ and\ \bibinfo {author} {\bibfnamefont {T.~M.}\ \bibnamefont
  {Stace}},\ }\bibfield  {title} {\bibinfo {title} {Tensor-network codes},\
  }\href@noop {} {\bibfield  {journal} {\bibinfo  {journal} {Physical Review
  Letters}\ }\textbf {\bibinfo {volume} {127}},\ \bibinfo {pages} {040507}
  (\bibinfo {year} {2021})}\BibitemShut {NoStop}%
\bibitem [{\citenamefont {Farrelly}\ \emph
  {et~al.}(2022{\natexlab{a}})\citenamefont {Farrelly}, \citenamefont
  {Tuckett},\ and\ \citenamefont {Stace}}]{farrelly2022local}%
  \BibitemOpen
  \bibfield  {author} {\bibinfo {author} {\bibfnamefont {T.}~\bibnamefont
  {Farrelly}}, \bibinfo {author} {\bibfnamefont {D.~K.}\ \bibnamefont
  {Tuckett}},\ and\ \bibinfo {author} {\bibfnamefont {T.~M.}\ \bibnamefont
  {Stace}},\ }\bibfield  {title} {\bibinfo {title} {Local tensor-network
  codes},\ }\href@noop {} {\bibfield  {journal} {\bibinfo  {journal} {New
  Journal of Physics}\ }\textbf {\bibinfo {volume} {24}},\ \bibinfo {pages}
  {043015} (\bibinfo {year} {2022}{\natexlab{a}})}\BibitemShut {NoStop}%
\bibitem [{\citenamefont {Cao}\ and\ \citenamefont
  {Lackey}(2022)}]{cao2022quantum}%
  \BibitemOpen
  \bibfield  {author} {\bibinfo {author} {\bibfnamefont {C.}~\bibnamefont
  {Cao}}\ and\ \bibinfo {author} {\bibfnamefont {B.}~\bibnamefont {Lackey}},\
  }\bibfield  {title} {\bibinfo {title} {Quantum lego: Building quantum error
  correction codes from tensor networks},\ }\href@noop {} {\bibfield  {journal}
  {\bibinfo  {journal} {PRX Quantum}\ }\textbf {\bibinfo {volume} {3}},\
  \bibinfo {pages} {020332} (\bibinfo {year} {2022})}\BibitemShut {NoStop}%
\bibitem [{\citenamefont {Cao}\ \emph {et~al.}(2024)\citenamefont {Cao},
  \citenamefont {Gullans}, \citenamefont {Lackey},\ and\ \citenamefont
  {Wang}}]{cao2024quantum}%
  \BibitemOpen
  \bibfield  {author} {\bibinfo {author} {\bibfnamefont {C.}~\bibnamefont
  {Cao}}, \bibinfo {author} {\bibfnamefont {M.~J.}\ \bibnamefont {Gullans}},
  \bibinfo {author} {\bibfnamefont {B.}~\bibnamefont {Lackey}},\ and\ \bibinfo
  {author} {\bibfnamefont {Z.}~\bibnamefont {Wang}},\ }\bibfield  {title}
  {\bibinfo {title} {Quantum lego expansion pack: Enumerators from tensor
  networks},\ }\href@noop {} {\bibfield  {journal} {\bibinfo  {journal} {PRX
  Quantum}\ }\textbf {\bibinfo {volume} {5}},\ \bibinfo {pages} {030313}
  (\bibinfo {year} {2024})}\BibitemShut {NoStop}%
\bibitem [{\citenamefont {Fan}\ \emph {et~al.}(2024{\natexlab{a}})\citenamefont
  {Fan}, \citenamefont {Steinberg}, \citenamefont {Jahn}, \citenamefont {Cao},
  \citenamefont {Sarkar},\ and\ \citenamefont {Feld}}]{fan2024analyzing}%
  \BibitemOpen
  \bibfield  {author} {\bibinfo {author} {\bibfnamefont {J.}~\bibnamefont
  {Fan}}, \bibinfo {author} {\bibfnamefont {M.}~\bibnamefont {Steinberg}},
  \bibinfo {author} {\bibfnamefont {A.}~\bibnamefont {Jahn}}, \bibinfo {author}
  {\bibfnamefont {C.}~\bibnamefont {Cao}}, \bibinfo {author} {\bibfnamefont
  {A.}~\bibnamefont {Sarkar}},\ and\ \bibinfo {author} {\bibfnamefont
  {S.}~\bibnamefont {Feld}},\ }\bibfield  {title} {\bibinfo {title} {Lego hqec:
  A software tool for analyzing holographic quantum codes},\ }\href@noop {}
  {\bibfield  {journal} {\bibinfo  {journal} {arXiv preprint arXiv:2410.22861}\
  } (\bibinfo {year} {2024}{\natexlab{a}})}\BibitemShut {NoStop}%
\bibitem [{\citenamefont {Pastawski}\ \emph {et~al.}(2015)\citenamefont
  {Pastawski}, \citenamefont {Yoshida}, \citenamefont {Harlow},\ and\
  \citenamefont {Preskill}}]{pastawski2015holographic}%
  \BibitemOpen
  \bibfield  {author} {\bibinfo {author} {\bibfnamefont {F.}~\bibnamefont
  {Pastawski}}, \bibinfo {author} {\bibfnamefont {B.}~\bibnamefont {Yoshida}},
  \bibinfo {author} {\bibfnamefont {D.}~\bibnamefont {Harlow}},\ and\ \bibinfo
  {author} {\bibfnamefont {J.}~\bibnamefont {Preskill}},\ }\bibfield  {title}
  {\bibinfo {title} {Holographic quantum error-correcting codes: Toy models for
  the bulk/boundary correspondence},\ }\href@noop {} {\bibfield  {journal}
  {\bibinfo  {journal} {Journal of High Energy Physics}\ }\textbf {\bibinfo
  {volume} {2015}},\ \bibinfo {pages} {1} (\bibinfo {year} {2015})}\BibitemShut
  {NoStop}%
\bibitem [{\citenamefont {Jahn}\ and\ \citenamefont
  {Eisert}(2021)}]{jahn2021holographic}%
  \BibitemOpen
  \bibfield  {author} {\bibinfo {author} {\bibfnamefont {A.}~\bibnamefont
  {Jahn}}\ and\ \bibinfo {author} {\bibfnamefont {J.}~\bibnamefont {Eisert}},\
  }\bibfield  {title} {\bibinfo {title} {Holographic tensor network models and
  quantum error correction: a topical review},\ }\href@noop {} {\bibfield
  {journal} {\bibinfo  {journal} {Quantum Science and Technology}\ }\textbf
  {\bibinfo {volume} {6}},\ \bibinfo {pages} {033002} (\bibinfo {year}
  {2021})}\BibitemShut {NoStop}%
\bibitem [{\citenamefont {Steinberg}\ \emph {et~al.}(2023)\citenamefont
  {Steinberg}, \citenamefont {Feld},\ and\ \citenamefont
  {Jahn}}]{steinberg2023holographic}%
  \BibitemOpen
  \bibfield  {author} {\bibinfo {author} {\bibfnamefont {M.}~\bibnamefont
  {Steinberg}}, \bibinfo {author} {\bibfnamefont {S.}~\bibnamefont {Feld}},\
  and\ \bibinfo {author} {\bibfnamefont {A.}~\bibnamefont {Jahn}},\ }\bibfield
  {title} {\bibinfo {title} {Holographic codes from hyperinvariant tensor
  networks},\ }\href@noop {} {\bibfield  {journal} {\bibinfo  {journal} {Nature
  Communications}\ }\textbf {\bibinfo {volume} {14}},\ \bibinfo {pages} {7314}
  (\bibinfo {year} {2023})}\BibitemShut {NoStop}%
\bibitem [{\citenamefont {Shen}\ \emph {et~al.}(2023)\citenamefont {Shen},
  \citenamefont {Wang},\ and\ \citenamefont {Cao}}]{shen2023quantum}%
  \BibitemOpen
  \bibfield  {author} {\bibinfo {author} {\bibfnamefont {R.}~\bibnamefont
  {Shen}}, \bibinfo {author} {\bibfnamefont {Y.}~\bibnamefont {Wang}},\ and\
  \bibinfo {author} {\bibfnamefont {C.}~\bibnamefont {Cao}},\ }\bibfield
  {title} {\bibinfo {title} {Quantum lego and xp stabilizer codes},\
  }\href@noop {} {\bibfield  {journal} {\bibinfo  {journal} {arXiv preprint
  arXiv:2310.19538}\ } (\bibinfo {year} {2023})}\BibitemShut {NoStop}%
\bibitem [{\citenamefont {Schotte}\ \emph {et~al.}(2022)\citenamefont
  {Schotte}, \citenamefont {Zhu}, \citenamefont {Burgelman},\ and\
  \citenamefont {Verstraete}}]{verstraete2022fibonacci}%
  \BibitemOpen
  \bibfield  {author} {\bibinfo {author} {\bibfnamefont {A.}~\bibnamefont
  {Schotte}}, \bibinfo {author} {\bibfnamefont {G.}~\bibnamefont {Zhu}},
  \bibinfo {author} {\bibfnamefont {L.}~\bibnamefont {Burgelman}},\ and\
  \bibinfo {author} {\bibfnamefont {F.}~\bibnamefont {Verstraete}},\ }\bibfield
   {title} {\bibinfo {title} {Quantum error correction thresholds for the
  universal fibonacci turaev-viro code},\ }\href
  {https://doi.org/10.1103/PhysRevX.12.021012} {\bibfield  {journal} {\bibinfo
  {journal} {Phys. Rev. X}\ }\textbf {\bibinfo {volume} {12}},\ \bibinfo
  {pages} {021012} (\bibinfo {year} {2022})}\BibitemShut {NoStop}%
\bibitem [{\citenamefont {Bettaque}\ and\ \citenamefont
  {Swingle}(2024)}]{bettaque2024nora}%
  \BibitemOpen
  \bibfield  {author} {\bibinfo {author} {\bibfnamefont {V.}~\bibnamefont
  {Bettaque}}\ and\ \bibinfo {author} {\bibfnamefont {B.}~\bibnamefont
  {Swingle}},\ }\bibfield  {title} {\bibinfo {title} {Nora: A tensor network
  ansatz for volume-law entangled equilibrium states of highly connected
  hamiltonians},\ }\href@noop {} {\bibfield  {journal} {\bibinfo  {journal}
  {Quantum}\ }\textbf {\bibinfo {volume} {8}},\ \bibinfo {pages} {1362}
  (\bibinfo {year} {2024})}\BibitemShut {NoStop}%
\bibitem [{\citenamefont {Cao}\ and\ \citenamefont
  {Lackey}(2021)}]{cao2021approximate}%
  \BibitemOpen
  \bibfield  {author} {\bibinfo {author} {\bibfnamefont {C.}~\bibnamefont
  {Cao}}\ and\ \bibinfo {author} {\bibfnamefont {B.}~\bibnamefont {Lackey}},\
  }\bibfield  {title} {\bibinfo {title} {Approximate bacon-shor code and
  holography},\ }\href@noop {} {\bibfield  {journal} {\bibinfo  {journal}
  {Journal of High Energy Physics}\ }\textbf {\bibinfo {volume} {2021}},\
  \bibinfo {pages} {1} (\bibinfo {year} {2021})}\BibitemShut {NoStop}%
\bibitem [{\citenamefont {Steinberg}\ \emph {et~al.}(2024)\citenamefont
  {Steinberg}, \citenamefont {Fan}, \citenamefont {Harris}, \citenamefont
  {Elkouss}, \citenamefont {Feld},\ and\ \citenamefont
  {Jahn}}]{steinberg2024far}%
  \BibitemOpen
  \bibfield  {author} {\bibinfo {author} {\bibfnamefont {M.}~\bibnamefont
  {Steinberg}}, \bibinfo {author} {\bibfnamefont {J.}~\bibnamefont {Fan}},
  \bibinfo {author} {\bibfnamefont {R.~J.}\ \bibnamefont {Harris}}, \bibinfo
  {author} {\bibfnamefont {D.}~\bibnamefont {Elkouss}}, \bibinfo {author}
  {\bibfnamefont {S.}~\bibnamefont {Feld}},\ and\ \bibinfo {author}
  {\bibfnamefont {A.}~\bibnamefont {Jahn}},\ }\bibfield  {title} {\bibinfo
  {title} {Far from perfect: Quantum error correction with (hyperinvariant)
  evenbly codes},\ }\href@noop {} {\bibfield  {journal} {\bibinfo  {journal}
  {arXiv preprint arXiv:2407.11926}\ } (\bibinfo {year} {2024})}\BibitemShut
  {NoStop}%
\bibitem [{\citenamefont {Pastawski}\ and\ \citenamefont
  {Preskill}(2017)}]{pastawski2017code}%
  \BibitemOpen
  \bibfield  {author} {\bibinfo {author} {\bibfnamefont {F.}~\bibnamefont
  {Pastawski}}\ and\ \bibinfo {author} {\bibfnamefont {J.}~\bibnamefont
  {Preskill}},\ }\bibfield  {title} {\bibinfo {title} {Code properties from
  holographic geometries},\ }\href@noop {} {\bibfield  {journal} {\bibinfo
  {journal} {Physical Review X}\ }\textbf {\bibinfo {volume} {7}},\ \bibinfo
  {pages} {021022} (\bibinfo {year} {2017})}\BibitemShut {NoStop}%
\bibitem [{\citenamefont {Su}\ \emph {et~al.}(2023)\citenamefont {Su},
  \citenamefont {Cao}, \citenamefont {Hu}, \citenamefont {Yanay}, \citenamefont
  {Tahan},\ and\ \citenamefont {Swingle}}]{su2023discovery}%
  \BibitemOpen
  \bibfield  {author} {\bibinfo {author} {\bibfnamefont {V.~P.}\ \bibnamefont
  {Su}}, \bibinfo {author} {\bibfnamefont {C.}~\bibnamefont {Cao}}, \bibinfo
  {author} {\bibfnamefont {H.-Y.}\ \bibnamefont {Hu}}, \bibinfo {author}
  {\bibfnamefont {Y.}~\bibnamefont {Yanay}}, \bibinfo {author} {\bibfnamefont
  {C.}~\bibnamefont {Tahan}},\ and\ \bibinfo {author} {\bibfnamefont
  {B.}~\bibnamefont {Swingle}},\ }\bibfield  {title} {\bibinfo {title}
  {Discovery of optimal quantum error correcting codes via reinforcement
  learning},\ }\href@noop {} {\bibfield  {journal} {\bibinfo  {journal} {arXiv
  preprint arXiv:2305.06378}\ } (\bibinfo {year} {2023})}\BibitemShut {NoStop}%
\bibitem [{\citenamefont {Mauron}\ \emph {et~al.}(2024)\citenamefont {Mauron},
  \citenamefont {Farrelly},\ and\ \citenamefont
  {Stace}}]{mauron2024optimization}%
  \BibitemOpen
  \bibfield  {author} {\bibinfo {author} {\bibfnamefont {C.}~\bibnamefont
  {Mauron}}, \bibinfo {author} {\bibfnamefont {T.}~\bibnamefont {Farrelly}},\
  and\ \bibinfo {author} {\bibfnamefont {T.~M.}\ \bibnamefont {Stace}},\
  }\bibfield  {title} {\bibinfo {title} {Optimization of tensor network codes
  with reinforcement learning},\ }\href@noop {} {\bibfield  {journal} {\bibinfo
   {journal} {New Journal of Physics}\ }\textbf {\bibinfo {volume} {26}},\
  \bibinfo {pages} {023024} (\bibinfo {year} {2024})}\BibitemShut {NoStop}%
\bibitem [{\citenamefont {Bravyi}\ \emph {et~al.}(2014)\citenamefont {Bravyi},
  \citenamefont {Suchara},\ and\ \citenamefont {Vargo}}]{bravyi2014efficient}%
  \BibitemOpen
  \bibfield  {author} {\bibinfo {author} {\bibfnamefont {S.}~\bibnamefont
  {Bravyi}}, \bibinfo {author} {\bibfnamefont {M.}~\bibnamefont {Suchara}},\
  and\ \bibinfo {author} {\bibfnamefont {A.}~\bibnamefont {Vargo}},\ }\bibfield
   {title} {\bibinfo {title} {Efficient algorithms for maximum likelihood
  decoding in the surface code},\ }\href@noop {} {\bibfield  {journal}
  {\bibinfo  {journal} {Physical Review A}\ }\textbf {\bibinfo {volume} {90}},\
  \bibinfo {pages} {032326} (\bibinfo {year} {2014})}\BibitemShut {NoStop}%
\bibitem [{\citenamefont {Chubb}\ and\ \citenamefont
  {Flammia}(2021)}]{chubb2021statistical}%
  \BibitemOpen
  \bibfield  {author} {\bibinfo {author} {\bibfnamefont {C.~T.}\ \bibnamefont
  {Chubb}}\ and\ \bibinfo {author} {\bibfnamefont {S.~T.}\ \bibnamefont
  {Flammia}},\ }\bibfield  {title} {\bibinfo {title} {Statistical mechanical
  models for quantum codes with correlated noise},\ }\href@noop {} {\bibfield
  {journal} {\bibinfo  {journal} {Annales de l’Institut Henri Poincar{\'e}
  D}\ }\textbf {\bibinfo {volume} {8}},\ \bibinfo {pages} {269} (\bibinfo
  {year} {2021})}\BibitemShut {NoStop}%
\bibitem [{\citenamefont {Darmawan}\ \emph {et~al.}(2022)\citenamefont
  {Darmawan}, \citenamefont {Nakata}, \citenamefont {Tamiya},\ and\
  \citenamefont {Yamasaki}}]{darmawan2022low}%
  \BibitemOpen
  \bibfield  {author} {\bibinfo {author} {\bibfnamefont {A.~S.}\ \bibnamefont
  {Darmawan}}, \bibinfo {author} {\bibfnamefont {Y.}~\bibnamefont {Nakata}},
  \bibinfo {author} {\bibfnamefont {S.}~\bibnamefont {Tamiya}},\ and\ \bibinfo
  {author} {\bibfnamefont {H.}~\bibnamefont {Yamasaki}},\ }\bibfield  {title}
  {\bibinfo {title} {{Low-depth random Clifford circuits for quantum coding
  against Pauli noise using a tensor-network decoder}},\ }\href@noop {}
  {\bibfield  {journal} {\bibinfo  {journal} {arXiv preprint arXiv:2212.05071}\
  } (\bibinfo {year} {2022})}\BibitemShut {NoStop}%
\bibitem [{\citenamefont {Farrelly}\ \emph
  {et~al.}(2022{\natexlab{b}})\citenamefont {Farrelly}, \citenamefont
  {Milicevic}, \citenamefont {Harris}, \citenamefont {McMahon},\ and\
  \citenamefont {Stace}}]{farrelly2022parallel}%
  \BibitemOpen
  \bibfield  {author} {\bibinfo {author} {\bibfnamefont {T.}~\bibnamefont
  {Farrelly}}, \bibinfo {author} {\bibfnamefont {N.}~\bibnamefont {Milicevic}},
  \bibinfo {author} {\bibfnamefont {R.~J.}\ \bibnamefont {Harris}}, \bibinfo
  {author} {\bibfnamefont {N.~A.}\ \bibnamefont {McMahon}},\ and\ \bibinfo
  {author} {\bibfnamefont {T.~M.}\ \bibnamefont {Stace}},\ }\bibfield  {title}
  {\bibinfo {title} {Parallel decoding of multiple logical qubits in
  tensor-network codes},\ }\href@noop {} {\bibfield  {journal} {\bibinfo
  {journal} {Physical Review A}\ }\textbf {\bibinfo {volume} {105}},\ \bibinfo
  {pages} {052446} (\bibinfo {year} {2022}{\natexlab{b}})}\BibitemShut
  {NoStop}%
\bibitem [{\citenamefont {{Google Quantum AI}}(2023)}]{google2023suppressing}%
  \BibitemOpen
  \bibfield  {author} {\bibinfo {author} {\bibnamefont {{Google Quantum AI}}},\
  }\bibfield  {title} {\bibinfo {title} {Suppressing quantum errors by scaling
  a surface code logical qubit},\ }\href@noop {} {\bibfield  {journal}
  {\bibinfo  {journal} {Nature}\ }\textbf {\bibinfo {volume} {614}},\ \bibinfo
  {pages} {676} (\bibinfo {year} {2023})}\BibitemShut {NoStop}%
\bibitem [{\citenamefont {Piveteau}\ \emph
  {et~al.}(2024{\natexlab{a}})\citenamefont {Piveteau}, \citenamefont {Chubb},\
  and\ \citenamefont {Renes}}]{piveteau2023tensor}%
  \BibitemOpen
  \bibfield  {author} {\bibinfo {author} {\bibfnamefont {C.}~\bibnamefont
  {Piveteau}}, \bibinfo {author} {\bibfnamefont {C.~T.}\ \bibnamefont
  {Chubb}},\ and\ \bibinfo {author} {\bibfnamefont {J.~M.}\ \bibnamefont
  {Renes}},\ }\bibfield  {title} {\bibinfo {title} {Tensor-network decoding
  beyond 2d},\ }\href@noop {} {\bibfield  {journal} {\bibinfo  {journal} {PRX
  Quantum}\ }\textbf {\bibinfo {volume} {5}},\ \bibinfo {pages} {040303}
  (\bibinfo {year} {2024}{\natexlab{a}})}\BibitemShut {NoStop}%
\bibitem [{\citenamefont {Shutty}\ \emph {et~al.}(2024)\citenamefont {Shutty},
  \citenamefont {Newman},\ and\ \citenamefont
  {Villalonga}}]{shutty2024efficient}%
  \BibitemOpen
  \bibfield  {author} {\bibinfo {author} {\bibfnamefont {N.}~\bibnamefont
  {Shutty}}, \bibinfo {author} {\bibfnamefont {M.}~\bibnamefont {Newman}},\
  and\ \bibinfo {author} {\bibfnamefont {B.}~\bibnamefont {Villalonga}},\
  }\bibfield  {title} {\bibinfo {title} {Efficient near-optimal decoding of the
  surface code through ensembling},\ }\href@noop {} {\bibfield  {journal}
  {\bibinfo  {journal} {arXiv preprint arXiv:2401.12434}\ } (\bibinfo {year}
  {2024})}\BibitemShut {NoStop}%
\bibitem [{\citenamefont {Kukliansky}\ and\ \citenamefont
  {Lackey}(2024)}]{kukliansky2024quantum}%
  \BibitemOpen
  \bibfield  {author} {\bibinfo {author} {\bibfnamefont {A.}~\bibnamefont
  {Kukliansky}}\ and\ \bibinfo {author} {\bibfnamefont {B.}~\bibnamefont
  {Lackey}},\ }\bibfield  {title} {\bibinfo {title} {Quantum circuit tensors
  and enumerators with applications to quantum fault tolerance},\ }\href@noop
  {} {\bibfield  {journal} {\bibinfo  {journal} {arXiv preprint
  arXiv:2405.19643}\ } (\bibinfo {year} {2024})}\BibitemShut {NoStop}%
\bibitem [{\citenamefont {Cao}\ and\ \citenamefont
  {Lackey}(2023)}]{cao2023quantum}%
  \BibitemOpen
  \bibfield  {author} {\bibinfo {author} {\bibfnamefont {C.}~\bibnamefont
  {Cao}}\ and\ \bibinfo {author} {\bibfnamefont {B.}~\bibnamefont {Lackey}},\
  }\bibfield  {title} {\bibinfo {title} {Quantum weight enumerators and tensor
  networks},\ }\href@noop {} {\bibfield  {journal} {\bibinfo  {journal} {IEEE
  Transactions on Information Theory}\ }\textbf {\bibinfo {volume} {70}},\
  \bibinfo {pages} {3512} (\bibinfo {year} {2023})}\BibitemShut {NoStop}%
\bibitem [{\citenamefont {Harris}\ \emph {et~al.}(2018)\citenamefont {Harris},
  \citenamefont {McMahon}, \citenamefont {Brennen},\ and\ \citenamefont
  {Stace}}]{harris2018calderbank}%
  \BibitemOpen
  \bibfield  {author} {\bibinfo {author} {\bibfnamefont {R.~J.}\ \bibnamefont
  {Harris}}, \bibinfo {author} {\bibfnamefont {N.~A.}\ \bibnamefont {McMahon}},
  \bibinfo {author} {\bibfnamefont {G.~K.}\ \bibnamefont {Brennen}},\ and\
  \bibinfo {author} {\bibfnamefont {T.~M.}\ \bibnamefont {Stace}},\ }\bibfield
  {title} {\bibinfo {title} {Calderbank-shor-steane holographic quantum
  error-correcting codes},\ }\href@noop {} {\bibfield  {journal} {\bibinfo
  {journal} {Physical Review A}\ }\textbf {\bibinfo {volume} {98}},\ \bibinfo
  {pages} {052301} (\bibinfo {year} {2018})}\BibitemShut {NoStop}%
\bibitem [{\citenamefont {Harris}\ \emph {et~al.}(2020)\citenamefont {Harris},
  \citenamefont {Coupe}, \citenamefont {McMahon}, \citenamefont {Brennen},\
  and\ \citenamefont {Stace}}]{harris2020decoding}%
  \BibitemOpen
  \bibfield  {author} {\bibinfo {author} {\bibfnamefont {R.~J.}\ \bibnamefont
  {Harris}}, \bibinfo {author} {\bibfnamefont {E.}~\bibnamefont {Coupe}},
  \bibinfo {author} {\bibfnamefont {N.~A.}\ \bibnamefont {McMahon}}, \bibinfo
  {author} {\bibfnamefont {G.~K.}\ \bibnamefont {Brennen}},\ and\ \bibinfo
  {author} {\bibfnamefont {T.~M.}\ \bibnamefont {Stace}},\ }\bibfield  {title}
  {\bibinfo {title} {Decoding holographic codes with an integer optimization
  decoder},\ }\href@noop {} {\bibfield  {journal} {\bibinfo  {journal}
  {Physical Review A}\ }\textbf {\bibinfo {volume} {102}},\ \bibinfo {pages}
  {062417} (\bibinfo {year} {2020})}\BibitemShut {NoStop}%
\bibitem [{\citenamefont {Bao}\ and\ \citenamefont
  {Naskar}(2022)}]{bao2022code}%
  \BibitemOpen
  \bibfield  {author} {\bibinfo {author} {\bibfnamefont {N.}~\bibnamefont
  {Bao}}\ and\ \bibinfo {author} {\bibfnamefont {J.}~\bibnamefont {Naskar}},\
  }\bibfield  {title} {\bibinfo {title} {Code properties of the holographic
  sierpinski triangle},\ }\href@noop {} {\bibfield  {journal} {\bibinfo
  {journal} {Physical Review D}\ }\textbf {\bibinfo {volume} {106}},\ \bibinfo
  {pages} {126006} (\bibinfo {year} {2022})}\BibitemShut {NoStop}%
\bibitem [{\citenamefont {Fan}\ \emph {et~al.}(2024{\natexlab{b}})\citenamefont
  {Fan}, \citenamefont {Steinberg}, \citenamefont {Jahn}, \citenamefont {Cao},\
  and\ \citenamefont {Feld}}]{fan2024overcoming}%
  \BibitemOpen
  \bibfield  {author} {\bibinfo {author} {\bibfnamefont {J.}~\bibnamefont
  {Fan}}, \bibinfo {author} {\bibfnamefont {M.}~\bibnamefont {Steinberg}},
  \bibinfo {author} {\bibfnamefont {A.}~\bibnamefont {Jahn}}, \bibinfo {author}
  {\bibfnamefont {C.}~\bibnamefont {Cao}},\ and\ \bibinfo {author}
  {\bibfnamefont {S.}~\bibnamefont {Feld}},\ }\bibfield  {title} {\bibinfo
  {title} {Overcoming the zero-rate hashing bound with holographic quantum
  error correction},\ }\href@noop {} {\bibfield  {journal} {\bibinfo  {journal}
  {arXiv preprint arXiv:2408.06232}\ } (\bibinfo {year}
  {2024}{\natexlab{b}})}\BibitemShut {NoStop}%
\bibitem [{\citenamefont {Tuckett}\ \emph {et~al.}(2019)\citenamefont
  {Tuckett}, \citenamefont {Darmawan}, \citenamefont {Chubb}, \citenamefont
  {Bravyi}, \citenamefont {Bartlett},\ and\ \citenamefont
  {Flammia}}]{tuckett2019tailoring}%
  \BibitemOpen
  \bibfield  {author} {\bibinfo {author} {\bibfnamefont {D.~K.}\ \bibnamefont
  {Tuckett}}, \bibinfo {author} {\bibfnamefont {A.~S.}\ \bibnamefont
  {Darmawan}}, \bibinfo {author} {\bibfnamefont {C.~T.}\ \bibnamefont {Chubb}},
  \bibinfo {author} {\bibfnamefont {S.}~\bibnamefont {Bravyi}}, \bibinfo
  {author} {\bibfnamefont {S.~D.}\ \bibnamefont {Bartlett}},\ and\ \bibinfo
  {author} {\bibfnamefont {S.~T.}\ \bibnamefont {Flammia}},\ }\bibfield
  {title} {\bibinfo {title} {Tailoring surface codes for highly biased noise},\
  }\href {https://doi.org/10.1103/PhysRevX.9.041031} {\bibfield  {journal}
  {\bibinfo  {journal} {Phys. Rev. X}\ }\textbf {\bibinfo {volume} {9}},\
  \bibinfo {pages} {041031} (\bibinfo {year} {2019})}\BibitemShut {NoStop}%
\bibitem [{\citenamefont {Xu}\ \emph {et~al.}(2023)\citenamefont {Xu},
  \citenamefont {Mannucci}, \citenamefont {Seif}, \citenamefont {Kubica},
  \citenamefont {Flammia},\ and\ \citenamefont {Jiang}}]{qian2023tailoredxzzx}%
  \BibitemOpen
  \bibfield  {author} {\bibinfo {author} {\bibfnamefont {Q.}~\bibnamefont
  {Xu}}, \bibinfo {author} {\bibfnamefont {N.}~\bibnamefont {Mannucci}},
  \bibinfo {author} {\bibfnamefont {A.}~\bibnamefont {Seif}}, \bibinfo {author}
  {\bibfnamefont {A.}~\bibnamefont {Kubica}}, \bibinfo {author} {\bibfnamefont
  {S.~T.}\ \bibnamefont {Flammia}},\ and\ \bibinfo {author} {\bibfnamefont
  {L.}~\bibnamefont {Jiang}},\ }\bibfield  {title} {\bibinfo {title} {Tailored
  xzzx codes for biased noise},\ }\href
  {https://doi.org/10.1103/PhysRevResearch.5.013035} {\bibfield  {journal}
  {\bibinfo  {journal} {Phys. Rev. Res.}\ }\textbf {\bibinfo {volume} {5}},\
  \bibinfo {pages} {013035} (\bibinfo {year} {2023})}\BibitemShut {NoStop}%
\bibitem [{\citenamefont {Tuckett}\ \emph {et~al.}(2018)\citenamefont
  {Tuckett}, \citenamefont {Bartlett},\ and\ \citenamefont
  {Flammia}}]{tuckett2018ultrahigh}%
  \BibitemOpen
  \bibfield  {author} {\bibinfo {author} {\bibfnamefont {D.~K.}\ \bibnamefont
  {Tuckett}}, \bibinfo {author} {\bibfnamefont {S.~D.}\ \bibnamefont
  {Bartlett}},\ and\ \bibinfo {author} {\bibfnamefont {S.~T.}\ \bibnamefont
  {Flammia}},\ }\bibfield  {title} {\bibinfo {title} {Ultrahigh error threshold
  for surface codes with biased noise},\ }\href@noop {} {\bibfield  {journal}
  {\bibinfo  {journal} {Physical review letters}\ }\textbf {\bibinfo {volume}
  {120}},\ \bibinfo {pages} {050505} (\bibinfo {year} {2018})}\BibitemShut
  {NoStop}%
\bibitem [{\citenamefont {Bonilla~Ataides}\ \emph {et~al.}(2021)\citenamefont
  {Bonilla~Ataides}, \citenamefont {Tuckett}, \citenamefont {Bartlett},
  \citenamefont {Flammia},\ and\ \citenamefont {Brown}}]{bonilla2021xzzx}%
  \BibitemOpen
  \bibfield  {author} {\bibinfo {author} {\bibfnamefont {J.~P.}\ \bibnamefont
  {Bonilla~Ataides}}, \bibinfo {author} {\bibfnamefont {D.~K.}\ \bibnamefont
  {Tuckett}}, \bibinfo {author} {\bibfnamefont {S.~D.}\ \bibnamefont
  {Bartlett}}, \bibinfo {author} {\bibfnamefont {S.~T.}\ \bibnamefont
  {Flammia}},\ and\ \bibinfo {author} {\bibfnamefont {B.~J.}\ \bibnamefont
  {Brown}},\ }\bibfield  {title} {\bibinfo {title} {The xzzx surface code},\
  }\href@noop {} {\bibfield  {journal} {\bibinfo  {journal} {Nature
  communications}\ }\textbf {\bibinfo {volume} {12}},\ \bibinfo {pages} {2172}
  (\bibinfo {year} {2021})}\BibitemShut {NoStop}%
\bibitem [{\citenamefont {Dua}\ \emph {et~al.}(2024)\citenamefont {Dua},
  \citenamefont {Kubica}, \citenamefont {Jiang}, \citenamefont {Flammia},\ and\
  \citenamefont {Gullans}}]{dua2024clifforddeformed}%
  \BibitemOpen
  \bibfield  {author} {\bibinfo {author} {\bibfnamefont {A.}~\bibnamefont
  {Dua}}, \bibinfo {author} {\bibfnamefont {A.}~\bibnamefont {Kubica}},
  \bibinfo {author} {\bibfnamefont {L.}~\bibnamefont {Jiang}}, \bibinfo
  {author} {\bibfnamefont {S.~T.}\ \bibnamefont {Flammia}},\ and\ \bibinfo
  {author} {\bibfnamefont {M.~J.}\ \bibnamefont {Gullans}},\ }\bibfield
  {title} {\bibinfo {title} {Clifford-deformed surface codes},\ }\href
  {https://doi.org/10.1103/PRXQuantum.5.010347} {\bibfield  {journal} {\bibinfo
   {journal} {PRX Quantum}\ }\textbf {\bibinfo {volume} {5}},\ \bibinfo {pages}
  {010347} (\bibinfo {year} {2024})}\BibitemShut {NoStop}%
\bibitem [{\citenamefont {Kobayashi}\ \emph {et~al.}(2024)\citenamefont
  {Kobayashi}, \citenamefont {Manabe}, \citenamefont {White}, \citenamefont
  {Farrelly}, \citenamefont {Modi},\ and\ \citenamefont
  {Stace}}]{kobayashi2024tensor}%
  \BibitemOpen
  \bibfield  {author} {\bibinfo {author} {\bibfnamefont {F.}~\bibnamefont
  {Kobayashi}}, \bibinfo {author} {\bibfnamefont {H.}~\bibnamefont {Manabe}},
  \bibinfo {author} {\bibfnamefont {G.~A.}\ \bibnamefont {White}}, \bibinfo
  {author} {\bibfnamefont {T.}~\bibnamefont {Farrelly}}, \bibinfo {author}
  {\bibfnamefont {K.}~\bibnamefont {Modi}},\ and\ \bibinfo {author}
  {\bibfnamefont {T.~M.}\ \bibnamefont {Stace}},\ }\bibfield  {title} {\bibinfo
  {title} {Tensor-network decoders for process tensor descriptions of
  non-markovian noise},\ }\href@noop {} {\bibfield  {journal} {\bibinfo
  {journal} {arXiv preprint arXiv:2412.13739}\ } (\bibinfo {year}
  {2024})}\BibitemShut {NoStop}%
\bibitem [{\citenamefont {Piveteau}\ \emph
  {et~al.}(2024{\natexlab{b}})\citenamefont {Piveteau}, \citenamefont {Chubb},\
  and\ \citenamefont {Renes}}]{piveteau2024tensor}%
  \BibitemOpen
  \bibfield  {author} {\bibinfo {author} {\bibfnamefont {C.}~\bibnamefont
  {Piveteau}}, \bibinfo {author} {\bibfnamefont {C.~T.}\ \bibnamefont
  {Chubb}},\ and\ \bibinfo {author} {\bibfnamefont {J.~M.}\ \bibnamefont
  {Renes}},\ }\bibfield  {title} {\bibinfo {title} {Tensor-network decoding
  beyond 2d},\ }\href@noop {} {\bibfield  {journal} {\bibinfo  {journal} {PRX
  Quantum}\ }\textbf {\bibinfo {volume} {5}},\ \bibinfo {pages} {040303}
  (\bibinfo {year} {2024}{\natexlab{b}})}\BibitemShut {NoStop}%
\bibitem [{\citenamefont {Battistel}\ \emph {et~al.}(2023)\citenamefont
  {Battistel}, \citenamefont {Chamberland}, \citenamefont {Johar},
  \citenamefont {Overwater}, \citenamefont {Sebastiano}, \citenamefont
  {Skoric}, \citenamefont {Ueno},\ and\ \citenamefont
  {Usman}}]{battistel2023real}%
  \BibitemOpen
  \bibfield  {author} {\bibinfo {author} {\bibfnamefont {F.}~\bibnamefont
  {Battistel}}, \bibinfo {author} {\bibfnamefont {C.}~\bibnamefont
  {Chamberland}}, \bibinfo {author} {\bibfnamefont {K.}~\bibnamefont {Johar}},
  \bibinfo {author} {\bibfnamefont {R.~W.}\ \bibnamefont {Overwater}}, \bibinfo
  {author} {\bibfnamefont {F.}~\bibnamefont {Sebastiano}}, \bibinfo {author}
  {\bibfnamefont {L.}~\bibnamefont {Skoric}}, \bibinfo {author} {\bibfnamefont
  {Y.}~\bibnamefont {Ueno}},\ and\ \bibinfo {author} {\bibfnamefont
  {M.}~\bibnamefont {Usman}},\ }\bibfield  {title} {\bibinfo {title} {Real-time
  decoding for fault-tolerant quantum computing: Progress, challenges and
  outlook},\ }\href@noop {} {\bibfield  {journal} {\bibinfo  {journal} {Nano
  Futures}\ }\textbf {\bibinfo {volume} {7}},\ \bibinfo {pages} {032003}
  (\bibinfo {year} {2023})}\BibitemShut {NoStop}%
\bibitem [{\citenamefont {Temme}\ \emph {et~al.}(2017)\citenamefont {Temme},
  \citenamefont {Bravyi},\ and\ \citenamefont {Gambetta}}]{temme2017error}%
  \BibitemOpen
  \bibfield  {author} {\bibinfo {author} {\bibfnamefont {K.}~\bibnamefont
  {Temme}}, \bibinfo {author} {\bibfnamefont {S.}~\bibnamefont {Bravyi}},\ and\
  \bibinfo {author} {\bibfnamefont {J.~M.}\ \bibnamefont {Gambetta}},\
  }\bibfield  {title} {\bibinfo {title} {Error mitigation for short-depth
  quantum circuits},\ }\href@noop {} {\bibfield  {journal} {\bibinfo  {journal}
  {Physical review letters}\ }\textbf {\bibinfo {volume} {119}},\ \bibinfo
  {pages} {180509} (\bibinfo {year} {2017})}\BibitemShut {NoStop}%
\bibitem [{\citenamefont {Li}\ and\ \citenamefont
  {Benjamin}(2017)}]{li2017efficient}%
  \BibitemOpen
  \bibfield  {author} {\bibinfo {author} {\bibfnamefont {Y.}~\bibnamefont
  {Li}}\ and\ \bibinfo {author} {\bibfnamefont {S.~C.}\ \bibnamefont
  {Benjamin}},\ }\bibfield  {title} {\bibinfo {title} {Efficient variational
  quantum simulator incorporating active error minimization},\ }\href@noop {}
  {\bibfield  {journal} {\bibinfo  {journal} {Physical Review X}\ }\textbf
  {\bibinfo {volume} {7}},\ \bibinfo {pages} {021050} (\bibinfo {year}
  {2017})}\BibitemShut {NoStop}%
\bibitem [{\citenamefont {Guo}\ and\ \citenamefont
  {Yang}(2022)}]{guo2022quantum}%
  \BibitemOpen
  \bibfield  {author} {\bibinfo {author} {\bibfnamefont {Y.}~\bibnamefont
  {Guo}}\ and\ \bibinfo {author} {\bibfnamefont {S.}~\bibnamefont {Yang}},\
  }\bibfield  {title} {\bibinfo {title} {Quantum error mitigation via matrix
  product operators},\ }\href@noop {} {\bibfield  {journal} {\bibinfo
  {journal} {PRX Quantum}\ }\textbf {\bibinfo {volume} {3}},\ \bibinfo {pages}
  {040313} (\bibinfo {year} {2022})}\BibitemShut {NoStop}%
\bibitem [{\citenamefont {Tepaske}\ and\ \citenamefont
  {Luitz}(2023)}]{tepaske2023compressed}%
  \BibitemOpen
  \bibfield  {author} {\bibinfo {author} {\bibfnamefont {M.~S.}\ \bibnamefont
  {Tepaske}}\ and\ \bibinfo {author} {\bibfnamefont {D.~J.}\ \bibnamefont
  {Luitz}},\ }\bibfield  {title} {\bibinfo {title} {Compressed quantum error
  mitigation},\ }\href@noop {} {\bibfield  {journal} {\bibinfo  {journal}
  {Physical Review B}\ }\textbf {\bibinfo {volume} {107}},\ \bibinfo {pages}
  {L201114} (\bibinfo {year} {2023})}\BibitemShut {NoStop}%
\bibitem [{\citenamefont {Filippov}\ \emph {et~al.}(2023)\citenamefont
  {Filippov}, \citenamefont {Leahy}, \citenamefont {Rossi},\ and\ \citenamefont
  {Garc{\'\i}a-P{\'e}rez}}]{filippov2023scalable}%
  \BibitemOpen
  \bibfield  {author} {\bibinfo {author} {\bibfnamefont {S.}~\bibnamefont
  {Filippov}}, \bibinfo {author} {\bibfnamefont {M.}~\bibnamefont {Leahy}},
  \bibinfo {author} {\bibfnamefont {M.~A.}\ \bibnamefont {Rossi}},\ and\
  \bibinfo {author} {\bibfnamefont {G.}~\bibnamefont {Garc{\'\i}a-P{\'e}rez}},\
  }\bibfield  {title} {\bibinfo {title} {Scalable tensor-network error
  mitigation for near-term quantum computing},\ }\href@noop {} {\bibfield
  {journal} {\bibinfo  {journal} {arXiv preprint arXiv:2307.11740}\ } (\bibinfo
  {year} {2023})}\BibitemShut {NoStop}%
\bibitem [{\citenamefont {Filippov}\ \emph {et~al.}(2024)\citenamefont
  {Filippov}, \citenamefont {Maniscalco},\ and\ \citenamefont
  {Garc{\'\i}a-P{\'e}rez}}]{filippov2024scalability}%
  \BibitemOpen
  \bibfield  {author} {\bibinfo {author} {\bibfnamefont {S.~N.}\ \bibnamefont
  {Filippov}}, \bibinfo {author} {\bibfnamefont {S.}~\bibnamefont
  {Maniscalco}},\ and\ \bibinfo {author} {\bibfnamefont {G.}~\bibnamefont
  {Garc{\'\i}a-P{\'e}rez}},\ }\bibfield  {title} {\bibinfo {title} {Scalability
  of quantum error mitigation techniques: from utility to advantage},\
  }\href@noop {} {\bibfield  {journal} {\bibinfo  {journal} {arXiv preprint
  arXiv:2403.13542}\ } (\bibinfo {year} {2024})}\BibitemShut {NoStop}%
\bibitem [{\citenamefont {Fischer}\ \emph {et~al.}(2024)\citenamefont {Fischer}
  \emph {et~al.}}]{fischer2024dynamical}%
  \BibitemOpen
  \bibfield  {author} {\bibinfo {author} {\bibfnamefont {L.~E.}\ \bibnamefont
  {Fischer}} \emph {et~al.},\ }\bibfield  {title} {\bibinfo {title} {Dynamical
  simulations of many-body quantum chaos on a quantum computer},\ }\href@noop
  {} {\bibfield  {journal} {\bibinfo  {journal} {arXiv preprint
  arXiv:2411.00765}\ } (\bibinfo {year} {2024})}\BibitemShut {NoStop}%
\bibitem [{\citenamefont {Piveteau}\ \emph {et~al.}(2021)\citenamefont
  {Piveteau}, \citenamefont {Sutter}, \citenamefont {Bravyi}, \citenamefont
  {Gambetta},\ and\ \citenamefont {Temme}}]{piveteau2021error}%
  \BibitemOpen
  \bibfield  {author} {\bibinfo {author} {\bibfnamefont {C.}~\bibnamefont
  {Piveteau}}, \bibinfo {author} {\bibfnamefont {D.}~\bibnamefont {Sutter}},
  \bibinfo {author} {\bibfnamefont {S.}~\bibnamefont {Bravyi}}, \bibinfo
  {author} {\bibfnamefont {J.~M.}\ \bibnamefont {Gambetta}},\ and\ \bibinfo
  {author} {\bibfnamefont {K.}~\bibnamefont {Temme}},\ }\bibfield  {title}
  {\bibinfo {title} {Error mitigation for universal gates on encoded qubits},\
  }\href@noop {} {\bibfield  {journal} {\bibinfo  {journal} {Physical review
  letters}\ }\textbf {\bibinfo {volume} {127}},\ \bibinfo {pages} {200505}
  (\bibinfo {year} {2021})}\BibitemShut {NoStop}%
\bibitem [{\citenamefont {Novikov}\ \emph {et~al.}(2015)\citenamefont
  {Novikov}, \citenamefont {Podoprikhin}, \citenamefont {Osokin},\ and\
  \citenamefont {Vetrov}}]{novikov2015tensorizing}%
  \BibitemOpen
  \bibfield  {author} {\bibinfo {author} {\bibfnamefont {A.}~\bibnamefont
  {Novikov}}, \bibinfo {author} {\bibfnamefont {D.}~\bibnamefont
  {Podoprikhin}}, \bibinfo {author} {\bibfnamefont {A.}~\bibnamefont
  {Osokin}},\ and\ \bibinfo {author} {\bibfnamefont {D.~P.}\ \bibnamefont
  {Vetrov}},\ }\bibfield  {title} {\bibinfo {title} {Tensorizing neural
  networks},\ }\href@noop {} {\bibfield  {journal} {\bibinfo  {journal}
  {Advances in neural information processing systems}\ }\textbf {\bibinfo
  {volume} {28}} (\bibinfo {year} {2015})}\BibitemShut {NoStop}%
\bibitem [{\citenamefont {Novikov}\ \emph {et~al.}(2016)\citenamefont
  {Novikov}, \citenamefont {Trofimov},\ and\ \citenamefont
  {Oseledets}}]{novikov2016exponential}%
  \BibitemOpen
  \bibfield  {author} {\bibinfo {author} {\bibfnamefont {A.}~\bibnamefont
  {Novikov}}, \bibinfo {author} {\bibfnamefont {M.}~\bibnamefont {Trofimov}},\
  and\ \bibinfo {author} {\bibfnamefont {I.}~\bibnamefont {Oseledets}},\
  }\bibfield  {title} {\bibinfo {title} {Exponential machines},\ }\href@noop {}
  {\bibfield  {journal} {\bibinfo  {journal} {arXiv preprint arXiv:1605.03795}\
  } (\bibinfo {year} {2016})}\BibitemShut {NoStop}%
\bibitem [{\citenamefont {Stoudenmire}\ and\ \citenamefont
  {Schwab}(2016)}]{stoudenmire2016supervised}%
  \BibitemOpen
  \bibfield  {author} {\bibinfo {author} {\bibfnamefont {E.}~\bibnamefont
  {Stoudenmire}}\ and\ \bibinfo {author} {\bibfnamefont {D.~J.}\ \bibnamefont
  {Schwab}},\ }\bibfield  {title} {\bibinfo {title} {Supervised learning with
  tensor networks},\ }\href@noop {} {\bibfield  {journal} {\bibinfo  {journal}
  {Advances in neural information processing systems}\ }\textbf {\bibinfo
  {volume} {29}} (\bibinfo {year} {2016})}\BibitemShut {NoStop}%
\bibitem [{\citenamefont {Chen}\ \emph
  {et~al.}(2018{\natexlab{b}})\citenamefont {Chen}, \citenamefont {Cheng},
  \citenamefont {Xie}, \citenamefont {Wang},\ and\ \citenamefont
  {Xiang}}]{chen2018equivalence}%
  \BibitemOpen
  \bibfield  {author} {\bibinfo {author} {\bibfnamefont {J.}~\bibnamefont
  {Chen}}, \bibinfo {author} {\bibfnamefont {S.}~\bibnamefont {Cheng}},
  \bibinfo {author} {\bibfnamefont {H.}~\bibnamefont {Xie}}, \bibinfo {author}
  {\bibfnamefont {L.}~\bibnamefont {Wang}},\ and\ \bibinfo {author}
  {\bibfnamefont {T.}~\bibnamefont {Xiang}},\ }\bibfield  {title} {\bibinfo
  {title} {{Equivalence of restricted Boltzmann machines and tensor network
  states}},\ }\href@noop {} {\bibfield  {journal} {\bibinfo  {journal}
  {Physical Review B}\ }\textbf {\bibinfo {volume} {97}},\ \bibinfo {pages}
  {085104} (\bibinfo {year} {2018}{\natexlab{b}})}\BibitemShut {NoStop}%
\bibitem [{\citenamefont {Li}\ \emph {et~al.}(2021)\citenamefont {Li},
  \citenamefont {Pan}, \citenamefont {Zhou},\ and\ \citenamefont
  {Zhang}}]{li2021boltzmann}%
  \BibitemOpen
  \bibfield  {author} {\bibinfo {author} {\bibfnamefont {S.}~\bibnamefont
  {Li}}, \bibinfo {author} {\bibfnamefont {F.}~\bibnamefont {Pan}}, \bibinfo
  {author} {\bibfnamefont {P.}~\bibnamefont {Zhou}},\ and\ \bibinfo {author}
  {\bibfnamefont {P.}~\bibnamefont {Zhang}},\ }\bibfield  {title} {\bibinfo
  {title} {Boltzmann machines as two-dimensional tensor networks},\ }\href@noop
  {} {\bibfield  {journal} {\bibinfo  {journal} {Physical Review B}\ }\textbf
  {\bibinfo {volume} {104}},\ \bibinfo {pages} {075154} (\bibinfo {year}
  {2021})}\BibitemShut {NoStop}%
\bibitem [{\citenamefont {Han}\ \emph {et~al.}(2018)\citenamefont {Han},
  \citenamefont {Wang}, \citenamefont {Fan}, \citenamefont {Wang},\ and\
  \citenamefont {Zhang}}]{han2018unsupervised}%
  \BibitemOpen
  \bibfield  {author} {\bibinfo {author} {\bibfnamefont {Z.-Y.}\ \bibnamefont
  {Han}}, \bibinfo {author} {\bibfnamefont {J.}~\bibnamefont {Wang}}, \bibinfo
  {author} {\bibfnamefont {H.}~\bibnamefont {Fan}}, \bibinfo {author}
  {\bibfnamefont {L.}~\bibnamefont {Wang}},\ and\ \bibinfo {author}
  {\bibfnamefont {P.}~\bibnamefont {Zhang}},\ }\bibfield  {title} {\bibinfo
  {title} {Unsupervised generative modeling using matrix product states},\
  }\href@noop {} {\bibfield  {journal} {\bibinfo  {journal} {Physical Review
  X}\ }\textbf {\bibinfo {volume} {8}},\ \bibinfo {pages} {031012} (\bibinfo
  {year} {2018})}\BibitemShut {NoStop}%
\bibitem [{\citenamefont {Liu}\ \emph {et~al.}(2023{\natexlab{b}})\citenamefont
  {Liu}, \citenamefont {Li}, \citenamefont {Zhang},\ and\ \citenamefont
  {Zhang}}]{liu2023tensor}%
  \BibitemOpen
  \bibfield  {author} {\bibinfo {author} {\bibfnamefont {J.}~\bibnamefont
  {Liu}}, \bibinfo {author} {\bibfnamefont {S.}~\bibnamefont {Li}}, \bibinfo
  {author} {\bibfnamefont {J.}~\bibnamefont {Zhang}},\ and\ \bibinfo {author}
  {\bibfnamefont {P.}~\bibnamefont {Zhang}},\ }\bibfield  {title} {\bibinfo
  {title} {Tensor networks for unsupervised machine learning},\ }\href@noop {}
  {\bibfield  {journal} {\bibinfo  {journal} {Physical Review E}\ }\textbf
  {\bibinfo {volume} {107}},\ \bibinfo {pages} {L012103} (\bibinfo {year}
  {2023}{\natexlab{b}})}\BibitemShut {NoStop}%
\bibitem [{\citenamefont {Glasser}\ \emph {et~al.}(2019)\citenamefont
  {Glasser}, \citenamefont {Sweke}, \citenamefont {Pancotti}, \citenamefont
  {Eisert},\ and\ \citenamefont {Cirac}}]{glasser2019expressive}%
  \BibitemOpen
  \bibfield  {author} {\bibinfo {author} {\bibfnamefont {I.}~\bibnamefont
  {Glasser}}, \bibinfo {author} {\bibfnamefont {R.}~\bibnamefont {Sweke}},
  \bibinfo {author} {\bibfnamefont {N.}~\bibnamefont {Pancotti}}, \bibinfo
  {author} {\bibfnamefont {J.}~\bibnamefont {Eisert}},\ and\ \bibinfo {author}
  {\bibfnamefont {I.}~\bibnamefont {Cirac}},\ }\bibfield  {title} {\bibinfo
  {title} {Expressive power of tensor-network factorizations for probabilistic
  modeling},\ }\href@noop {} {\bibfield  {journal} {\bibinfo  {journal}
  {Advances in neural information processing systems}\ }\textbf {\bibinfo
  {volume} {32}} (\bibinfo {year} {2019})}\BibitemShut {NoStop}%
\bibitem [{\citenamefont {Cheng}\ \emph {et~al.}(2019)\citenamefont {Cheng},
  \citenamefont {Wang}, \citenamefont {Xiang},\ and\ \citenamefont
  {Zhang}}]{cheng2019tree}%
  \BibitemOpen
  \bibfield  {author} {\bibinfo {author} {\bibfnamefont {S.}~\bibnamefont
  {Cheng}}, \bibinfo {author} {\bibfnamefont {L.}~\bibnamefont {Wang}},
  \bibinfo {author} {\bibfnamefont {T.}~\bibnamefont {Xiang}},\ and\ \bibinfo
  {author} {\bibfnamefont {P.}~\bibnamefont {Zhang}},\ }\bibfield  {title}
  {\bibinfo {title} {Tree tensor networks for generative modeling},\
  }\href@noop {} {\bibfield  {journal} {\bibinfo  {journal} {Physical Review
  B}\ }\textbf {\bibinfo {volume} {99}},\ \bibinfo {pages} {155131} (\bibinfo
  {year} {2019})}\BibitemShut {NoStop}%
\bibitem [{\citenamefont {Vieijra}\ \emph {et~al.}(2022)\citenamefont
  {Vieijra}, \citenamefont {Vanderstraeten},\ and\ \citenamefont
  {Verstraete}}]{vieijra2022generative}%
  \BibitemOpen
  \bibfield  {author} {\bibinfo {author} {\bibfnamefont {T.}~\bibnamefont
  {Vieijra}}, \bibinfo {author} {\bibfnamefont {L.}~\bibnamefont
  {Vanderstraeten}},\ and\ \bibinfo {author} {\bibfnamefont {F.}~\bibnamefont
  {Verstraete}},\ }\bibfield  {title} {\bibinfo {title} {Generative modeling
  with projected entangled-pair states},\ }\href@noop {} {\bibfield  {journal}
  {\bibinfo  {journal} {arXiv preprint arXiv:2202.08177}\ } (\bibinfo {year}
  {2022})}\BibitemShut {NoStop}%
\bibitem [{\citenamefont {Reyes}\ and\ \citenamefont
  {Stoudenmire}(2021)}]{reyes2021multi}%
  \BibitemOpen
  \bibfield  {author} {\bibinfo {author} {\bibfnamefont {J.~A.}\ \bibnamefont
  {Reyes}}\ and\ \bibinfo {author} {\bibfnamefont {E.~M.}\ \bibnamefont
  {Stoudenmire}},\ }\bibfield  {title} {\bibinfo {title} {Multi-scale tensor
  network architecture for machine learning},\ }\href@noop {} {\bibfield
  {journal} {\bibinfo  {journal} {Machine Learning: Science and Technology}\
  }\textbf {\bibinfo {volume} {2}},\ \bibinfo {pages} {035036} (\bibinfo {year}
  {2021})}\BibitemShut {NoStop}%
\bibitem [{\citenamefont {Huggins}\ \emph {et~al.}(2019)\citenamefont
  {Huggins}, \citenamefont {Patil}, \citenamefont {Mitchell}, \citenamefont
  {Whaley},\ and\ \citenamefont {Stoudenmire}}]{huggins2019towards}%
  \BibitemOpen
  \bibfield  {author} {\bibinfo {author} {\bibfnamefont {W.}~\bibnamefont
  {Huggins}}, \bibinfo {author} {\bibfnamefont {P.}~\bibnamefont {Patil}},
  \bibinfo {author} {\bibfnamefont {B.}~\bibnamefont {Mitchell}}, \bibinfo
  {author} {\bibfnamefont {K.~B.}\ \bibnamefont {Whaley}},\ and\ \bibinfo
  {author} {\bibfnamefont {E.~M.}\ \bibnamefont {Stoudenmire}},\ }\bibfield
  {title} {\bibinfo {title} {Towards quantum machine learning with tensor
  networks},\ }\href@noop {} {\bibfield  {journal} {\bibinfo  {journal}
  {Quantum Science and technology}\ }\textbf {\bibinfo {volume} {4}},\ \bibinfo
  {pages} {024001} (\bibinfo {year} {2019})}\BibitemShut {NoStop}%
\bibitem [{\citenamefont {Wall}\ \emph {et~al.}(2021)\citenamefont {Wall},
  \citenamefont {Abernathy},\ and\ \citenamefont
  {Quiroz}}]{wall2021generative}%
  \BibitemOpen
  \bibfield  {author} {\bibinfo {author} {\bibfnamefont {M.~L.}\ \bibnamefont
  {Wall}}, \bibinfo {author} {\bibfnamefont {M.~R.}\ \bibnamefont
  {Abernathy}},\ and\ \bibinfo {author} {\bibfnamefont {G.}~\bibnamefont
  {Quiroz}},\ }\bibfield  {title} {\bibinfo {title} {Generative machine
  learning with tensor networks: Benchmarks on near-term quantum computers},\
  }\href@noop {} {\bibfield  {journal} {\bibinfo  {journal} {Physical Review
  Research}\ }\textbf {\bibinfo {volume} {3}},\ \bibinfo {pages} {023010}
  (\bibinfo {year} {2021})}\BibitemShut {NoStop}%
\bibitem [{\citenamefont {Grant}\ \emph {et~al.}(2018)\citenamefont {Grant},
  \citenamefont {Benedetti}, \citenamefont {Cao}, \citenamefont {Hallam},
  \citenamefont {Lockhart}, \citenamefont {Stojevic}, \citenamefont {Green},\
  and\ \citenamefont {Severini}}]{grant2018hierarchical}%
  \BibitemOpen
  \bibfield  {author} {\bibinfo {author} {\bibfnamefont {E.}~\bibnamefont
  {Grant}}, \bibinfo {author} {\bibfnamefont {M.}~\bibnamefont {Benedetti}},
  \bibinfo {author} {\bibfnamefont {S.}~\bibnamefont {Cao}}, \bibinfo {author}
  {\bibfnamefont {A.}~\bibnamefont {Hallam}}, \bibinfo {author} {\bibfnamefont
  {J.}~\bibnamefont {Lockhart}}, \bibinfo {author} {\bibfnamefont
  {V.}~\bibnamefont {Stojevic}}, \bibinfo {author} {\bibfnamefont {A.~G.}\
  \bibnamefont {Green}},\ and\ \bibinfo {author} {\bibfnamefont
  {S.}~\bibnamefont {Severini}},\ }\bibfield  {title} {\bibinfo {title}
  {Hierarchical quantum classifiers},\ }\href@noop {} {\bibfield  {journal}
  {\bibinfo  {journal} {npj Quantum Information}\ }\textbf {\bibinfo {volume}
  {4}},\ \bibinfo {pages} {65} (\bibinfo {year} {2018})}\BibitemShut {NoStop}%
\bibitem [{\citenamefont {Lazzarin}\ \emph {et~al.}(2022)\citenamefont
  {Lazzarin}, \citenamefont {Galli},\ and\ \citenamefont
  {Prati}}]{lazzarin2022multi}%
  \BibitemOpen
  \bibfield  {author} {\bibinfo {author} {\bibfnamefont {M.}~\bibnamefont
  {Lazzarin}}, \bibinfo {author} {\bibfnamefont {D.~E.}\ \bibnamefont
  {Galli}},\ and\ \bibinfo {author} {\bibfnamefont {E.}~\bibnamefont {Prati}},\
  }\bibfield  {title} {\bibinfo {title} {Multi-class quantum classifiers with
  tensor network circuits for quantum phase recognition},\ }\href@noop {}
  {\bibfield  {journal} {\bibinfo  {journal} {Physics Letters A}\ }\textbf
  {\bibinfo {volume} {434}},\ \bibinfo {pages} {128056} (\bibinfo {year}
  {2022})}\BibitemShut {NoStop}%
\bibitem [{\citenamefont {Rieser}\ \emph {et~al.}(2023)\citenamefont {Rieser},
  \citenamefont {K{\"o}ster},\ and\ \citenamefont {Raulf}}]{rieser2023tensor}%
  \BibitemOpen
  \bibfield  {author} {\bibinfo {author} {\bibfnamefont {H.-M.}\ \bibnamefont
  {Rieser}}, \bibinfo {author} {\bibfnamefont {F.}~\bibnamefont {K{\"o}ster}},\
  and\ \bibinfo {author} {\bibfnamefont {A.~P.}\ \bibnamefont {Raulf}},\
  }\bibfield  {title} {\bibinfo {title} {Tensor networks for quantum machine
  learning},\ }\href@noop {} {\bibfield  {journal} {\bibinfo  {journal}
  {Proceedings of the Royal Society A}\ }\textbf {\bibinfo {volume} {479}},\
  \bibinfo {pages} {20230218} (\bibinfo {year} {2023})}\BibitemShut {NoStop}%
\bibitem [{\citenamefont {Zhao}\ and\ \citenamefont
  {Gao}(2021)}]{zhao2021analyzing}%
  \BibitemOpen
  \bibfield  {author} {\bibinfo {author} {\bibfnamefont {C.}~\bibnamefont
  {Zhao}}\ and\ \bibinfo {author} {\bibfnamefont {X.-S.}\ \bibnamefont {Gao}},\
  }\bibfield  {title} {\bibinfo {title} {Analyzing the barren plateau
  phenomenon in training quantum neural networks with the zx-calculus},\
  }\href@noop {} {\bibfield  {journal} {\bibinfo  {journal} {Quantum}\ }\textbf
  {\bibinfo {volume} {5}},\ \bibinfo {pages} {466} (\bibinfo {year}
  {2021})}\BibitemShut {NoStop}%
\bibitem [{\citenamefont {Pesah}\ \emph {et~al.}(2021)\citenamefont {Pesah},
  \citenamefont {Cerezo}, \citenamefont {Wang}, \citenamefont {Volkoff},
  \citenamefont {Sornborger},\ and\ \citenamefont {Coles}}]{pesah2021absence}%
  \BibitemOpen
  \bibfield  {author} {\bibinfo {author} {\bibfnamefont {A.}~\bibnamefont
  {Pesah}}, \bibinfo {author} {\bibfnamefont {M.}~\bibnamefont {Cerezo}},
  \bibinfo {author} {\bibfnamefont {S.}~\bibnamefont {Wang}}, \bibinfo {author}
  {\bibfnamefont {T.}~\bibnamefont {Volkoff}}, \bibinfo {author} {\bibfnamefont
  {A.~T.}\ \bibnamefont {Sornborger}},\ and\ \bibinfo {author} {\bibfnamefont
  {P.~J.}\ \bibnamefont {Coles}},\ }\bibfield  {title} {\bibinfo {title}
  {Absence of barren plateaus in quantum convolutional neural networks},\
  }\href@noop {} {\bibfield  {journal} {\bibinfo  {journal} {Physical Review
  X}\ }\textbf {\bibinfo {volume} {11}},\ \bibinfo {pages} {041011} (\bibinfo
  {year} {2021})}\BibitemShut {NoStop}%
\bibitem [{\citenamefont {Liao}\ \emph {et~al.}(2023)\citenamefont {Liao},
  \citenamefont {Convy}, \citenamefont {Yang},\ and\ \citenamefont
  {Whaley}}]{liao2023decohering}%
  \BibitemOpen
  \bibfield  {author} {\bibinfo {author} {\bibfnamefont {H.}~\bibnamefont
  {Liao}}, \bibinfo {author} {\bibfnamefont {I.}~\bibnamefont {Convy}},
  \bibinfo {author} {\bibfnamefont {Z.}~\bibnamefont {Yang}},\ and\ \bibinfo
  {author} {\bibfnamefont {K.~B.}\ \bibnamefont {Whaley}},\ }\bibfield  {title}
  {\bibinfo {title} {Decohering tensor network quantum machine learning
  models},\ }\href@noop {} {\bibfield  {journal} {\bibinfo  {journal} {Quantum
  Machine Intelligence}\ }\textbf {\bibinfo {volume} {5}},\ \bibinfo {pages}
  {7} (\bibinfo {year} {2023})}\BibitemShut {NoStop}%
\bibitem [{\citenamefont {Cong}\ \emph {et~al.}(2019)\citenamefont {Cong},
  \citenamefont {Choi},\ and\ \citenamefont {Lukin}}]{cong2019quantum}%
  \BibitemOpen
  \bibfield  {author} {\bibinfo {author} {\bibfnamefont {I.}~\bibnamefont
  {Cong}}, \bibinfo {author} {\bibfnamefont {S.}~\bibnamefont {Choi}},\ and\
  \bibinfo {author} {\bibfnamefont {M.~D.}\ \bibnamefont {Lukin}},\ }\bibfield
  {title} {\bibinfo {title} {Quantum convolutional neural networks},\
  }\href@noop {} {\bibfield  {journal} {\bibinfo  {journal} {Nature Physics}\
  }\textbf {\bibinfo {volume} {15}},\ \bibinfo {pages} {1273} (\bibinfo {year}
  {2019})}\BibitemShut {NoStop}%
\bibitem [{\citenamefont {Bermejo}\ \emph {et~al.}(2024)\citenamefont
  {Bermejo}, \citenamefont {Braccia}, \citenamefont {Rudolph}, \citenamefont
  {Holmes}, \citenamefont {Cincio},\ and\ \citenamefont
  {Cerezo}}]{bermejo2024quantum}%
  \BibitemOpen
  \bibfield  {author} {\bibinfo {author} {\bibfnamefont {P.}~\bibnamefont
  {Bermejo}}, \bibinfo {author} {\bibfnamefont {P.}~\bibnamefont {Braccia}},
  \bibinfo {author} {\bibfnamefont {M.~S.}\ \bibnamefont {Rudolph}}, \bibinfo
  {author} {\bibfnamefont {Z.}~\bibnamefont {Holmes}}, \bibinfo {author}
  {\bibfnamefont {L.}~\bibnamefont {Cincio}},\ and\ \bibinfo {author}
  {\bibfnamefont {M.}~\bibnamefont {Cerezo}},\ }\bibfield  {title} {\bibinfo
  {title} {Quantum convolutional neural networks are (effectively) classically
  simulable},\ }\href@noop {} {\bibfield  {journal} {\bibinfo  {journal} {arXiv
  preprint arXiv:2408.12739}\ } (\bibinfo {year} {2024})}\BibitemShut {NoStop}%
\bibitem [{\citenamefont {Dilip}\ \emph {et~al.}(2022)\citenamefont {Dilip},
  \citenamefont {Liu}, \citenamefont {Smith},\ and\ \citenamefont
  {Pollmann}}]{dilip2022data}%
  \BibitemOpen
  \bibfield  {author} {\bibinfo {author} {\bibfnamefont {R.}~\bibnamefont
  {Dilip}}, \bibinfo {author} {\bibfnamefont {Y.-J.}\ \bibnamefont {Liu}},
  \bibinfo {author} {\bibfnamefont {A.}~\bibnamefont {Smith}},\ and\ \bibinfo
  {author} {\bibfnamefont {F.}~\bibnamefont {Pollmann}},\ }\bibfield  {title}
  {\bibinfo {title} {Data compression for quantum machine learning},\
  }\href@noop {} {\bibfield  {journal} {\bibinfo  {journal} {Physical Review
  Research}\ }\textbf {\bibinfo {volume} {4}},\ \bibinfo {pages} {043007}
  (\bibinfo {year} {2022})}\BibitemShut {NoStop}%
\bibitem [{\citenamefont {Iaconis}\ and\ \citenamefont
  {Johri}(2023)}]{iaconis2023tensor}%
  \BibitemOpen
  \bibfield  {author} {\bibinfo {author} {\bibfnamefont {J.}~\bibnamefont
  {Iaconis}}\ and\ \bibinfo {author} {\bibfnamefont {S.}~\bibnamefont
  {Johri}},\ }\bibfield  {title} {\bibinfo {title} {Tensor network based
  efficient quantum data loading of images},\ }\href@noop {} {\bibfield
  {journal} {\bibinfo  {journal} {arXiv preprint arXiv:2310.05897}\ } (\bibinfo
  {year} {2023})}\BibitemShut {NoStop}%
\bibitem [{\citenamefont {Dborin}\ \emph {et~al.}(2022)\citenamefont {Dborin},
  \citenamefont {Barratt}, \citenamefont {Wimalaweera}, \citenamefont
  {Wright},\ and\ \citenamefont {Green}}]{dborin2022matrix}%
  \BibitemOpen
  \bibfield  {author} {\bibinfo {author} {\bibfnamefont {J.}~\bibnamefont
  {Dborin}}, \bibinfo {author} {\bibfnamefont {F.}~\bibnamefont {Barratt}},
  \bibinfo {author} {\bibfnamefont {V.}~\bibnamefont {Wimalaweera}}, \bibinfo
  {author} {\bibfnamefont {L.}~\bibnamefont {Wright}},\ and\ \bibinfo {author}
  {\bibfnamefont {A.~G.}\ \bibnamefont {Green}},\ }\bibfield  {title} {\bibinfo
  {title} {Matrix product state pre-training for quantum machine learning},\
  }\href@noop {} {\bibfield  {journal} {\bibinfo  {journal} {Quantum Science
  and Technology}\ }\textbf {\bibinfo {volume} {7}},\ \bibinfo {pages} {035014}
  (\bibinfo {year} {2022})}\BibitemShut {NoStop}%
\bibitem [{\citenamefont {Rudolph}\ \emph {et~al.}(2022)\citenamefont
  {Rudolph}, \citenamefont {Miller}, \citenamefont {Motlagh}, \citenamefont
  {Chen}, \citenamefont {Acharya},\ and\ \citenamefont
  {Perdomo-Ortiz}}]{rudolph2022synergy}%
  \BibitemOpen
  \bibfield  {author} {\bibinfo {author} {\bibfnamefont {M.~S.}\ \bibnamefont
  {Rudolph}}, \bibinfo {author} {\bibfnamefont {J.}~\bibnamefont {Miller}},
  \bibinfo {author} {\bibfnamefont {D.}~\bibnamefont {Motlagh}}, \bibinfo
  {author} {\bibfnamefont {J.}~\bibnamefont {Chen}}, \bibinfo {author}
  {\bibfnamefont {A.}~\bibnamefont {Acharya}},\ and\ \bibinfo {author}
  {\bibfnamefont {A.}~\bibnamefont {Perdomo-Ortiz}},\ }\bibfield  {title}
  {\bibinfo {title} {Synergy between quantum circuits and tensor networks:
  Short-cutting the race to practical quantum advantage},\ }\href@noop {}
  {\bibfield  {journal} {\bibinfo  {journal} {arXiv preprint arXiv:2208.13673}\
  } (\bibinfo {year} {2022})}\BibitemShut {NoStop}%
\bibitem [{\citenamefont {Khan}\ \emph {et~al.}(2023)\citenamefont {Khan},
  \citenamefont {Clark},\ and\ \citenamefont {Tubman}}]{khan2023pre}%
  \BibitemOpen
  \bibfield  {author} {\bibinfo {author} {\bibfnamefont {A.}~\bibnamefont
  {Khan}}, \bibinfo {author} {\bibfnamefont {B.~K.}\ \bibnamefont {Clark}},\
  and\ \bibinfo {author} {\bibfnamefont {N.~M.}\ \bibnamefont {Tubman}},\
  }\bibfield  {title} {\bibinfo {title} {Pre-optimizing variational quantum
  eigensolvers with tensor networks},\ }\href@noop {} {\bibfield  {journal}
  {\bibinfo  {journal} {arXiv preprint arXiv:2310.12965}\ } (\bibinfo {year}
  {2023})}\BibitemShut {NoStop}%
\bibitem [{\citenamefont {Shin}\ \emph {et~al.}(2024)\citenamefont {Shin},
  \citenamefont {Teo},\ and\ \citenamefont {Jeong}}]{shin2024dequantizing}%
  \BibitemOpen
  \bibfield  {author} {\bibinfo {author} {\bibfnamefont {S.}~\bibnamefont
  {Shin}}, \bibinfo {author} {\bibfnamefont {Y.~S.}\ \bibnamefont {Teo}},\ and\
  \bibinfo {author} {\bibfnamefont {H.}~\bibnamefont {Jeong}},\ }\bibfield
  {title} {\bibinfo {title} {Dequantizing quantum machine learning models using
  tensor networks},\ }\href@noop {} {\bibfield  {journal} {\bibinfo  {journal}
  {Physical Review Research}\ }\textbf {\bibinfo {volume} {6}},\ \bibinfo
  {pages} {023218} (\bibinfo {year} {2024})}\BibitemShut {NoStop}%
\bibitem [{\citenamefont {de~Beaudrap}\ \emph {et~al.}(2020)\citenamefont
  {de~Beaudrap}, \citenamefont {Kissinger},\ and\ \citenamefont
  {Meichanetzidis}}]{de2020tensor}%
  \BibitemOpen
  \bibfield  {author} {\bibinfo {author} {\bibfnamefont {N.}~\bibnamefont
  {de~Beaudrap}}, \bibinfo {author} {\bibfnamefont {A.}~\bibnamefont
  {Kissinger}},\ and\ \bibinfo {author} {\bibfnamefont {K.}~\bibnamefont
  {Meichanetzidis}},\ }\bibfield  {title} {\bibinfo {title} {Tensor network
  rewriting strategies for satisfiability and counting},\ }\href@noop {}
  {\bibfield  {journal} {\bibinfo  {journal} {arXiv preprint arXiv:2004.06455}\
  } (\bibinfo {year} {2020})}\BibitemShut {NoStop}%
\bibitem [{\citenamefont {Rakovszky}\ \emph {et~al.}(2022)\citenamefont
  {Rakovszky}, \citenamefont {Von~Keyserlingk},\ and\ \citenamefont
  {Pollmann}}]{rakovszky2022dissipation}%
  \BibitemOpen
  \bibfield  {author} {\bibinfo {author} {\bibfnamefont {T.}~\bibnamefont
  {Rakovszky}}, \bibinfo {author} {\bibfnamefont {C.}~\bibnamefont
  {Von~Keyserlingk}},\ and\ \bibinfo {author} {\bibfnamefont {F.}~\bibnamefont
  {Pollmann}},\ }\bibfield  {title} {\bibinfo {title} {Dissipation-assisted
  operator evolution method for capturing hydrodynamic transport},\ }\href@noop
  {} {\bibfield  {journal} {\bibinfo  {journal} {Physical Review B}\ }\textbf
  {\bibinfo {volume} {105}},\ \bibinfo {pages} {075131} (\bibinfo {year}
  {2022})}\BibitemShut {NoStop}%
\bibitem [{\citenamefont {Tarabunga}\ \emph {et~al.}(2024)\citenamefont
  {Tarabunga}, \citenamefont {Tirrito}, \citenamefont {Ba{\~n}uls},\ and\
  \citenamefont {Dalmonte}}]{tarabunga2024nonstabilizerness}%
  \BibitemOpen
  \bibfield  {author} {\bibinfo {author} {\bibfnamefont {P.~S.}\ \bibnamefont
  {Tarabunga}}, \bibinfo {author} {\bibfnamefont {E.}~\bibnamefont {Tirrito}},
  \bibinfo {author} {\bibfnamefont {M.~C.}\ \bibnamefont {Ba{\~n}uls}},\ and\
  \bibinfo {author} {\bibfnamefont {M.}~\bibnamefont {Dalmonte}},\ }\bibfield
  {title} {\bibinfo {title} {Nonstabilizerness via matrix product states in the
  pauli basis},\ }\href@noop {} {\bibfield  {journal} {\bibinfo  {journal}
  {Physical Review Letters}\ }\textbf {\bibinfo {volume} {133}},\ \bibinfo
  {pages} {010601} (\bibinfo {year} {2024})}\BibitemShut {NoStop}%
\bibitem [{\citenamefont {Masot-Llima}\ and\ \citenamefont
  {Garcia-Saez}(2024)}]{masot2024stabilizer}%
  \BibitemOpen
  \bibfield  {author} {\bibinfo {author} {\bibfnamefont {S.}~\bibnamefont
  {Masot-Llima}}\ and\ \bibinfo {author} {\bibfnamefont {A.}~\bibnamefont
  {Garcia-Saez}},\ }\bibfield  {title} {\bibinfo {title} {Stabilizer tensor
  networks: universal quantum simulator on a basis of stabilizer states},\
  }\href@noop {} {\bibfield  {journal} {\bibinfo  {journal} {Physical Review
  Letters}\ }\textbf {\bibinfo {volume} {133}},\ \bibinfo {pages} {230601}
  (\bibinfo {year} {2024})}\BibitemShut {NoStop}%
\bibitem [{\citenamefont {Breuckmann}\ and\ \citenamefont
  {Eberhardt}(2021)}]{breuckmann2021}%
  \BibitemOpen
  \bibfield  {author} {\bibinfo {author} {\bibfnamefont {N.~P.}\ \bibnamefont
  {Breuckmann}}\ and\ \bibinfo {author} {\bibfnamefont {J.~N.}\ \bibnamefont
  {Eberhardt}},\ }\bibfield  {title} {\bibinfo {title} {Quantum low-density
  parity-check codes},\ }\href {https://doi.org/10.1103/PRXQuantum.2.040101}
  {\bibfield  {journal} {\bibinfo  {journal} {PRX Quantum}\ }\textbf {\bibinfo
  {volume} {2}},\ \bibinfo {pages} {040101} (\bibinfo {year}
  {2021})}\BibitemShut {NoStop}%
\bibitem [{\citenamefont {Panteleev}\ and\ \citenamefont
  {Kalachev}(2022)}]{panteleev2022asymptotically}%
  \BibitemOpen
  \bibfield  {author} {\bibinfo {author} {\bibfnamefont {P.}~\bibnamefont
  {Panteleev}}\ and\ \bibinfo {author} {\bibfnamefont {G.}~\bibnamefont
  {Kalachev}},\ }\bibfield  {title} {\bibinfo {title} {{Asymptotically good
  quantum and locally testable classical LDPC codes}},\ }in\ \href@noop {}
  {\emph {\bibinfo {booktitle} {Proceedings of the 54th Annual ACM SIGACT
  Symposium on Theory of Computing}}}\ (\bibinfo {year} {2022})\ pp.\ \bibinfo
  {pages} {375--388}\BibitemShut {NoStop}%
\bibitem [{\citenamefont {Bausch}\ \emph {et~al.}(2024)\citenamefont {Bausch},
  \citenamefont {Senior}, \citenamefont {Heras}, \citenamefont {Edlich},
  \citenamefont {Davies}, \citenamefont {Newman}, \citenamefont {Jones},
  \citenamefont {Satzinger}, \citenamefont {Niu}, \citenamefont {Blackwell}
  \emph {et~al.}}]{bausch2024learning}%
  \BibitemOpen
  \bibfield  {author} {\bibinfo {author} {\bibfnamefont {J.}~\bibnamefont
  {Bausch}}, \bibinfo {author} {\bibfnamefont {A.~W.}\ \bibnamefont {Senior}},
  \bibinfo {author} {\bibfnamefont {F.~J.}\ \bibnamefont {Heras}}, \bibinfo
  {author} {\bibfnamefont {T.}~\bibnamefont {Edlich}}, \bibinfo {author}
  {\bibfnamefont {A.}~\bibnamefont {Davies}}, \bibinfo {author} {\bibfnamefont
  {M.}~\bibnamefont {Newman}}, \bibinfo {author} {\bibfnamefont
  {C.}~\bibnamefont {Jones}}, \bibinfo {author} {\bibfnamefont
  {K.}~\bibnamefont {Satzinger}}, \bibinfo {author} {\bibfnamefont {M.~Y.}\
  \bibnamefont {Niu}}, \bibinfo {author} {\bibfnamefont {S.}~\bibnamefont
  {Blackwell}}, \emph {et~al.},\ }\bibfield  {title} {\bibinfo {title}
  {Learning high-accuracy error decoding for quantum processors},\ }\href@noop
  {} {\bibfield  {journal} {\bibinfo  {journal} {Nature}\ ,\ \bibinfo {pages}
  {1}} (\bibinfo {year} {2024})}\BibitemShut {NoStop}%
\bibitem [{\citenamefont {Zhou}\ \emph {et~al.}(2025)\citenamefont {Zhou},
  \citenamefont {Zhao}, \citenamefont {Nakaji}, \citenamefont {Lietz},
  \citenamefont {McCaskey}, \citenamefont {Patti},\ and\ \citenamefont
  {Chamberland}}]{zhou2025nvidia}%
  \BibitemOpen
  \bibfield  {author} {\bibinfo {author} {\bibfnamefont {H.~H.}\ \bibnamefont
  {Zhou}}, \bibinfo {author} {\bibfnamefont {C.}~\bibnamefont {Zhao}}, \bibinfo
  {author} {\bibfnamefont {K.}~\bibnamefont {Nakaji}}, \bibinfo {author}
  {\bibfnamefont {J.}~\bibnamefont {Lietz}}, \bibinfo {author} {\bibfnamefont
  {A.}~\bibnamefont {McCaskey}}, \bibinfo {author} {\bibfnamefont {T.~L.}\
  \bibnamefont {Patti}},\ and\ \bibinfo {author} {\bibfnamefont
  {C.}~\bibnamefont {Chamberland}},\ }\bibfield  {title} {\bibinfo {title}
  {{NVIDIA and QuEra Decode Quantum Errors with AI}},\ }\href
  {https://developer.nvidia.com/blog/nvidia-and-quera-decode-quantum-errors-with-ai/}
  {\bibfield  {journal} {\bibinfo  {journal} {NVIDIA Technical Blog}\ }
  (\bibinfo {year} {2025})}\BibitemShut {NoStop}%
\bibitem [{\citenamefont {Cerezo}\ \emph {et~al.}(2023)\citenamefont {Cerezo},
  \citenamefont {Larocca}, \citenamefont {Garc{\'\i}a-Mart{\'\i}n},
  \citenamefont {Diaz}, \citenamefont {Braccia}, \citenamefont {Fontana},
  \citenamefont {Rudolph}, \citenamefont {Bermejo}, \citenamefont {Ijaz},
  \citenamefont {Thanasilp} \emph {et~al.}}]{cerezo2023does}%
  \BibitemOpen
  \bibfield  {author} {\bibinfo {author} {\bibfnamefont {M.}~\bibnamefont
  {Cerezo}}, \bibinfo {author} {\bibfnamefont {M.}~\bibnamefont {Larocca}},
  \bibinfo {author} {\bibfnamefont {D.}~\bibnamefont
  {Garc{\'\i}a-Mart{\'\i}n}}, \bibinfo {author} {\bibfnamefont {N.~L.}\
  \bibnamefont {Diaz}}, \bibinfo {author} {\bibfnamefont {P.}~\bibnamefont
  {Braccia}}, \bibinfo {author} {\bibfnamefont {E.}~\bibnamefont {Fontana}},
  \bibinfo {author} {\bibfnamefont {M.~S.}\ \bibnamefont {Rudolph}}, \bibinfo
  {author} {\bibfnamefont {P.}~\bibnamefont {Bermejo}}, \bibinfo {author}
  {\bibfnamefont {A.}~\bibnamefont {Ijaz}}, \bibinfo {author} {\bibfnamefont
  {S.}~\bibnamefont {Thanasilp}}, \emph {et~al.},\ }\bibfield  {title}
  {\bibinfo {title} {Does provable absence of barren plateaus imply classical
  simulability? or, why we need to rethink variational quantum computing},\
  }\href@noop {} {\bibfield  {journal} {\bibinfo  {journal} {arXiv preprint
  arXiv:2312.09121}\ } (\bibinfo {year} {2023})}\BibitemShut {NoStop}%
\bibitem [{\citenamefont {Gil-Fuster}\ \emph {et~al.}(2024)\citenamefont
  {Gil-Fuster}, \citenamefont {Gyurik}, \citenamefont {P{\'e}rez-Salinas},\
  and\ \citenamefont {Dunjko}}]{gil2024relation}%
  \BibitemOpen
  \bibfield  {author} {\bibinfo {author} {\bibfnamefont {E.}~\bibnamefont
  {Gil-Fuster}}, \bibinfo {author} {\bibfnamefont {C.}~\bibnamefont {Gyurik}},
  \bibinfo {author} {\bibfnamefont {A.}~\bibnamefont {P{\'e}rez-Salinas}},\
  and\ \bibinfo {author} {\bibfnamefont {V.}~\bibnamefont {Dunjko}},\
  }\bibfield  {title} {\bibinfo {title} {On the relation between trainability
  and dequantization of variational quantum learning models},\ }\href@noop {}
  {\bibfield  {journal} {\bibinfo  {journal} {arXiv preprint arXiv:2406.07072}\
  } (\bibinfo {year} {2024})}\BibitemShut {NoStop}%
\end{thebibliography}%

\section*{Disclaimer}
This paper was prepared for informational purposes with contributions from the Global Technology Applied Research center of JPMorganChase. This paper is not a product of the Research Department of JPMorganChase or its affiliates. Neither JPMorganChase nor any of its affiliates makes any explicit or implied representation or warranty and none of them accept any liability in connection with this paper, including, without limitation, with respect to the completeness, accuracy, or reliability of the information contained herein and the potential legal, compliance, tax, or accounting effects thereof. This document is not intended as investment research or investment advice, or as a recommendation, offer, or solicitation for the purchase or sale of any security, financial instrument, financial product or service, or to be used in any way for evaluating the merits of participating in any transaction. The United States Government retains, and by accepting the article for publication, the publisher acknowledges that the United States Government retains, a nonexclusive, paid-up, irrevocable, worldwide license to publish or reproduce the published form of this work, or allow others to do so, for United States Government purposes.

\end{document}